\documentclass[aps,preprint,superscriptaddress,showpacs,nofootinbib,floatfix]{revtex4-1}

\usepackage{amsmath}
\usepackage{amssymb}
\usepackage{mathtools}
\usepackage{graphics}

\usepackage{natbib}

\usepackage{wrapfig}
\usepackage[caption=false]{subfig}

\usepackage[colorlinks=true,linkcolor=blue,linktoc=page,citecolor=blue,urlcolor=black]{hyperref}

\begin{document}

\title{Late-time quantum backreaction of a very light nonminimally coupled scalar}

\author{Dra\v{z}en~Glavan}
\email[]{Drazen.Glavan@fuw.edu.pl}

\affiliation{Institute for Theoretical Physics and Spinoza Institute,
Center for Extreme Matter and Emergent Phenomena, 
Science Faculty, Utrecht University,
Postbus 80.195, 3508 TD Utrecht, The Netherlands}

\affiliation{Institute of Theoretical Physics, Faculty of Physics, 
University of Warsaw, Pasteura 5, 02-093 Warsaw, Poland}

\author{Tomislav~Prokopec}
\email[]{T.Prokopec@uu.nl}

\affiliation{Institute for Theoretical Physics and Spinoza Institute,
Center for Extreme Matter and Emergent Phenomena, 
Science Faculty, Utrecht University,
Postbus 80.195, 3508 TD Utrecht, The Netherlands}

\author{Tomo~Takahashi}
\email[]{tomot@cc.saga-u.ac.jp}

\affiliation{Department of Physics, Saga University, \\ Saga 840-8502, Japan}

\date{\today}


\begin{abstract}

We investigate the backreaction of the quantum fluctuations of a very light
($m \!\lesssim\! H_{\text{today}}$) nonminimally coupled spectator scalar field
on the expansion dynamics of the Universe. The one-loop expectation
value of the energy momentum tensor of these fluctuations, as a measure
of the backreaction, is computed
throughout the expansion history from the early inflationary universe until
the onset of recent acceleration today. We show that, when the nonminimal 
coupling $\xi$
to Ricci curvature is negative ($\xi_c \!=\! 1/6$ corresponding to conformal 
coupling), the quantum backreaction grows exponentially during inflation,
such that it can grow large enough rather quickly (within a few hundred e-foldings)
to survive until late time and constitute a contribution of the
cosmological constant type of the right magnitude to appreciably alter 
the expansion dynamics.
The unique feature of this model is
in that, under rather generic assumptions, inflation provides natural 
explanation for the initial
conditions needed to explain the late-time accelerated expansion
of the Universe, making it a particularly attractive model of
dark energy.

\end{abstract}



\maketitle

\tableofcontents


\section{Introduction}
\label{sec: Introduction}

The recent accelerated expansion of the Universe and its cause is one of the 
most puzzling mysteries in cosmology and physics today. Since the observations 
of type Ia supernovae (SNeIa) reported by two 
groups~\cite{Perlmutter:1998np,Riess:1998cb},
a lot of observations have been accumulating to indicate that the Universe is 
well described by the $\Lambda$CDM model (for recent observational results, 
see {\it e.g.} Ref.~\cite{Adam:2015rua,Ade:2015xua}). Although the 
cosmological constant (CC) gives a good fit to the data and explains well the 
current cosmic acceleration,  future observations promise to provide much 
tighter constraints of the models. That motivated theorists to explore other 
possibilities, which generically came to be known as {\it dark energy} (DE)
(for a review on dark energy see {\it e.g.} Ref.~\cite{Amendola:2011bo}). 
The origin of dark energy could be matter whose 
properties mimic those of cosmological constant, but it could be also in 
the modification of gravity on cosmological scales~\cite{Clifton:2011jh}.
In fact, due to the intricate coupling between gravity and matter in some 
theories, it is not always possible to tell whether dark 
energy comes from a new kind of matter or from a modification of gravity.

In this work we examine the idea that dark energy originates
from the backreaction of quantum fluctuations originating in the primordial
inflationary universe.
The idea that the origin of dark energy can be linked to primordial inflation 
has not been widely explored (for an early attempt to link the cause of inflation with 
the cause of dark energy see Ref.~\cite{Peebles:1998qn}, for a more recent attempt
see~\cite{GarciaBellido:2011de}). Also, early
attempts~\cite{Kolb:2005da,Kolb:2005me} to link quantum fluctuations of 
the {\it inflaton} to dark energy turned out not to be correct \cite{Hirata:2005ei}.
Indeed, a careful one-loop calculation of the energy-momentum tensor from inflationary 
gravitons~\cite{Glavan:2013mra}, and an educated estimate
of the corresponding energy-momentum tensor from scalar cosmological 
perturbations~\cite{Glavan:2013mra} shows that, rather than contributing to 
dark energy, these inflationary perturbations contribute a tiny amount 
(about $10^{-13}$ of the critical density today) to dark matter.

Recently, the idea of relating quantum backreaction to dark energy has 
again drawn some attention, and here we give a brief overview.
In \cite{Glavan:2013mra} the minimally coupled massless spectator scalar was 
studied where it was found that the quantum backreaction in late-time matter era
scales just as nonrelativistic matter fluid driving the background expansion,
but its fraction is tiny, about $10^{-13}$ of the critical density. 
This ratio is determined by the same ratio reached by
the end of inflation, and it effectively freezes afterwards. 
The same was concluded for the contribution from gravitons.
The result for scalars was subsequently confirmed 
in \cite{Aoki:2014ita}. There it was also investigated what are the influences of 
possible pre-inflationary periods on the magnitude of late-time
quantum backreaction, and it was found it could be increased if additional
inflationary periods at a much higher (Planck) energy scale existed prior to
the standard one, but the late-time scaling cannot be changed.

However, when one considers the quantum backreaction from inflationary 
quantum fluctuations of a massless non-minimally 
coupled spectator scalar field, and when the 
relevant non-minimal coupling is negative (such that it gives a tachyonic 
mass to the scalar during inflation), then under reasonable conditions on 
inflation and non-minimal coupling, the scalar field can yield a large quantum 
backreaction at late times~\cite{Glavan:2014uga}, making it potentially a 
candidate for dark energy (the analysis performed in~\cite{Glavan:2014uga} 
is perturbative, and hence any statements about whether that is a reliable candidate 
for dark energy cannot be made). In this model the backreaction can grow
considerably during inflation (even to nonperturbative values), how fast depending
on the nonminimal coupling (this was already found in 
\cite{Finelli:2008zg,Finelli:2010sh} for more general slow-roll inflation),
 so its ratio to the background fluid
is greatly enhanced compared to the minimally coupled case. This ratio 
again freezes during radiation period, but starts to evolve again in matter period.

Another idea involves the late-time quantum 
backreaction from inflationary quantum fluctuations of a very light spectator scalar 
field~\cite{Ringeval:2010hf}, where the backreaction contributes
like a CC at late-time matter era. In order for this idea to work, the authors of 
Ref.~\cite{Ringeval:2010hf} needed to lower the scale of inflation and furthermore 
they needed a humongously large number of e-foldings of inflation 
($N_I\!\gtrsim\!10^{60}$).
In a more recent paper~\cite{Aoki:2014dqa} it was pointed out that the mechanism 
works as well for inflation at the grand-unified scale and that the required 
number of e-foldings is `only' about $10^{13}$. When compared with 
the original work~\cite{Ringeval:2010hf}, 
that was a significant improvement. In~\cite{Aoki:2014dqa} it was also studied how
some pre-inflationary periods can lead to lowering the requirements on the number
of e-foldings of this model where they managed to get it down to $N_I\!\sim\!240$
with the assumption of another Planck scale inflationary period preceding the 
GUT scale one.

Apart from various technical improvements, the goal of this paper is to 
study the late-time quantum backreaction from inflationary quantum 
fluctuations of scalar fields and its relation to dark energy in more 
general models, without relying on any pre-inflationary physics, 
and one of the important result of this work is the 
realization that the simple addition of a small non-minimal coupling to 
the Ricci scalar can completely alleviate the constraint on the length of inflation.
We show that the late time one-loop quantum backreaction from a 
nonminimally coupled, light spectator scalar field can be a good candidate for 
dark energy. 
In our model conditions on inflation are completely relaxed: inflation can occur 
at the grand unified scale and it can last as little as hundreds of e-foldings.

The model of a very light, nonminimally coupled spectator scalar was studied 
and proposed as a dark energy model very soon after the discovery of the 
recent accelerated expansion of the universe by Parker and 
Raval~\cite{Parker:1999td,Parker:1999ac,Parker:1999fc,Parker:2000pr,
Parker:2001ws}. In those works the main effect derives from the ultraviolet 
quantum fluctuations, as opposed to the work presented here where the 
main contribution to the energy-momentum tensor comes from the 
infrared (super-Hubble) quantum fluctuations. We provide a more detailed 
comparison of our work with 
that of Parker and Raval in section~\ref{sec: Discussion and outlook}.

Quantum fluctuations of fields are generally nonvanishing, so we expect
them to contribute as corrections to the classical Einstein's equations. 
This statement can be neatly summarized by the following equation,
\begin{equation}
G_{\mu\nu} = 8\pi G_{\! N} 
	\Bigl[ T_{\mu\nu}^{\text{cl}} 
	+ \langle \hat{T}_{\mu\nu}^{\text{Q}} \rangle \Bigr] \, ,
\label{semiclassicalGR}
\end{equation}
which can be recognized as a quantum corrected Einstein's equation
The two source terms on the right are the classical energy-momentum
tensor of matter fields, and the {\it quantum backreaction}, respectively.
The quantum backreaction is the expectation value of the energy-momentum
tensor operator of quantum fluctuations of matter fields, and the metric field.
The evidence that quantum fluctuations indeed interact gravitationally comes
from studying the fluctuations in the spectrum of the CMB, 
and ultimately attributing it to the 
spectrum of primordial inflationary quantum scalar fluctuations. As opposed
to the spatial variations of the quantum fluctuations, which ultimately
contribute to the CMB temperature fluctuations, here we study
the homogeneous nonvanishing contribution of quantum fluctuations.
We want to study the effects of the backreaction term in cosmology, in 
particular its influence on the dynamics of the large scale expansion
of the universe, and its possible connection to the dark energy problem. 
That is a rather ambitious task, and in this work we opt to address
less ambitious, but still very important questions:
\begin{itemize}
\item
Can the backreaction
every become large enough to influence the expansion dynamics?
\item
When does it become large and how it depends on the parameters
of the model and the expansion history?
\item
What is the behaviour of the
backreaction when it becomes large and does it tend to accelerate the expansion?
\end{itemize}

Since we are interested in modeling DE, we want the backreaction not
to spoil the previous expansion history. Therefore, its influence on the expansion
has to be negligibly small up until the recent onset of accelerated expansion.
Therefore, we study equation \eqref{semiclassicalGR} perturbatively,
in the sense that we do not consider the backreaction actually backreacting
on the classical evolution, since it is small by assumption
(we thoroughly check whether this assumption is satisfied
in different epochs in the history of the Universe). The formalism
appropriate for this study is the quantum field theory in curved space-time,
originating in the late 60's and early 
70's~\cite{Chernikov:1968zm,Parker:1969au,Parker:1971pt,{Zeldovich:1971mw}}, and by now a 
well established subject, covered in standard references
\cite{Birrell:1982ix,Fulling:1989nb,Mukhanov:2007zz,Parker:2009uva}. Of course, when the
backreaction becomes large, this approach breaks down, and a full
self-consistent solution is needed.

Here we study the backreaction from quantum fluctuations of a very light, 
nonminimally coupled, spectator scalar
field as they evolve from an initial state specified at the beginning of inflationary
period of our Universe, through the radiation and matter dominated
era, up until the onset of the late-time acceleration period 
(see Fig.~\ref{expansion history} for the schematic depiction of the 
expansion history). 
The leading order contribution to one-loop
expectation value of the energy-momentum tensor of these quantum fluctuations
is computed in each era, as a controlled expansion in small ratios of physical
parameters.


This paper is organized as follows. The following section introduces definitions
and conventions for the cosmological space-time. The third section presents
the scalar field model, outlines its quantization, and defines the main 
quantities to be computed -- the scalar field mode function and the
expectation value of the energy-momentum tensor. The representation
and approximations of the evolution of the mode function is given in 
Section \ref{sec: Evolution of mode function}. In Section 
\ref{sec: Energy density and pressure integrals}
the quantum backreaction energy density and pressure integrals
are analyzed from a general point of view, and the relevant contributions are identified. Section \ref{sec: Mode functions} is devoted to
calculating approximate mode functions on constant epsilon backgrounds,
in particular their expansion in the small mass limit. In 
section~\ref{sec: Energy density and pressure}
the dominant contributions to energy density and pressure of the backreaction
are evaluated. In the concluding section~\ref{sec: Discussion and outlook} 
we summarize the results and discuss their connection
to dark energy. An outline of the future work is also given.


\section{FLRW background}
\label{sec: FLRW background}

The line element of a $D$-dimensional, spatially flat 
Friedmann-Lema\^{i}tre-Robertson-Walker
(FLRW) space-time is given by
\begin{equation}
ds^2 = g_{\mu\nu} dx^\mu dx^\nu = -dt^2 + a^2(t) d\vec{x}^{\,2} 
	= a^2(\eta) \Bigl[ -d\eta^2 + d\vec{x}^{\,2} \Bigr] \, ,
\end{equation}
where $g_{\mu\nu}$ is the metric, $a$ is the scale factor, 
$t$ denotes physical (cosmological) time,
and $\eta$ conformal time. 
The physical and conformal time are related
via $dt = a d\eta$. In this work we prefer to perform computations
in conformal time, for which the metric is conformally flat, 
$g_{\mu\nu}=a^2(\eta) \eta_{\mu\nu}$, and
$\eta_{\mu\nu}=\text{diag}(-1,1,\dots,1)$ is the $D$-dimensional
Minkowski metric. All the expressions are written in $D$ dimensions
in order to facilitate dimensional regularization utilized in computing
quantum expectation values ($D\!=\!4$ is the number of physical 
space-time dimensions). We adopt the
natural units convention ($c=\hbar=1$), unless explicitly stated. The 
geometric conventions we use are
$\Gamma^\alpha_{\mu\nu} = \frac{1}{2}g^{\alpha\beta}
(\partial_\mu g_{\nu\beta} + \partial_\nu g_{\mu\beta} 
- \partial_\beta g_{\mu\nu})$ for Christoffel symbols, 
${R^\alpha}_{\mu\beta\nu} = \partial_\beta \Gamma^\alpha_{\mu\nu}
- \partial_\nu \Gamma^\alpha_{\mu\beta}
+ \Gamma^\alpha_{\beta\rho} \Gamma^\rho_{\mu\nu}
- \Gamma^\alpha_{\nu\rho} \Gamma^\rho_{\mu\beta}$
for the Riemann tensor, $R_{\mu\nu} = {R^\alpha}_{\mu\alpha\nu}$
for the Ricci tensor, and $R = {R^\mu}_{\mu}$ for the Ricci scalar.

The dynamics of the scale factor is governed by the Friedmann equations,
\begin{equation}
\Bigl( \frac{\mathcal{H}}{a} \Bigr)^2 
	= \frac{6}{(D \!-\! 2)(D \!-\! 1)} \times 
	\frac{8\pi G_{\!N}}{3} \sum_i \rho_i \, ,
\end{equation}
\begin{equation}
\frac{\mathcal{H}' \!-\! \mathcal{H}^2}{a^2}
	= \frac{2}{D \!-\! 2} \times (-4\pi G_{\!N})
		\sum_i (\rho_i + p_i) \, ,
\end{equation}
where $\rho_i$ and $p_i$ are the energy density and pressure of the 
$i$-th matter fluid, respectively, $\mathcal{H}\!=\!a'/a$ is the conformal
Hubble rate related to the physical one $H$ via $\mathcal{H}\!=\!aH$, and 
a prime denotes differentiation
with respect to conformal time. The (noninteracting) 
matter fluids each satisfy the conservation equation,
\begin{equation}
\rho_i' + (D\!-\!1)\mathcal{H}(\rho_i+p_i) = 0 \, .
\label{ConservationEq}
\end{equation}
They are usually assumed to
be ideal fluids, characterized by a linear equation of state,
\begin{equation}
p_i = w_i \rho_i \, ,
\end{equation}
with a constant equation of state parameter $w_i$. Using this equation
of state the conservation equation \eqref{ConservationEq} can be easily 
integrated to yield
the scaling of the fluid's energy density and pressure,
\begin{equation}
\rho_i = \frac{p_i}{w_i} = \rho_{i,0} \Bigl( \frac{a}{a_0} \Bigr)^{3(1+w_i)} \, ,
\end{equation}
where $\rho_{i,0} \!=\! \rho_i(\eta_0)$ and $a_0 \!=\! a(\eta_0).$

\begin{figure}[h!]
\includegraphics[width=10cm]{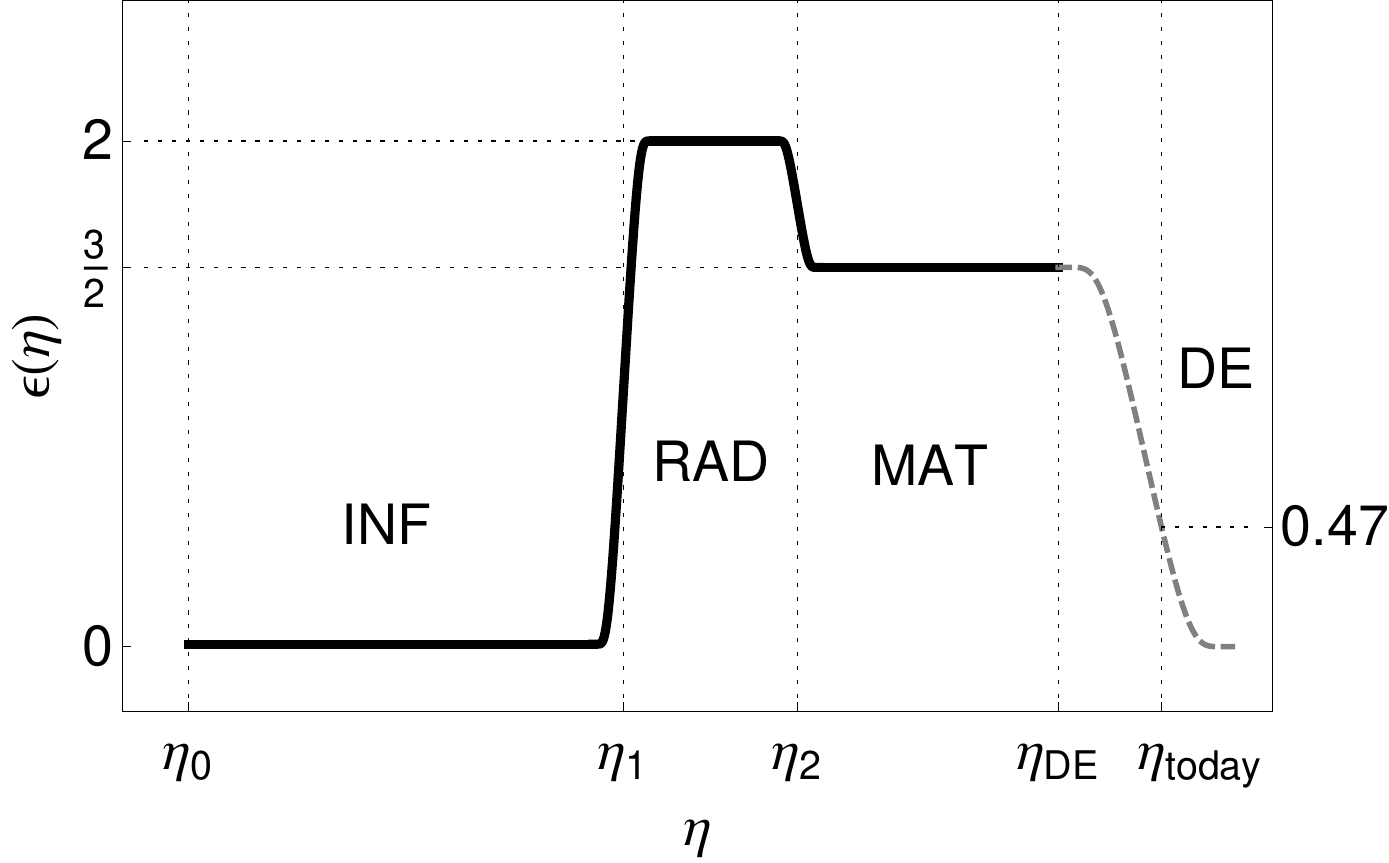}
\vskip-0.5cm
\caption{Schematic depiction of the expansion history in terms of 
$\epsilon \!=\! -\dot{H}/H^2$ 
parameter, consisting of three past cosmological eras -- inflation period,
radiation dominated period, and matter dominated period -- and the 
today we are in a dark energy dominated period.
Time $\eta_0$ corresponds to beginning of inflation, $n_1$ to
the end of inflation, $\eta_2$ to
the time of radiation-matter equality, and 
$\eta_{\text{DE}}$ to the onset of
dark energy domination period. Today we are at 
at $\epsilon_{\text{today}}\!=\!0.47$. The first two transitions are assumed to be
fast in the sense that the scale of the duration of the transition $\tau$ 
satisfies $\tau_i \!\ll\!1/\mathcal{H}_i$. 
We approximate the inflationary period by an
exact de Sitter one, $\epsilon_I\!=\!0$.} 
\label{expansion history}
\end{figure}

\begin{figure}[h!]
\includegraphics[width=10cm]{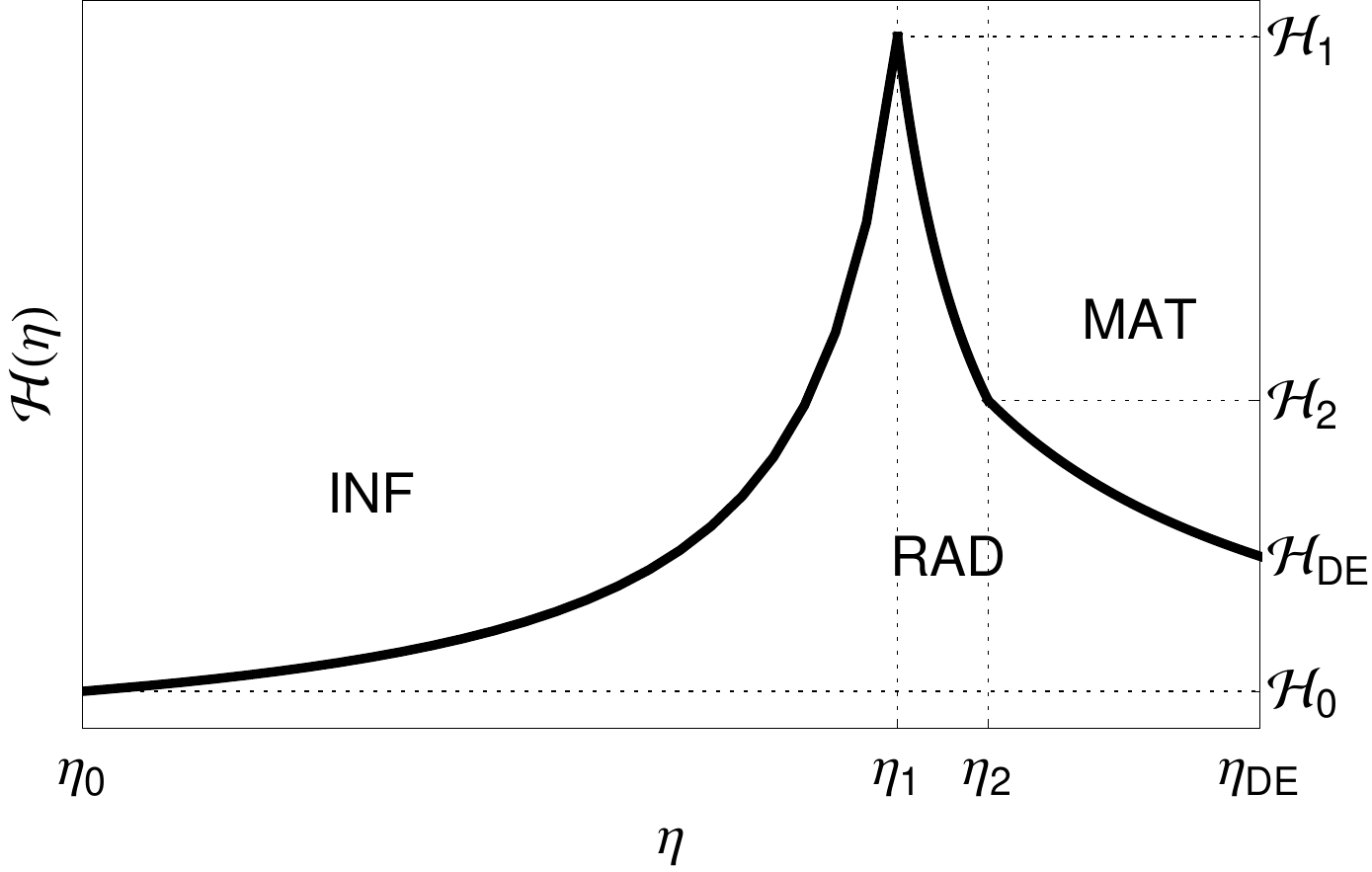}
\vskip-0.5cm
\caption{Schematic depiction of the evolution of the conformal Hubble rate
$\mathcal{H}$ throughout the expansion history of the universe.
 Hubble rates at the times of transitions satisfy
the hierarchy $\mathcal{H}_0 \!\ll\! \mathcal{H}_{\text{DE}} \!\ll\! \mathcal{H}_2 \!\ll\!
\mathcal{H}_1$. As in Fig.~\ref{expansion history}, time $\eta_0$ denotes
the beginning of inflation, time $\eta_1$ the end of inflation, 
$\eta_2$ the end of radiation and beginning of matter 
dominated period (radiation-matter equality), and time $\eta_{\text{DE}}$ denotes 
the onset of dark energy domination.}
\label{Hubble rate}
\end{figure}

The expansion history of our universe consists of a few eras in which 
one fluid dominates over the others that can be neglected.
In such regimes Friedmann equations are readily solved for the scale factor 
and the Hubble rate,
\begin{align}
a(\eta) ={}& a_0 \Bigl[ 1 + (\epsilon \!-\! 1) \mathcal{H}_0 (\eta \!-\! \eta_0) 
	\Bigr]^{\frac{1}{\epsilon-1}} \ , \qquad
\mathcal{H}(\eta) = \mathcal{H}_0 \Bigl( \frac{a}{a_0} \Bigr)^{1-\epsilon} \, ,
\end{align}
where $\mathcal{H}(\eta_0) \!=\! \mathcal{H}_0$, and the principal 
slow-roll parameter $\epsilon$ (which is a measure of the acceleration of the 
universe), generally defined as
\begin{equation}
\epsilon = - \frac{\dot{H}}{H^2} = 1 - \frac{\mathcal{H}'}{\mathcal{H}^2} \, ,
\end{equation}
is a constant during single fluid dominated era, related to the equation of state
parameter,
\begin{equation}
\epsilon = \frac{(D \!-\! 1)}{2}(1 \!+\! w) \, ,
\end{equation}
and we use it to characterize different cosmological eras.
A schematic depiction of the expansion history of the Universe assumed in this
work, in terms of $\epsilon$ parameter and the conformal Hubble rate is given
in Fig.~\ref{expansion history} and Fig.~\ref{Hubble rate}, respectively.
%




\section{Scalar field model}
\label{sec: Scalar field model}

This section introduces the model of a nonminimally coupled massive scalar
field on cosmological space-time studied in this work. The quantization of 
the scalar on FLRW backgrounds is outlined, and the main quantities to be calculated
-- expectation values of the energy-momentum tensor components -- 
are defined. The choice of initial state is discussed.

The action for the massive nonminimally coupled scalar on curved space-time is
\begin{equation}
S_\phi = \int\! d^D\!x \, \sqrt{-g} \, 
	\Biggl[ -\frac{1}{2} g^{\mu\nu} \partial_\mu \phi \partial_\nu \phi
		- \frac{1}{2}m^2\phi^2 - \frac{1}{2}\xi R \phi^2 \Biggr] \, ,
\label{action:massive nonminimal scalar}
\end{equation}
where $m$ is the scalar field mass, and $\xi$ is the nonminimal coupling constant.
Note that the sign convention we use implies 
$\xi_c = (D \!-\! 2) / [4(D \!-\! 1)] \xrightarrow{D\rightarrow4}1/6$ is the
conformal coupling.


\subsection{Quantization on FLRW}
\label{subsec: Quantization on FLRW}

On a FLRW background the Lagrangian density takes the form
\begin{equation}
\mathcal{L}_\phi = \frac{a^{D-2}}{2} \bigg[ 
	\bigl( \phi' \bigr)^2 - \bigl( \vec{\nabla}\phi \bigr)^2 
	- (ma)^2 \phi^2 - \xi a^2 R \, \phi^2 \bigg] \, .
\end{equation}
In order to quantize this field we need to switch to the Hamilton formalism.
First we define a canonically conjugate momentum,
\begin{equation}
\pi(x) = \frac{\partial \mathcal{L}_{\phi}}{\partial \phi'(x)} =
	a^{D-2}(\eta) \phi'(x) \, ,
\end{equation}
and then the Hamiltonian via the Legendre transform,
\begin{align}
H[\phi,\pi;\eta) ={}& \int \! d^{D-1}x \Bigl[ \pi(x) \phi'(x)
	- \mathcal{L}_\phi(x) \Bigr]
\nonumber \\
={}&	\frac{a^{D-2}}{2} \int\! d^{D-1}x
	\Bigl[ a^{4-2D} \pi^2 + \bigl( \vec{\nabla} \phi \bigr)^2
	+ (ma)^2 \phi^2 + \xi a^2 R \, \phi^2 \Bigr] \, .
\end{align}
Next we promote $\phi$ and $\pi$ to operators, and their Poisson brackets to 
commutators,
\begin{align}
& \bigl[ \hat{\phi}(\eta,\vec{x}) , \hat{\pi}(\eta,\vec{x}^{\,\prime}) \bigr]
	= i \delta^{D-1}(\vec{x} \!-\! \vec{x}^{\,\prime}) \, ,
\label{commPhi1}
\\
& \bigl[ \hat{\phi}(\eta,\vec{x}) , \hat{\phi}(\eta,\vec{x}^{\,\prime}) \bigr] = 0
	= \bigl[ \hat{\pi}(\eta,\vec{x}) , \hat{\pi}(\eta,\vec{x}^{\,\prime}) \bigr] \, .
\label{commPhi2}
\end{align}
The Hamiltonian operator defined as
\begin{equation}
\hat{H}(\eta) = H \bigl[ \hat{\phi}, \hat{\pi} ; \eta \bigr)
\end{equation}
now determines the dynamics via Heisenberg equations for the field operators,
\begin{align}
\hat{\phi}'(\eta,\vec{x}) ={}& i \bigl[ \hat{H}(\eta) , \hat{\pi}(\eta,\vec{x}) \bigr]
	= a^{2-D}(\eta) \, \hat{\pi}(\eta,\vec{x}) \, ,
\\
\hat{\pi}'(\eta,\vec{x}) ={}&  i \bigl[ \hat{H}(\eta) , \hat{\phi}(\eta,\vec{x}) \bigr]
	= a^{D-2}(\eta) \Bigl[ \vec{\nabla}^2 
	- (ma)^2 - \xi a^2(\eta) R(\eta) \Bigr] \hat{\phi}(\eta,\vec{x}) \, .
\end{align}
These two combined together yield the equation of motion for the field operator
\begin{equation}
\Biggl\{ \frac{\partial^2}{\partial\eta^2} 
	+ (D \!-\! 2)\mathcal{H} \frac{\partial}{\partial\eta}
	- \vec{\nabla}^2 + (ma)^2
	+ (D \!-\! 1) \xi \Bigl[ 2\mathcal{H}' + (D \!-\! 2)\mathcal{H}^2 \Bigr] \Biggr\}
	\hat{\phi}(x) = 0 \, ,
\label{Klein-Gordon equation:1}
\end{equation}
where $\vec{\nabla}^2 \!=\! \sum_i \partial_i^2$ is the
Laplace operator. This is just a Klein-Gordon equation,
\begin{equation}
\Bigl[ \square - m^2 - \xi R \Bigr] \hat{\phi} = 0 \, ,
\label{Klein-Gordon equation:2}
\end{equation}
specialized to FLRW, where $\square \!=\! g^{\mu\nu}\nabla_\mu \nabla_\nu
\!=\! (-g)^{-1/2}\partial_\mu [ (-g)^{1/2}g^{\mu\nu}\partial_\nu]$
is the d'Alembert operator. This equation is standardly analyzed in Fourier 
(comoving momentum) space,
\begin{equation}
\hat{\phi}(\eta,\vec{x}) =
	a^{\frac{2-D}{2}} \int \! 
	\frac{d^{D-1}k}{(2\pi)^{\frac{D-1}{2}}}
	\Biggl[ e^{i\vec{k}\cdot\vec{x}} U(\eta,k) \, \hat{b}(\vec{k})
	+ e^{-i\vec{k}\cdot\vec{x}} U^*(\eta,k) \, \hat{b}^{\dag}(\vec{k}) \Biggr]
	\, ,
\label{FourierTransform}
\end{equation}
where $\hat{b}(\vec{k})$ and $\hat{b}^\dag(\vec{k})$ are the annihilation
and creation operators, respectively, which satisfy the following commutation
relations,
\begin{align}
& \bigl[ \hat{b}(\vec{k}) , \hat{b}^{\dag}(\vec{k}^{\,\prime}) \bigr]
	= \delta^{D-1}(\vec{k} \!-\! \vec{k}^{\,\prime}) \, ,
\label{commB1}
\\
& \bigl[ \hat{b}(\vec{k}) , \hat{b}(\vec{k}^{\,\prime}) \bigr] = 0 
	= \bigl[ \hat{b}^{\dag}(\vec{k}) , \hat{b}^{\dag}(\vec{k}^{\,\prime}) \bigr] \, ,
\label{commB2}
\end{align}
and $U(k,\eta)$ is the mode function. Note the $a^{\frac{2-D}{2}}$
factor taken out in the definition of the Fourier transform \eqref{FourierTransform}.
The commutation relations (\ref{commPhi1}-\ref{commPhi2}) and (\ref{commB1}-\ref{commB2})
require the Wronskian of the mode function to be
normalized as,
\begin{equation}
U(k,\eta) U'^*(k,\eta) - U'(k,\eta) U^*(k,\eta) = i \, .
\label{Wronskian}
\end{equation}
The equation of motion satisfied by the mode function is the one for 
a harmonic oscillator with a time dependent frequency,
\begin{equation}
U''(k,\eta) + \Bigl[ k^2 + \mathcal{M}^2(\eta) \Bigr] U(k,\eta) = 0 \, ,
\label{modeEOM}
\end{equation}
where
\begin{equation}
\mathcal{M}^2 = m^2a^2 - \frac{1}{4} \Bigl[ D \!-\! 2 \!-\! 4\xi(D \!-\! 1) \Bigr]
	\Bigl[ 2\mathcal{H}' + (D\!-\!2) \mathcal{H}^2 \Bigr] \, .
\label{M2}
\end{equation}

The state $|\Omega\rangle$ that we choose to examine we pick to be 
the one annihilated by the annihilation operator, 
$\hat{b}(\vec{k})|\Omega\rangle=0$, which implies there is no classical 
condensate (hence the name spectator),
\begin{equation}
\langle\Omega|\hat{\phi}(t,\vec{x}) |\Omega\rangle = 0 \, .
\end{equation}
This state respects the symmetries of the background space-time, namely
homogeneity and isotropy, which is evident from requiring the mode function
to depend only on the modulus of the comoving momentum. In order to
completely specify this state one needs to specify the initial conditions
for the mode function (initial state), which we comment upon in 
subsection~\ref{subsec: Choice of state}.


\subsection{Energy-momentum tensor}
\label{subsec: Energy-momentum tensor}

The energy-momentum tensor operator is defined as
\begin{align}
\hat{T}_{\mu\nu}(x) ={}& 
	\left. \frac{-2}{\sqrt{-g}} 
	\frac{\delta S_{\phi}[\phi,g^{\mu\nu}]}{\delta g^{\mu\nu}(x)} 
		\right|_{\phi\rightarrow\hat{\phi}}
\nonumber \\
={}&	\partial_{\mu} \hat{\phi}(x) \, \partial_{\nu} \hat{\phi}(x)
	- \frac{1}{2} g_{\mu\nu}(x) \Bigl[ g^{\alpha\beta(x)}
		\partial_{\alpha} \hat{\phi}(x) \, \partial_{\beta} \hat{\phi}(x) \Bigr]
	- \frac{m^2}{2} g_{\mu\nu}(x) \hat{\phi}^2(x)
\nonumber \\
&	+ \xi \Bigl[ G_{\mu\nu}(x) - \nabla_{\mu} \nabla_{\nu} 
		+ g_{\mu\nu}(x) \square \Bigr] \hat{\phi}^2(x) \, ,
\end{align}
where $G_{\mu\nu} = R_{\mu\nu} \!-\! \frac{1}{2}g_{\mu\nu}R$
is the Einstein tensor, $\nabla$ denotes the covariant derivative, 
and $\square = g^{\mu\nu}\nabla_{\mu}\nabla_{\nu}$ is
the covariant d'Alembertian operator. The expectation value of the energy 
momentum tensor operator with respect to the state defined in the
previous section is diagonal, and is conveniently expressed in terms
of energy density and pressure,
\begin{align}
\rho_Q ={}&  \frac{1}{a^2}\langle \Omega | \hat{T}_{00} | \Omega \rangle 
\nonumber \\
={}&
	\frac{a^{-D}}{(4\pi)^{\frac{D-1}{2}} 
			\, \Gamma \! \left( \frac{D-1}{2} \right)}
	\int\limits_{0}^{\infty} \!\! dk  \, k^{D-2}
	\Biggl\{ 2k^2 |U|^2 
	- \frac{1}{2} \bigl[ D \!-\! 2 \!-\! 4\xi(D \!-\! 1) \bigr]
		\mathcal{H}' |U|^2
\nonumber \\
&	+ 2m^2a^2 |U|^2
	- \frac{1}{2} \bigl[ D \!-\! 2 \!-\! 4\xi(D \!-\! 1) \bigr] \mathcal{H}
		\frac{\partial}{\partial\eta} |U|^2
	+ \frac{1}{2} \frac{\partial^2}{\partial\eta^2} |U|^2 \Biggr\} \, ,
\label{RHOintegral}
\end{align}
\begin{align}
\delta_{ij} p_Q ={}&
\frac{1}{a^2} \langle \Omega | \hat{T}_{ij} | \Omega \rangle 
\nonumber \\
={}&
	\frac{\delta_{ij} a^{-D}}{(4\pi)^{\frac{D-1}{2}} 
			\, \Gamma \! \left( \frac{D-1}{2} \right)}
	\int\limits_{0}^{\infty} \!\! dk  \, k^{D-2}
	\Biggl\{ \frac{2k^2}{(D \!-\! 1)} |U|^2 
	- \frac{1}{2} \bigl[ D \!-\! 2 \!-\! 4\xi(D \!-\! 1) \bigr]
		\mathcal{H}' |U|^2
\nonumber \\
&	- \frac{1}{2} \bigl[ D \!-\! 2 \!-\! 4\xi(D \!-\! 1) \bigr] \mathcal{H}
		\frac{\partial}{\partial\eta} |U|^2
	+ \frac{(1 \!-\! 4\xi)}{2} \frac{\partial^2}{\partial\eta^2} |U|^2 \Biggr\} \, ,
\label{Pintegral}
\end{align}
where we have used the equation of motion \eqref{modeEOM} to write
\begin{equation}
|U'|^2 = \bigl( k^2 + \mathcal{M}^2 \bigr) |U|^2
	+ \frac{1}{2} \frac{\partial^2}{\partial\eta^2} |U|^2 \, ,
\end{equation}
and eliminate $|U'|^2$ in favor of $|U|^2$ and its derivatives. We can take some 
derivatives out of the integral to write the \eqref{RHOintegral} and
\eqref{Pintegral} in a convenient way
\begin{align}
\rho_Q ={}& \frac{a^{-D}}{(4\pi)^{\frac{D-1}{2}} \, 
		\Gamma\!\left( \frac{D-1}{2} \right)} 
	\Biggl\{ 2 \, \mathcal{I}_1 + \Biggl[ 2(ma)^2
	- \frac{1}{2}\bigl[ D \!-\! 2 \!-\! 4\xi(D \!-\! 1) \bigr] 
	\bigl[ \mathcal{H}' + \mathcal{H}\partial_\eta \bigr]
	+ \frac{1}{2} \partial_\eta^2 \Biggr] \mathcal{I}_0 \Biggr\} \, ,
\label{RHOinI}
\\
p_Q ={}& \frac{a^{-D}}{(4\pi)^{\frac{D-1}{2}} \, 
		\Gamma\!\left( \frac{D-1}{2} \right)} 
	\Biggl\{ \frac{2 \, \mathcal{I}_1}{(D\!-\!1)} - \Biggl[
	\frac{1}{2}\bigl[ D \!-\! 2 \!-\! 4\xi(D \!-\! 1) \bigr] 
	\bigl[ \mathcal{H}' + \mathcal{H}\partial_\eta \bigr]
	+ \frac{(1\!-\!4\xi)}{2} \partial_\eta^2 \Biggr] \mathcal{I}_0 \Biggr\} \, ,
\label{PinI}
\end{align}
where we have defined the two integrals,
\begin{equation}
\mathcal{I}_n = \int\limits_{0}^{\infty} \! dk\, k^{D-2+2n} \, |U(k,\eta)|^2 
	\, , \qquad n=0,1 \, .
\label{integralsI}
\end{equation}
Finding a good approximation for these integrals is one of the two main technical 
tasks of this work.


\subsection{Choice of state}
\label{subsec: Choice of state}

 Understanding how the choice of the initial state affects our final results is important, and this is what we discuss next
at some length.
 We assume that the Universe starts in a natural state defined on a global equal-time hyper-surface $\Sigma_0$ 
as the Chernikov-Tagirov-Bunch-Davies (CTBD) 
vacuum state in the ultraviolet (UV). This means that the mode function
reduces to the flat space form in the deep UV,
\begin{equation}
U(k,\eta) \rightarrow \frac{e^{-ik\eta}}{\sqrt{2k}} \, ,
\end{equation}
(a more precise statement is given in Appendix~\ref{app: Regularization and renormalization}).
Furthermore, we assume the state is suitably regulated in the infrared (on super-Hubble wavelengths).
That is namely necessary to regulate the infrared (IR) modes since 
attempting to impose the usual Bunch-Davies condition on the infrared modes would produce 
unphysical infrared divergences in the initial one-loop energy-momentum tensor. 
Infrared states can be regulated in various ways:  
(i) by choosing the global CTBD vacuum state associated with the epoch
that precedes inflation in which the CTBD state is regular, 
(ii) by making the Universe's (initial) equal-time hyper-surface 
compact, or (iii) by introducing a comoving IR cutoff. 
The first prescription can be achieved by {\it e.g.} 
assuming a pre-inflationary radiation 
epoch~\cite{Janssen:2009nz,Glavan:2014uga}, 
while the second on
by imposing a positive constant spatial curvature 
($\kappa\!>\!0$) on  $\Sigma_0$ or by 
making  $\Sigma_0$ compact by imposing periodic boundary conditions 
(in the former case 
$\Sigma_0$ is a three-dimensional sphere  $S^3$ while in the latter case $\Sigma_0$ is a 3-dimensional torus $T^3$). 
The third option is technically perhaps the simplest, and we employ it in this work.
It should be stressed that the point of view on this regularization is not to
throw away the deep IR modes below the cutoff on principal grounds, but rather
that they are smoothly suppressed under this scale and contribute negligibly
to the observables. Then the leading approximation to this case is to introduce
a sharp cutoff. The deep IR suppression can be attributed to some physical process 
during or before inflation, or can be viewed as an approximation to the state 
obtained by placing the Universe in a co-moving box. One can 
show~\cite{Tsamis:1992xa}
that, to leading order in powers of the IR comoving cutoff $k_0$, (expectation 
values of) physical observables are correctly reproduced by the sudden cutoff 
approximation.

All these methods qualitatively agree. For (i) and (ii) this was shown
in~\cite{Janssen:2009nz}
in the sense that qualitative dependence on the relevant physical scale is the same, where in the case of 
a pre-inflationary radiation epoch the relevant physical scale is the Hubble parameter at the radiation-inflation transition,
in the case when $\Sigma_0\equiv S^3$ the relevant physical scale is $\sqrt{\kappa}$
and when $\Sigma_0\equiv T^3$  the relevant physical scale is the comoving length of the torus $L$.
Note that the three aforementioned ways of regulating the infrared correspond to three (very) different physical situations.

Here we use the simple cutoff regularization -- we {\it effectively} remove the modes below certain pivotal mode $k_0$. In practice this is implemented by cutting
of the integration of $\mathcal{I}_n$ integrals \eqref{integralsI} at $k_0$,
\begin{equation}
\mathcal{I}_n \approx \int\limits_{k_0}^{\infty} \! dk\, k^{D-2+2n}
	|U(k,\eta)|^2 \, .
\end{equation}
In the limit of very small $k_0$ this can be shown to be equivalent  (up to 
corrections suppressed as $k_0^2$)
to (i) mentioned above, with $k_0$ identified with $2\pi/L$,
and is shown to be equivalent to (ii) here in 
Sec.~\ref{sec: Energy density and pressure}
by comparison to~\cite{Janssen:2009nz,Glavan:2014uga}.
The main point we are trying to make here is that, for a large class 
of initial states that are regular in the infrared
one will get answers that qualitatively agree with the results 
obtained in this work, hence making 
the results of our analysis {\it quite generic}, {\it i.e.} to a 
large extent independent on the choice of the initial state.


\section{Evolution of the mode function}
\label{sec: Evolution of mode function}

The two main technical tasks of this work are (i) to solve for the time evolution
of the mode function \eqref{modeEOM} as it evolves through cosmological eras, 
and (ii) to perform the integrals \eqref{RHOintegral} 
and \eqref{Pintegral} over these mode functions 
to obtain the backreaction energy density and pressure. This section
discusses these two issues from a more general point of view. 
The mode function is organized in a convenient way. 
Relevant integration interval is identified, and a sudden transition approximation
introduced for the contributing modes.
These considerations simplify following computations significantly.


\subsection{Bogolyubov coefficients}
\label{subsec: Bogolyubov coefficients}

When it comes to the evolution of the mode function, unfortunately, exact
results are known only for a handful of FLRW backgrounds. There is
a way to write down a general solution for an arbitrary FLRW background,
valid for all momentum scales~\cite{Tsamis:2002qk}, but it is difficult
to make use of it practically.
Luckily, for periods of constant $\epsilon$
(out of which most of the history of expansion consists,
Fig.~\ref{expansion history}) the exact solutions are known
in the massless limit,
%
%
%
A convenient way to express them is
in therms of Chernikov-Tagirov-Bunch-Davies (CTBD) mode function
\cite{Chernikov:1968zm,Bunch:1978yq},
\begin{equation}
u_\epsilon(k,\eta) = \sqrt{\frac{\pi}{4|1 \!-\! \epsilon|\mathcal{H}}}
	\, H_\nu^{(1)} \Bigl( \frac{k}{(1 \!-\! \epsilon) \mathcal{H}} \Bigr) \, ,
\label{CTBDmassless}
\end{equation}
where the index of the Hankel function of the first kind $H_\nu^{(1)}$ is
\begin{equation}
\nu = \sqrt{\frac{1}{4} 
	+ \frac{(D \!-\! 2\epsilon)}{4(1 \!-\! \epsilon)^2}
	\bigl[ D \!-\! 2 \!-\! 4\xi(D\!-\!1) \bigr]} \, .
\end{equation}
These functions are defined to reduce to the positive-frequency form
\eqref{UansatzUV} in the UV, and the IR is defined as an analytic continuation
of the UV.
The other linearly independent solution is a complex conjugate of 
\eqref{CTBDmassless}.
In the massive case, exact solutions are not known for constant $\epsilon$
periods, except in a few notable cases (de Sitter, and radiation dominated universe).
Nevertheless, there is a way of making a controlled expansion of this
function in the small ratio $m/H$ with which we will be concerned, which is
sufficient for our purposes. This expansion is presented in Section 
\ref{sec: Mode functions}.

Generally, the full mode function during a given constant $\epsilon$ period
will be a linear combination of the CTBD mode functions,
\begin{equation}
U_\epsilon(k,\eta) = \alpha_\epsilon(k) u_\epsilon(k,\eta)
	+ \beta_\epsilon(k) u_\epsilon^*(k,\eta) \, .
\end{equation}
Coefficients $\alpha(k)$ and $\beta(k)$ are
called the Bogolyubov coefficients, and they have to satisfy
\begin{equation}
|\alpha(k)|^2 - |\beta(k)|^2 = 1 \, ,
\label{BogWronskian}
\end{equation}
as dictated by the Wronskian normalization \eqref{Wronskian}.
They are determined for each era by the initial
conditions at the beginning of the given era, which are in turn given by the 
details of the transition from one era to another. 

If $\tau$ is a small time scale of the transition between periods, then for momenta
above this scale the Bogolyubov coefficients must reduce to 
\begin{equation}
\alpha(k) \xrightarrow{k\rightarrow\infty} 1 \, , \qquad
	\beta(k) \xrightarrow{k\rightarrow\infty} 0 \, ,
\label{BogUVlimit}
\end{equation}
nonadiabatically, meaning faster than any power of $1/k$.
This is provided that the initial condition is of adiabatic order $\infty$.
Otherwise, if the initial state is of adiabatic order $n$, it retains that property
during the evolution~\cite{Kulsrud:1957zz}.
This we can also infer
from considerations of Appendix \ref{app: Regularization and renormalization}. 
The physical reason behind this conclusion is that the deep UV modes oscillate
so fast so that their evolution is adiabatic.

In the IR, Bogolyubov coefficients are not universal as the deep UV are. 
On the contrary, they do depend on the details
of the transition between the two periods of constant $\epsilon$. In case of a
fast transition they are not so sensitive (to leading order) 
to the details of the transition, but rather
depend just on the two periods connected by the transition. This we show in the 
next subsection.


\subsection{Sudden transition approximation}
\label{subsec: Sudden transition approximation}

In case of fast transitions between constant $\epsilon$ periods,
$\tau\!\ll\!1/\mathcal{H}$, the evolution of the IR modes, $k\!\ll\!1/\tau$, through
the transition is well described by the so-called sudden transition approximation,
where the $\epsilon$ parameter jumps discontinuously from one constant
value to another.
Physically, these modes are very slow compared to the transition scale,
and the transition is effectively instantaneous for them. More precisely,
the transition is instantaneous for the IR modes to leading order in
the expansion in the transition scale $\tau$. We stress that this is an approximation
for the evolution of the IR modes, not a model for the background evolution,
and should not be extrapolated to UV modes\footnote{Taking the sudden
transition approximation too seriously as a model for the background, and
applying it to the UV modes leads to unphysical mode mixing in the deep UV.
This in turn results in additional divergences in the energy-momentum
tensor which cannot be absorbed into counterterms. These issues are discussed
in \cite{Glavan:2013mra}. Considering the sudden jumps in $\epsilon$ as
a model for the background makes sense only if one takes the continuum limit
of a series of such small transitions, as was done in \cite{Tsamis:2002qk}.}.

Here we illustrate the sudden transition approximation on a specific example 
of transition between two periods $\epsilon_0$ and $\epsilon_1$.
Let the evolution of $\epsilon$ between two periods be given by
\begin{equation}
\epsilon(\eta) = 
\left\{ \begin{matrix}
	\epsilon_0 & , & \eta<\eta_0 - \frac{\tau}{2} \\
	\epsilon_{\text{tr}}(\eta) & , 
		& \eta_0 - \frac{\tau}{2} < \eta < \eta_0 + \frac{\tau}{2} \\
	\epsilon_1 & , & \eta>\eta_0 + \frac{\tau}{2}
	\end{matrix} \right.
\end{equation}
where
\begin{equation}
\epsilon_{\text{tr}}(\eta) = 
	\frac{\epsilon_0}{2} \Biggl[ 1 + 
	\tanh\Biggl( \frac{1}{1 + \frac{\eta - \eta_0}{\tau/2}}
		- \frac{1}{1 - \frac{\eta - \eta_0}{\tau/2}} \Biggr) \Biggr]
	+ \frac{\epsilon_1}{2} \Biggl[ 1 + 
	\tanh\Biggl( - \frac{1}{1 + \frac{\eta - \eta_0}{\tau/2}}
		+ \frac{1}{1 - \frac{\eta - \eta_0}{\tau/2}} \Biggr) \Biggr] \, .
\label{epsilon transition}
\end{equation}
Before the transition let the full mode function be
\begin{equation}
U(k,\eta<\eta_0-\tfrac{\tau}{2})
	= \alpha_0(k) u_0(k,\eta) + \beta_0 u_0^*(k,\eta) \equiv U_0(k,\eta) \, ,
\label{Uleft}
\end{equation}
with some known Bogolyubov coefficients $\alpha_0$ and $\beta_0$.
After the transition the mode function is
\begin{equation}
U(k,\eta>\eta_0+\tfrac{\tau}{2})
	= \alpha_1(k) u_1(k,\eta) + \beta_1 u_1^*(k,\eta) \equiv U_1(k,\eta) \, .
\label{Uright}
\end{equation}
The equation \eqref{modeEOM} we are trying to solve is the harmonic oscillator 
one with a time dependent frequency,
\begin{equation}
U'' + \omega^2(\eta) U = 0 \, ,
\end{equation}
where
\begin{equation}
\omega^2(\eta) = k^2 + \mathcal{M}^2(\eta) \, ,
\end{equation}
and $\mathcal{M}^2$ is defined in \eqref{M2}.
One can check the WKB applicability condition $\omega'/\omega^2\!\ll\!1$, 
and for which ranges of momenta is it satisfied,
\begin{equation}
\frac{\omega'}{\omega^2}
	\sim \left\{ 
	\begin{matrix} 
		\displaystyle
		\frac{\mathcal{H}^2}{k^2} \frac{1}{(k\tau)^2} & , & k\gg\mathcal{H}
		\\
		\displaystyle
		1/(\mathcal{H}\tau) & , & k\lesssim \mathcal{H}
	 \end{matrix} \right. \ .
\end{equation}
Since $\tau \!\ll\! 1/\mathcal{H}$ by assumption, the only modes that
evolve adiabatically are the ones for which at least $k>\mu\gg\mathcal{H}$.

For the modes $k\!<\!\mu$ another approximation applies. There $1/\tau$ is
the largest scale in the hierarchy, and we can expand the evolution in 
powers of $\tau$. We do this by expanding $\epsilon(\eta)$ function 
\eqref{epsilon transition},
\begin{equation}
\epsilon(\eta) \approx \epsilon_0 \, \theta(\eta_0 \!-\! \eta)
	+ \epsilon_1 \, \theta(\eta \!-\! \eta_0) + \mathcal{O}(\tau) \, .
\end{equation}
Now it is straightforward to match the two solutions \eqref{Uleft} and 
\eqref{Uright}, which are just the continuity conditions for the mode function
and its derivative,
\begin{equation}
U_0(k,\eta_0) = U_1(k,\eta_0) \, , \qquad 
	U_0'(k,\eta_0) = U_1'(k,\eta_0) \, .
\end{equation}
Solving these conditions for Bogolyubov coefficients yields
\begin{align}
\alpha_1(k) ={}& -i \Bigl[ U_0(k,\eta_0) u_1'^*(k,\eta_0)
	- U_0'(k,\eta_0) u_1^*(k,\eta_0) \Bigr] + \mathcal{O}(\tau) \, ,
\label{suddenAlpha}
\\
\beta_1(k) ={}& i \Bigl[ U_0(k,\eta_0) u_1'(k,\eta_0)
	- U_0'(k,\eta_0) u_1(k,\eta_0) \Bigr] + \mathcal{O}(\tau) \, .
\label{suddenBeta}
\end{align}
These two formulas comprise the sudden transition approximation for the 
Bogolyubov coefficients and the evolution of the mode function.


\section{Energy density and pressure integrals}
\label{sec: Energy density and pressure integrals}

In this section we analyze the integrals \eqref{integralsI} on general grounds.
They cannot be evaluated exactly (except for very simple mode 
functions~\cite{Glavan:2013mra}), and we have to resort to approximation 
schemes. 
We first organize the integrand in a way which separates the part containing
all the UV divergences (among other contributions), 
and the UV finite part containing possibly relevant IR contributions. The
contributions from different scales, {\it i.e.} different integration intervals are
examined, and the relevant interval where the dominant contribution
comes from is identified. The analysis presented here greatly facilitates 
the evaluation of integrals \eqref{integralsI}, especially since we get away
with not evaluating certain parts of integrals explicitly (as was done in 
\cite{Glavan:2014uga}).


\subsection{Organizing the integrand}
\label{subsec: Organizing the integrand}

The integrands of integrals \eqref{integralsI} contain the mode function
only as a modulus $|U(k,\eta)|^2$, which can be written out in terms
of the CTBD mode functions of a given constant $\epsilon$ period,
\begin{equation}
|U(k,\eta)|^2 = |u(k,\eta)|^2 + 2|\beta(k)|^2 |u(k,\eta)|^2
	+ \alpha(k) \beta^*(k) u^2(k,\eta) 
	+ \alpha^*(k)\beta(k) [u^*(k,\eta)]^2 \, ,
\label{U2organization}
\end{equation}
where \eqref{BogWronskian} was used. In the deep UV Bogolyubov coefficients
reduce to \eqref{BogUVlimit} faster than power law (at least exponentially). 
Therefore, the UV divergent structure of integrals \eqref{integralsI} is completely
captured by the $|u(k,\eta)|^2$ on the right in \eqref{U2organization}.
We split the integrals into two parts,
\begin{equation}
\mathcal{I}_n = \mathcal{I}^{\text{CTBD}}_n 
	+ \mathcal{I}^{\text{Bog.}}_n \, ,
\end{equation}
where the CTBD part is
\begin{equation}
\mathcal{I}^{\text{CTBD}}_n = \int\limits_{k_0}^{\infty} \! dk
	k^{D-2+2n} |u(k,\eta)|^2 \, ,
\label{IntCTBD}
\end{equation}
which contains all the UV divergences, and the Bogolyubov part,
\begin{equation}
\mathcal{I}_n^{\text{Bog.}} = \int\limits_{k_0}^{\infty}
	\! dk \, k^{2+2n} \, Z_{\text{Bog.}}(k,\eta) \, ,
\end{equation}
where the integrand is
\begin{equation}
Z_{\text{Bog.}}(k,\eta) = 2|\beta(k)|^2 |u(k,\eta)|^2
	+ \alpha(k) \beta^*(k) u^2(k,\eta) 
	+ \alpha^*(k)\beta(k) [u^2(k,\eta)]^* \, .
\label{IntBog}
\end{equation}
Note that we take $D\!=\!4$ limit in the Bogolyubov part, since it is 
manifestly UV finite because of the properties of the Bogolyubov coefficients
\eqref{BogUVlimit}, which simplifies its evaluation.

We can immediately say a lot about the CTBD contribution \eqref{IntCTBD}.
The way to compute it is to first split it in the UV and IR parts by introducing
a UV cutoff $\mu\!\gg\!\mathcal{H}$,
\begin{equation}
\mathcal{I}_n^{\text{CTBD}} = \mathcal{I}_n^{\text{CTBD,UV}}
	+ \mathcal{I}_n^{\text{CTBD,IR}}
	= \int\limits_{\mu}^{\infty} \! dk
	k^{D-2+2n} |u(k,\eta)|^2
	+ \int\limits_{k_0}^{\mu} \! dk
	k^{2+2n} |u(k,\eta)|^2 \, .
\end{equation}
Its UV part needs to be regularized and then renormalized as is
outlined in Appendix~\ref{app: Regularization and renormalization},
and its contribution is given in (\ref{rhoQUV}-\ref{pQUV}) and \eqref{CA}.
Note that this UV contribution is dependent on the fiducial cutoff $\mu$
introduced by hand. That dependence cancels exactly with the opposite one
coming from the IR contribution (which can be evaluated in $D\!=\!4$ from the 
start). This computation is performed explicitly in 
subsection~\ref{subsec: Inflationary era} for inflationary era.

The only dimensionful quantities the CTBD part can depend on 
are the evolving Hubble rate and mass, and possibly
on the IR regulator. The dependence on the regulator actually must cancel
out with same contribution from the Bogolyubov part. The total dependence
on the regulator in the final answer can only come from the CTBD part and it
must have the same structure as the initial state~\cite{Ford:1977in}.
Therefore, in the small mass limit we can represent 
possible contributions to energy density and pressure as
\begin{equation}
\frac{\mathcal{H}^4}{a^4} \biggl\{ \ln(a) \Bigl[ \#_1
	+ \#_2 \Bigl( \frac{ma}{\mathcal{H}} \Bigr)^{\!2}
	+ \#_3 \Bigl( \frac{ma}{\mathcal{H}} \Bigr)^{\!4} + \dots \Bigr]
	+\Bigl[ \#_4 
	+ \#_5 \Bigl( \frac{ma}{\mathcal{H}} \Bigr)^{\!2}
	+ \#_6 \Bigl( \frac{ma}{\mathcal{H}} \Bigr)^{\!4} + \dots \Bigr] \biggr\} \, ,
\end{equation}
where $\#_i$'s stand for some numerical coefficients of order one. This contribution
is not relevant during the radiation or matter period. Its magnitude is tiny
compared to the background. In matter period it also redshifts faster than
the background. In radiation period it does not, rather it redshifts at the same
rate as the background fluid ($\mathcal{H}^4a^{-4}\ln(a)$ term is absent in 
this case), but its magnitude is tiny. The contribution from this CTBD part is
always negligible compared to the background, and therefore we can neglect it. If there
is a large effect, it must lie in the Bogolyubov part. That contribution
we analyze generally in the next subsection.


\subsection{Bogolyubov part}
\label{subsec: Bogolyubov part}

We would like to argue on general grounds about the
contributions from different momentum scales to this integral.
 For the sake of simplicity, consider the transition
between two periods of constant $\epsilon$. Before the
transition, during period $\epsilon_0$, 
the state was the CTBD one ($\alpha_0\!=\!1$, $\beta_0\!=\!0$).
The state after the transition, during period $\epsilon_1$
is dictated by the transition between periods, which is assumed to
be fast ($\tau_0 \!\ll\! 1/\mathcal{H}_0$).
We examine the two cases for the second period separately
 -- the decelerated case ($\epsilon_1 \!>\! 1$), and the accelerated case
($\epsilon_1 \!<\! 1$).


\subsubsection{Decelerating period}
\label{subsubsec: Decelerating period}

\begin{wrapfigure}[12]{l}{5cm}
  \begin{center}
\begin{minipage}{4cm}
\begin{center}
\vskip-0.5cm
    \includegraphics[width=3cm]{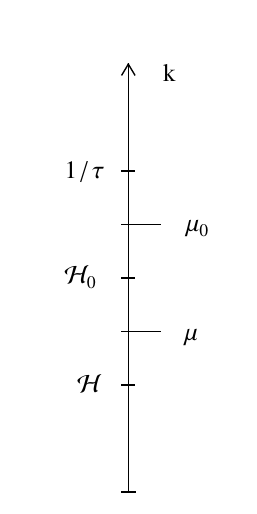}
\vskip-0.5cm
  \caption{Hierarchy of scales after the transition to a decelerating period}
\label{hierarchy3}
\end{center}
\end{minipage}
  \end{center}
\end{wrapfigure}

After the transition to the decelerating period the time evolving 
Hubble rate $\mathcal{H}$ drops below the one at the transition point
$\mathcal{H}_0$ (see Fig.~\ref{Hubble rate}) 
and eventually the hierarchy of scales depicted
in Figure \ref{hierarchy3} is reached. This happens some time
after the transition, which is the regime we are interested in. 
We split the integration into three intervals, separated by $\mu_0$ and $\mu$,
according to this hierarchy. The modes in the
highest interval $k>\mu_0$ contribute negligibly since the Bogolyubov
coefficients are nonadiabatically suppressed there ($\beta\sim e^{-\tau k}$).

The contribution of the middle interval can be estimated rather generally.
We start by noting that
\begin{align}
&\bigl| \mathcal{I}_n^{\text{Bog.,mid.}} \bigr|
	= \left| \int\limits_{\mu}^{\mu_0}\! dk\, k^{2+2n}\, 
	Z_{\text{Bog.}}(k,\eta) \right|
\nonumber \\
& \hspace{0.5cm}
	\le 2 \int\limits_{\mu}^{\mu_0}\! dk\, k^{2+2n}
	\Bigl[ \bigl| \beta_1(k) \bigr|^2 + \bigl| \alpha_1(k) \bigr| 
	\bigl| \beta_1(k) \bigr| \Bigr] \bigl| u_1(k,\eta) \bigr|^2 \, .
\label{middleInt}
\end{align}
The time dependent BD mode function may be expanded asymptotically
as in \eqref{U2expansion} since $k\gg\mathcal{H}\gg ma$ on this interval,
\begin{equation}
|u_1(k,\eta)|^2 \approx \frac{1}{2k} 
	\Biggl\{ 1 
	+ \frac{\mathcal{H}^2}{4k^2} \Bigl[ 1 
		+ \mathcal{O}\Bigl( \frac{ma}{\mathcal{H}} \Bigr)^{\!2} \Bigr]
	+ \mathcal{O}\Bigl( \frac{\mathcal{H}}{k} \Bigr)^{\!4} \Biggr\}
\end{equation}
The momentum scales in question are much smaller than the scale of the transition
$1/\tau$, so the Bogolyubov coefficients are well approximated by the sudden
transition ones, and depend on three quantities: $k$, $\mathcal{H}_0$, $ma_0$.
Since we are interested in the small mass limit we may expand the Bogolyubov
coefficients in powers of $ma_0/\mathcal{H}_0$,
\begin{equation}
\beta_1(k,\mathcal{H}_0,ma_0) \approx 
	\beta_1\Bigl( \frac{k}{\mathcal{H}_0} \Bigr)
	\Bigl[ 1 + \mathcal{O}\Bigl( \frac{ma_0}{\mathcal{H}_0} \Bigr) \Bigr] \, ,
\end{equation}
and analogously for $\alpha_1$, so to leading order they depend only on
the ratio $k/\mathcal{H}_0$. Therefore, we can approximate \eqref{middleInt}
with
\begin{equation}
\bigl| \mathcal{I}^{\text{Bog.,mid.}}_n \bigr| \lesssim
	 \int\limits_{\mu}^{\mu_0}\! dk \, k^{1+2n} \Biggl[ 
	 \Bigl| \beta_1\Bigl( \frac{k}{\mathcal{H}_0} \Bigr) \Bigr|^2
	+ \Bigl| \alpha_1\Bigl( \frac{k}{\mathcal{H}_0} \Bigr) \Bigr| 
	\Bigl| \beta_1\Bigl( \frac{k}{\mathcal{H}_0} \Bigr) \Bigr| \Biggr] \, .
\end{equation}
Making a variable substitution $K=k/\mathcal{H}_0$ puts this integral
into a form which is suitable for further approximation,
\begin{equation}
\bigl| \mathcal{I}^{\text{Bog.,mid.}}_n \bigr| \lesssim
	 \mathcal{H}_0^{2+2n}	
	\int\limits_{\mu/\mathcal{H}_0}^{\mu_0/\mathcal{H}_0}
	\!\! dK \, K^{1+2n} \Bigl[ 
	 \bigl| \beta_1(K) \bigr|^2
	+ \bigl| \alpha_1(K) \bigr| 
	\bigl| \beta_1(K) \bigr| \Bigr] \, .
\end{equation}
The integrand is now dimensionless, and the limits of integration are
$\mu/\mathcal{H}_0 \!\ll\! 1$, and $\mu_0/\mathcal{H}_0 \!\gg\! 1$,
so we may perform (asymptotic) expansions in these limits.
It might happen so that the leading order contribution is dominated by one
of the cutoffs, but this contribution (and in fact any other cutoff dependent one)
must cancel with the opposite contribution from another part of the full integral.
Although tedious, this can be checked explicitly as was done in 
\cite{Glavan:2014uga}.
Therefore, what we are interested in is the
contribution independent off $\mu_0$ and $\mu$, since that is the only one that 
remains after all the parts are added up.
That contribution has the following form,
\begin{equation}
\bigl| \mathcal{I}_n^{\text{Bog.,mid.}} \bigr| \sim \# \mathcal{H}_0^{2+2n} \, .
\end{equation}
This gives the contribution to the energy density (and momentum) of the form
\begin{equation}
\rho \sim \# \frac{\mathcal{H}_0^4}{a^4} \, .
\end{equation}
It is just a radiation-like contribution, and redshifts away faster than the background 
(or at the same rate in the case of radiation era). Therefore we may safely neglect it
as long as it is not too big before the start of radiation period.

The lower part of the integral has a chance to contribute something that does not
redshift away faster than the background. Its exact contribution is not so 
straightforward to estimate, but if there is an interesting effect to be found
it derives from this contribution. Therefore, it is the only one we need to examine,
which we do in Sec.~\ref{sec: Energy density and pressure}. 
For completeness, next we examine the 
accelerated case, $\epsilon_1\!<\!1$.


\subsubsection{Accelerating period}
\label{subsubsec: Accelerating period}

\begin{figure}[h]
\vskip-0.6cm
\captionsetup[subfigure]{labelformat=empty}
\begin{center}
\subfloat[Case A]{\includegraphics[width=3cm]{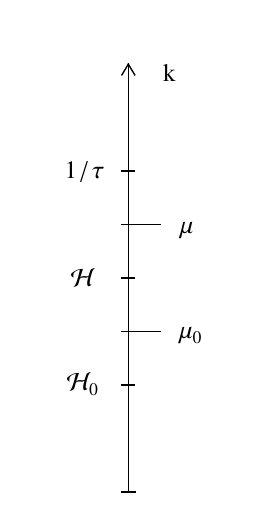}} 
\hspace{3cm}
\subfloat[Case B]{\includegraphics[width=3cm]{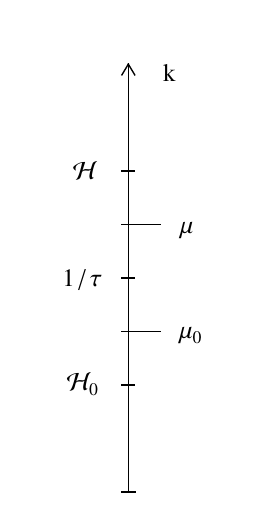}} 
\end{center}
\vskip-0.4cm
\caption{Hierarchy of scales after the transition to accelerating period;
Case A: some time after the transition, Case B: very long after the transition.}
\label{AcceleratingScales}
\end{figure}

In the case of universe transitioning from one period of constant $\epsilon_0$
where the scalar was in a CTBD state, to an accelerating period of constant
$\epsilon_1\!<\!1$, where the transition was fast, we soon reach a hierarchy of
scales shown in Case A of Fig. \ref{AcceleratingScales}, 
and afterwards the one in Case B, as the conformal Hubble rate continues to grow.
We treat these two cases separately.

\vskip+5mm

Case A
\vskip+3mm

\noindent 
The hierarchy of scales in this case is depicted on the left of 
Fig.~\ref{AcceleratingScales}.
The contribution from the modes $k\!>\!\mu_0$ 
is negligible because of the 
nonadiabatic suppression of Bogolyubov coefficients, just as in the decelerating 
case.

The contribution from the middle interval can be estimated as
\begin{equation}
\bigl| \mathcal{I}_n^{\text{Bog.,mid.}} \bigr|
	\le 2 \int\limits_{\mu_0}^{\mu}\! dk\, k^{2+2n}
	\Bigl[ \bigl| \beta_1(k) \bigr|^2 
	+ \bigl| \alpha_1(k) \bigr| \bigl| \beta_1(k) \bigr| \Bigr] 
	\bigl| u_1(k,\eta) \bigr|^2
\label{midIntAc}
\end{equation}
in a similar way as in the decelerating case. Here, because of the
hierarchy of scales, we may expand the 
sudden transition Bogolyubov coefficients for large momenta
(see \cite{Glavan:2013mra} for the expansion),
\begin{align}
\alpha(k) \approx{}&
	 1 + \mathcal{O}\Bigl( \frac{\mathcal{H}_0}{k} \Bigr)
\\
\beta(k) \approx{}&
	\frac{\mathcal{H}_0^2}{k^2} 
	+ \mathcal{O}\Bigl( \frac{\mathcal{H}_0}{k} \Bigr) \, .
\end{align}
The time dependent mode function depends on $k$, $\mathcal{H}$
and $ma$. Since $ma\ll k, \mathcal{H}$ we can expand away its $m$-dependence
so that the leading order term depends only on $k/\mathcal{H}$ and $\mathcal{H}$. Then, using the variable substitution $K=k/\mathcal{H}$
in the middle integral \label{midIntAc} we can estimate it as
\begin{equation}
|\mathcal{I}_n^{\text{Bog.,mid.}}| \le
	\# \mathcal{H}_0^2
	\mathcal{H}^{2n+1} \int\limits_{\mu_0/\mathcal{H}}^{\mu/\mathcal{H}}
	\! dK\, K^{2n} |u_1(k,\mathcal{H})|^2
	\Bigl[ 1 + \mathcal{O} 
		\Bigl( \frac{\mathcal{H}}{\mathcal{H}_0} \Bigr) \Bigr] \, .
\end{equation}
As in the decelerating case, this integral can be expanded for small lower
and large upper limit. Neglecting as before terms dependent on the 
artificially introduced cutoffs, what can remain is a contribution 
$\# \mathcal{H}_0^2 \mathcal{H}^{2n}$. Now, these contribute
to energy density and pressure as $\# \mathcal{H}_0^2\mathcal{H}^2/a^4$.
If the numerical coefficient is not exponentially large, this contribution is negligible 
compared to the background fluid energy density, and it only redshifts away
faster than the background. Only the lower integrals remains to be evaluated.

\vskip+5mm

Case B
\vskip+3mm

In this case the reasoning of Case A applies, we just need not examine the 
middle integral, since here it is shifted into the region where
Bogolyubov coefficients are nonadiabatically suppressed
(as depicted on the right of Fig.~\ref{AcceleratingScales}), and hence
contributes negligibly. One needs to examine the same (lower) interval
to look for the dominant contribution.


\section{Mode functions}
\label{sec: Mode functions}

In this section we derive the CTBD mode functions for each of the 
cosmological periods. Exact solutions are known for the inflationary and
radiation periods. While we can perform integrals \eqref{integralsI}
over the exact inflationary mode functions, the radiation ones are too complicated,
and need to be approximated. An expansion in small $m/H \!=\! ma/\mathcal{H}$
to first subleading order suffices for our goals. This expansion is performed in two
ways. Firstly, the exact radiation period mode function is expanded, 
and the approximation valid for 
all momenta obtained. Secondly, the method for obtaining the approximation 
directly from the equation of motion \eqref{modeEOM} 
(without referring to the exact solution)
is introduced, and the approximation for the radiation period mode function
derived. This method is shown to reproduce the expansion of the exact result,
which lends support for applying it in cases where the exact mode functions are
not known.
The matter period mode functions are not known exactly, and we apply this
method in order to find its expansion to first subleading order in small mass,
valid for all momenta. 
All the approximated mode functions derived in this section are simple enough
so that integrals \eqref{integralsI} can be performed, and the energy density
and pressure of quantum backreaction computed.


\subsection{Inflationary era}
\label{subsec: Inflationary period}

Fortunately, the de Sitter inflationary period ($\epsilon_I\!=\!0$) 
CTBD mode functions are known
exactly even in the massive case. The equation of motion for the mode function
is
\begin{equation}
u'' + \Bigl[ k^2 + (ma)^2 - 2(1 \!-\! 6\xi) \mathcal{H}^2 \Bigr]u = 0 \, .
\end{equation}
The positive frequency solution to this equation is
\begin{equation}
u_I(k,\eta) = \sqrt{\frac{\pi}{4\mathcal{H}}}\, 
	H_{\nu_I}^{(1)} \Bigl( \frac{k}{\mathcal{H}} \Bigr) \, ,
\label{CTBDinf}
\end{equation}
where
\begin{equation}
\nu_I = \sqrt{\frac{1}{4} + 2(1\!-\!6\xi) - \Bigl(\frac{m}{H}\Bigr)^2 } \, .
\label{nuI}
\end{equation}
In this work we will be considering mass of the order of Hubble rate today
(meaning that the ratio $m/H_I$ in inflation is extremely small), and nonminimal
couplings $0\!>\!\xi\!\gtrsim\!-0.05$. More negative nonminimal couplings would
lead to a too rapid growth of quantum fluctuations during inflation (as will be shown 
by the end of this subsection), and a larger mass would mean that the field becomes
very massive at some point during the cosmological evolution and starts contributing
like dust to the expansion (precluding it from having anything to do with DE).
For most of the range of these two parameters
$\nu_I \!>\! 3/2$, which leads to an IR divergence. This divergence has to be regulated
somehow, since it signals that the state chosen is unphysical. The practical method
of regularization we choose is introducing a comoving IR cutoff $k_0$.
The point of view we take on it is that it is a shortcut way of choosing a physical
mode function, since the contribution of the modes below $k_0$ will be
suppressed, and we neglect it from the start.

We will also need the small momentum 
($k \!\ll\! \mathcal{H}$) expansion of the mode function
in \eqref{CTBDinf},
\begin{equation}
u_I(k,\eta) \approx - \frac{i}{\sqrt{\pi}} 2^{\nu_I-1} \Gamma(\nu_I)
	\mathcal{H}_1^{\nu_I-1/2} k^{-\nu_I} 
	\Bigl[ 1 + \mathcal{O}\Bigl( \frac{k}{\mathcal{H}} \Bigr)^2 \Bigr] \, .
\label{uI_IR}
\end{equation}
%


\subsection{Radiation era}

In this subsection first the exact the radiation period CTBD mode function is derived,
and then an expansion for $m/H \!\ll\! 1$ performed (valid for all momenta).
Secondly, this small mass expansion is derived directly from the equation
of motion without referring to the exact solution with the help of Frobenius method.
This serves to introduce the method which can be applied to cases where an
exact solution is not known.

\subsubsection{Exact CTBD mode function}

The CTBD mode functions for radiation period are also known in the massive 
case, but are unfortunately too complicated for practical analytical computations.
The equation of motion for the modes is
\begin{equation}
u'' + \Bigl[ k^2 + (ma)^2 \Bigr] u = 0 \, .
\label{EOMrad1}
\end{equation}
In radiation period the scale factor and the conformal Hubble rate are related as
\begin{equation}
a\mathcal{H} = a_1 \mathcal{H}_1 \, ,
\end{equation}
where the quantities with index 1 refer to the values at the 
beginning of radiation period, so the equation \eqref{EOMrad1} can be written as
\begin{equation}
u'' + \Bigl[ k^2 + \frac{\overline{m}^4}{\mathcal{H}^2} \Bigr] u = 0 \, ,
\label{EOMrad2}
\end{equation}
where we have defined 
\begin{equation}
\overline{m} = \sqrt{m a_1 \mathcal{H}_1} \, .
\end{equation}
By making a variable substitution
\begin{equation}
s = \frac{i \, \overline{m}^2}{\mathcal{H}^2} \, ,
\end{equation}
the equation can be put in the form
\begin{equation}
\Biggl\{ \frac{\partial^2}{\partial s^2} 
	+ \biggl[ - \frac{1}{4} 
		- \frac{i}{4s} \Bigl( \frac{k}{\overline{m}} \Bigr)^2 
	+ \frac{3}{16s^2} \biggr] \Biggr\} \bigl( s^{1/4} u \bigr) = 0 \, ,
\end{equation}
which is a Whittaker equation\footnote{This equation can also be put
into the form of Weber's differential equation and solutions expressed in terms
of parabolic cylinder functions (as was done in \cite{Aoki:2014dqa}), and also as
Kummer's differential equation with solutions expressed in terms of 
confluent hyperbolic functions.} \cite{gradshteyn2007},
\begin{equation}
\Biggl\{ \frac{\partial^2}{\partial s^2} 
	+ \biggl[ - \frac{1}{4} + \frac{\lambda}{s} 
	+ \frac{\frac{1}{4} - \mu^2}{s^2} \biggr] \Biggr\} 
	\bigl( s^{-1/4} u \bigr) = 0 \, ,
\end{equation}
with coefficients 
\begin{equation}
\lambda = - \frac{i}{4} \Bigl( \frac{k}{\overline{m}} \Bigr)^2 \ , \qquad
	\mu = \frac{1}{4} \, .
\end{equation}
The properly normalized (using the Wronskian 13.14.30
from \cite{NIST:DLMF}) CTBD mode function is
\begin{equation}
u_R(k,\eta) = \sqrt{\frac{\mathcal{H}}{2\overline{m}^2}} \
	e^{ - \frac{\pi k^2}{8 \overline{m}^2} } \times
	W_{\!- \frac{i k^2}{4 \overline{m}^2}, \frac{1}{4}}
		\Bigl( \frac{i \, \overline{m}^2}{\mathcal{H}^2} \Bigr) \, ,
\label{radCTBD}
\end{equation}
where $W$ is the Whittaker function. By examining the UV expansion
$k\!\rightarrow\!\infty$ of \eqref{radCTBD} (corresponding to the large parameter 
expansion 9.229 from \cite{gradshteyn2007}),
\begin{equation}
u_R(k,\eta) \xrightarrow{k\rightarrow\infty}
	\frac{e^{-\frac{ik}{\mathcal{H}}}}{\sqrt{2k}} \times
	\exp\Bigl[ \frac{i\pi}{2} + \frac{i k^2}{4\overline{m}^2}
		- \frac{i k^2}{4\overline{m}^2} \ln\Bigl( \frac{i k^2}{4\overline{m}^2} 
	\Bigr) \Bigr] \, ,
\label{CTBDrad}
\end{equation}
we see that it indeed is the positive-frequency mode function (the time-independent
phase is irrelevant since it cancels out in all the physical quantities, the mode
function is defined up to such a phase).

Next we want to find the expansion of this function in small parameter
$ma/\mathcal{H}=\overline{m}^2/\mathcal{H}^2$, but valid for all momenta.
In order to accomplish this we view the function as a function of momenta.
In particular we represent it as a uniformly convergent power series in momenta.
The coefficients in this expansion are functions of $m$ and $\mathcal{H}$,
and them we expand in this small ratio.

It is more convenient to express the Whittaker function in terms
of confluent hypergeometric functions,
\begin{align}
W_{\! -\frac{ik^2}{4\overline{m}^2},\frac{1}{4}} 
	\Bigl( \frac{i\overline{m}^2}{\mathcal{H}^2} \Bigr)
	={}& \frac{-2\sqrt{\pi} \, e^{-\frac{i\overline{m}^2}{2\mathcal{H}^2}}}
	{\Gamma\left( \frac{1}{4} \!+\! \frac{ik^2}{4\overline{m}^2} \right)}
	\Bigl( \frac{i\overline{m}^2}{\mathcal{H}^2} \Bigr)^{\!\frac{3}{4}}
	{}_1F_1\Bigl( \frac{3}{4} \!+\! \frac{ik^2}{4\overline{m}^2};
	\frac{3}{2} ; \frac{i\overline{m}^2}{\mathcal{H}^2} \Bigr)
\nonumber \\
&	+ \frac{\sqrt{\pi} \, e^{-\frac{i\overline{m}^2}{2\mathcal{H}^2}}}
	{\Gamma \! \left( \frac{3}{4} \!+\! \frac{ik^2}{4\overline{m}^2} \right)}
	\Bigl( \frac{i\overline{m}^2}{\mathcal{H}^2} \Bigr)^{\!\frac{1}{4}}
	{}_1F_1\Bigl( \frac{1}{4} \!+\! \frac{ik^2}{4\overline{m}^2};
	\frac{1}{2} ; \frac{i\overline{m}^2}{\mathcal{H}^2} \Bigr) \, .
\label{Whittaker1F1}
\end{align}
The radius of convergence of the power series representation of the confluent 
hypergeometric function is infinite so we may safely examine it, and do 
manipulations of it,
\begin{equation}
{}_1F_1 \Bigl( \frac{1 \!+\! \sigma}{2} \!+\! \frac{ik^2}{4\overline{m}^2} ;
	1 \!+\! \sigma ; \frac{i\overline{m}^2}{\mathcal{H}^2} \Bigr)
	= \sum_{n=0}^{\infty} 
	\frac{\left( \frac{1 \!+\! \sigma}{2} 
				\!+\! \frac{ik^2}{4\overline{m}^2} \right)^{\!(n)}}
	{(1 \!+\! \sigma)^{(n)}} \frac{1}{n!} 
	\Bigl( \frac{i\overline{m}^2}{\mathcal{H}^2} \Bigr)^n \, ,
	\qquad \Bigl( \sigma \!=\! \pm \frac{1}{2} \Bigr) \, ,
\label{1F1expansion}
\end{equation}
where $(x)^{(n)} \!=\! x(x\!+\!1)(x\!+\!2)\dots(x\!+\!n\!-\!1)$ is
the Pochhammer symbol.
What we aim to do is to rewrite it as the power series in 
$k/\mathcal{H}$.\footnote{We could also start by rewriting it 
as a function of $k/\overline{m}$,
and ultimately arrive at the same result. Expressing it as a function of 
$k/\mathcal{H}$ is convenient though, since it is easy to take the massless
limit for which mode functions are known and considerably simpler.}
In order to do this we write out the Pochhammer symbol,
\begin{equation}
\Bigl( \frac{1 \!+\! \sigma}{2} + \frac{ik^2}{4\overline{m}^2} \Bigr)^{\!(n)} = 
	\sum_{s=0}^{n} d_{n,s}^{(\sigma)}
	\Bigl( \frac{ik^2}{4\overline{m}^2} \Bigr)^{\!n-s} \, ,
\label{PochExpansion}
\end{equation}
where $d$-coefficients can be determined by writing out the Pochhammer
symbol.
Using this the power series representation \eqref{1F1expansion} of the
confluent hypergeometric function can be rewritten as
\begin{equation}
{}_1F_1 \Bigl( \frac{1\!+\!\sigma}{2} \!+\! \frac{ik^2}{4\overline{m}^2} ;
	1 \!+\! \sigma ; \frac{i\overline{m}^2}{\mathcal{H}^2} \Bigr)
	= \sum_{s=0}^{\infty} \Bigl( \frac{ima}{\mathcal{H}} \Bigr)^s \times
	\sum_{n=0}^{\infty} \frac{d^{(\sigma)}_{n+s,s}}
		{(n\!+\!s)! \, (1\!+\!\sigma)^{(n+s)}} 
	\Bigl( \frac{-k^2}{4\mathcal{H}^2} \Bigr)^n \, .
\label{1F1reorganized}
\end{equation}
This power series is now straightforward to approximate this power series for 
small $ma/\mathcal{H}$ -- we simply throw away all the terms except the
first three ones,
\begin{equation}
{}_1F_1 \Bigl( \frac{1\!+\!\sigma}{2} \!+\! \frac{ik^2}{4\overline{m}^2} ;
	1 \!+\! \sigma ; \frac{i\overline{m}^2}{\mathcal{H}^2} \Bigr)
	\approx \sum_{s=0}^{3} \Bigl( \frac{ima}{\mathcal{H}} \Bigr)^s \times
	\sum_{n=0}^{\infty} \frac{d^{(\sigma)}_{n+s,s}}
		{(n\!+\!s)! \, (1\!+\!\sigma)^{(n+s)}} 
	\Bigl( \frac{-k^2}{4\mathcal{H}^2} \Bigr)^n \, .
\end{equation}
This approximation can be seen to be valid for all the
ranges of momenta. For $k\gg ma$ this is just an expansion in $ma$ which
is the smallest scale in the hierarchy. For $k\lesssim ma$ the function is well
described by a double expansion, in $ma/\mathcal{H}$, and $k/\mathcal{H}$,
so \label{1F1reorganized} is good, it just retains more terms than necessary 
in this limit, which we neglect anyway.
Now, we need only the first three $d$-coefficients introduced in
 \eqref{PochExpansion},
\begin{align}
& d_{n,0}^{(\sigma)} = 1 \, , \qquad 
	d_{n,1}^{(\sigma)} = \sum_{l=0}^{n-1} 
		\Bigl( \frac{1}{2} \!+\! \sigma \!+\! l \Bigr)
	= \frac{n(n \!+\! \sigma)}{2} \, ,
\\
& d_{n,2}^{(\sigma)} = \sum_{l=0}^{n-2} \sum_{j=l+1}^{n-1}
	\Bigl( \frac{1}{2} \!+\! \sigma \!+\! l \Bigr)
	\Bigl( \frac{1}{2} \!+\! \sigma \!+\! j \Bigr)
	= \frac{n(n \!-\! 1)}{24} \Bigl[ 3n^2 \!+\! (6\sigma \!-\! 1)n 
		\!-\! \frac{1}{4} \, \Bigr] \, .
\end{align}
The approximations for the confluent hypergeometric functions are then
\begin{align}
& {}_1F_1 \Bigl( \frac{3}{4} \!+\! \frac{ik^2}{4\overline{m}^2} ;
	\frac{3}{2} ; \frac{i\overline{m}^2}{\mathcal{H}^2} \Bigr)
	\approx  \frac{\mathcal{H}}{k} \sin\Bigl( \frac{k}{\mathcal{H}} \Bigr)
	+ \Bigl( \frac{ma}{\mathcal{H}} \Bigr) \times
		\frac{i \mathcal{H}}{2k} \sin\Bigl( \frac{k}{\mathcal{H}} \Bigr)
\nonumber \\
&	\hspace{2cm}	
	+ \Bigl( \frac{ma}{\mathcal{H}} \Bigr)^2 \times
	\Biggl\{ \Bigl[ -\frac{1}{8} \!-\! \frac{\mathcal{H}^2}{4k^2}
		\!+\! \frac{\mathcal{H}^4}{4k^4} \Bigr] \frac{\mathcal{H}}{k}
		\sin\Bigl( \frac{k}{H} \Bigr) 
	+ \Bigl[ \frac{\mathcal{H}^2}{6k^2} - \frac{\mathcal{H}^4}{4k^4} \Bigr]
		\cos\Bigl( \frac{k}{\mathcal{H}} \Bigr) \Biggr\} \, ,
\\
& {}_1F_1 \Bigl( \frac{1}{4} \!+\! \frac{ik^2}{4\overline{m}^2} ;
	\frac{1}{2} ; \frac{i\overline{m}^2}{\mathcal{H}^2} \Bigr)
	\approx  \cos\Bigl( \frac{k}{\mathcal{H}} \Bigr)
	+ \Bigl( \frac{ma}{\mathcal{H}} \Bigr) \times
		\frac{i}{2} \cos\Bigl( \frac{k}{\mathcal{H}} \Bigr)
\nonumber \\
&	\hspace{2cm}	
	+ \Bigl( \frac{ma}{\mathcal{H}} \Bigr)^2 \times
	\Biggl\{ \Bigl[ -\frac{1}{8} \!+\! \frac{\mathcal{H}^2}{4k^2}
		\Bigr] \cos\Bigl( \frac{k}{\mathcal{H}} \Bigr) 
	+ \Bigl[ \frac{1}{6} - \frac{\mathcal{H}^2}{4k^2} \Bigr]
		\frac{\mathcal{H}}{k}
		\sin \Bigl( \frac{k}{\mathcal{H}} \Bigr) \Biggr\} \, .
\end{align}
After approximating the confluent hypergeometric functions in the small mass limit,
we only need to approximate the exponential in \eqref{Whittaker1F1},
\begin{equation}
e^{-\frac{i\overline{m}^2}{2\mathcal{H}^2}}\approx
	1 - \frac{ima}{2\mathcal{H}}
	- \frac{(ma)^2}{8\mathcal{H}^2} \, ,
\end{equation}
to arrive at an approximation for the full CTBD mode function in radiation period,
\begin{align}
u_R(k,\eta) ={}& - i^{3/4} \frac{\sqrt{2\pi\overline{m}} \
			e^{-\frac{\pi k^2}{8\overline{m}^2}}}
		{\Gamma\!\left( \frac{1}{4} \!+\! \frac{ik^2}{4\overline{m}^2} \right)} 
	\Biggl\{ \frac{1}{k} \sin\Bigl( \frac{k}{\mathcal{H}} \Bigr)
\nonumber \\
&	\hspace{1cm}
	+ \Bigl( \frac{ma}{\mathcal{H}} \Bigr)^2 \Biggl[ 
	- \Bigl( 1 \!-\! \frac{\mathcal{H}^2}{k^2} \Bigr) \frac{\mathcal{H}^2}{4k^3}
		\sin\Bigl( \frac{k}{\mathcal{H}} \Bigr)
	+ \Bigl( 1 \!-\! \frac{3\mathcal{H}^2}{2k^2} \Bigr)
		\frac{\mathcal{H}}{6k^2} \cos\Bigl( \frac{k}{\mathcal{H}} \Bigr) \Biggr] 
	\Biggr\}
\nonumber \\
&	+ i^{1/4} \sqrt{\frac{\pi}{2\overline{m}}} \,
	\frac{e^{-\frac{\pi k^2}{8\overline{m}^2}}}
		{\Gamma\!\left( \frac{3}{4} \!+\! \frac{ik^2}{4\overline{m}^2} \right)} 
	\Biggl\{ \cos\Bigl( \frac{k}{\mathcal{H}} \Bigr)
\nonumber \\
&	\hspace{1cm}
	+ \Bigl( \frac{ma}{\mathcal{H}} \Bigr)^2 \Biggl[ 
	- \frac{\mathcal{H}^2}{4k^2}\cos\Bigl( \frac{k}{\mathcal{H}} \Bigr)
	- \Bigl( 1 \!-\! \frac{3\mathcal{H}^2}{2k^2} \Bigr)
		\frac{\mathcal{H}}{6k} \sin\Bigl( \frac{k}{\mathcal{H}} \Bigr) \Biggr] 
	\Biggr\} \, .
\label{CTBDradApprox}
\end{align}
Numerical comparisons with the exact CTBD mode function \eqref{radCTBD}
show this is a very good approximation for small mass, $ma\!\ll\!\mathcal{H}$,
for all the ranges of momenta. Note that we cannot expand the 
time-independent coefficients multiplying the curly brackets if we want this 
approximation to be valid for both small and large momenta. One important 
property that these coefficients satisfy is
\begin{equation}
\Im \Biggl\{ \Biggl[- i^{3/4} \frac{\sqrt{2\pi\overline{m}} \
		e^{-\frac{\pi k^2}{8\overline{m}^2}}}
	{\Gamma\!\left( \frac{1}{4} \!+\! \frac{ik^2}{4\overline{m}^2} \right)}
	\Biggr]^*
	\times \Biggl[
	 i^{1/4} \sqrt{\frac{\pi}{2\overline{m}}} \,
	\frac{e^{-\frac{\pi k^2}{8\overline{m}^2}}}
		{\Gamma\!\left( \frac{3}{4} \!+\! \frac{ik^2}{4\overline{m}^2} \right)} 
		\Biggr]
	 \Biggr\} = \frac{1}{2} \, .
\label{CoeffRelation}
\end{equation}

In the next subsection we develop a method to obtain this approximation directly 
from the equation of motion \eqref{EOMrad2}. We will be able to determine
the time dependent functions in the curly brackets in \eqref{CTBDradApprox},
but not the time-independent coefficients in front of the brackets. However,
the important property \eqref{CoeffRelation} will follow.


\subsubsection{Approximate CTBD mode functions from the equation of motion}

Here we wish to derive the approximation \eqref{CTBDradApprox}
directly from the equation of motion \eqref{EOMrad2}. The reason for doing 
this alongside having an exact solution \eqref{CTBDrad} is to establish the
approximation method on an example where we can compare and test it.
Then afterwards we will apply this method to the matter period case where
exact solution is not available.

The method is in the spirit of the way in which we derived the small mass
expansion of the mode function in the previous subsection. We use the
Frobenius method \cite{Arfken} to find the power series solution
to the equation of motion, and then reorganize it to write it as a double
power series in $ma/\mathcal{H}$ and $k/\mathcal{H}$, which is then 
straightforward to approximate.

Starting from the equation of motion \eqref{EOMrad2}, and making a variable
substitution
\begin{equation}
z = \frac{k}{\mathcal{H}} \, ,
\end{equation}
puts the equation in the form
\begin{equation}
\Biggl[ \frac{\partial^2}{\partial z^2} + 1 
	+ \frac{\overline{m}^4}{k^4} z^2 \Biggr] U = 0 \, .
\label{radEOM3}
\end{equation}
The Frobenius method consists of assuming a power series solution,
\begin{equation}
U^{(\lambda)} = \sum_{n=0}^{\infty} C_n^{(\lambda)} z^{n+\lambda} \, ,
\label{radFrobeniusExpansion}
\end{equation}
plugging it in the equation \eqref{radEOM3}, and solving order by order
for the coefficients. The resulting equation, ordered in powers of $z$ is
\begin{align}
&
0 = \lambda ( \lambda \!-\! 1 ) C^{(\lambda)}_0 z^{\lambda-2}
	+ ( 1 \!+\! \lambda) \lambda \, C^{(\lambda)}_1 z^{\lambda-1}
	+ \Bigl[ ( 2 \!+\! \lambda ) ( 1 \!+\! \lambda ) C^{(\lambda)}_2 
			+ C^{\lambda}_0 \Bigr] z^{\lambda}
\nonumber \\
&	+ \Bigl[ ( 3 \!+\! \lambda ) ( 2 \!+\! \lambda ) C^{(\lambda)}_3
			+ C^{(\lambda)}_1 \Bigr] z^{\lambda+1}
	+ \sum_{n=0}^{\infty} 
		\Biggl[ ( n \!+\! \lambda \!+\! 4) ( n \!+\! \lambda \!+\! 3) C^{(\lambda)}_{n+4}
			+ C^{(\lambda)}_{n+2}
			+ \frac{\overline{m}^4}{k^4} C^{(\lambda)}_n \Biggr]
		 z^{\lambda+2+n} \, .
\end{align}
Coefficients multiplying different powers of $z$ must vanish
independently, which gives us an infinite set of equations.
The leading order gives the so-called indical polynomial
\begin{equation}
0 = \lambda ( \lambda \!-\! 1 ) \, ,
\end{equation}
whose roots 
\begin{equation}
\lambda_1 = 1 \, , \qquad \lambda_2 = 0 \, ,
\end{equation}
distinguish between the two linearly independent solutions. The leading order
coefficient $C^{(\sigma)}_0$ is the overall normalization of the function, 
and can not be determined by this method (stemming from the fact that equation 
\eqref{radEOM3} is linear and homogeneous).

The second order requires
\begin{equation}
0 = ( 1 \!+\! \lambda ) \lambda \, C^{(\lambda)}_1 \, ,
\end{equation}
which is satisfied by setting $C^{(\sigma)}_1=0$ \footnote{Strictly speaking,
for $\sigma=\sigma_0=0$ coefficient $C^{(0)}_1$ is undetermined from this 
equation, and can be chosen arbitrarily, so we set it to zero for convenience. 
In fact, picking a nonzero value of it corresponds to choosing a different
linear combination of independent solutions.}. It is also straightforward to see that
all the rest of the odd coefficients must vanish as well,
\begin{equation}
C^{(\lambda)}_{2n+1} = 0 \, , \qquad (n \in \mathbb{N}) \, .
\end{equation}
This leaves the even coefficients to be determined. Order $z^{\lambda}$
gives
\begin{equation}
C^{(\lambda)}_2 = 
	\frac{-1}{(2 \!+\! \lambda)(1 \!+\! \lambda)} C^{(\lambda)}_0 \, ,
\end{equation}
and the remaining coefficients are determined by the recurrence relation,
\begin{equation}
C^{(\lambda)}_{2n+4} = 
	- \frac{C^{(\lambda)}_{2n+2} + \frac{\overline{m}^4}{k^4} C^{(\lambda)}_{2n}}
		{(2n \!+\! \lambda \!+\! 4)(2n \!+\! \lambda \!+\! 3)} \, ,
	\qquad (n \in \mathbb{N}_0) \, .
\label{radRecurrence}
\end{equation}
We do not bother to solve this recurrence relation exactly, since the order
of approximation we are after does not require it. Instead, we note that
the coefficients have the following form,
\begin{equation}
C_{4n}^{(\lambda)} = C_0^{(\lambda)} \sum_{s=0}^{n}
	\ell^{(\lambda)}_{2n,s} \Bigl( \frac{\overline{m}}{k} \Bigr)^{\!4s} \, , \qquad
C_{4n+2}^{(\lambda)} = C_0^{(\lambda)} \sum_{s=0}^{n}
	\ell^{(\lambda)}_{2n+1,s} \Bigl( \frac{\overline{m}}{k} \Bigr)^{\!4s} \, .
\label{Cexpansion}
\end{equation}
Plugging in this into the initial power series \eqref{radFrobeniusExpansion},
and reorganizing, gives the desired double power series (remember that
$\overline{m}^2/\mathcal{H}^2= ma/\mathcal{H}$),
\begin{equation}
U^{(\lambda)} = C_0^{(\lambda)} 
	\Bigl( \frac{k}{\mathcal{H}} \Bigr)^\lambda
	\sum_{n=0}^{\infty} \sum_{s=0}^{n} 
	\Biggl\{ \ell^{(\lambda)}_{2n,s}
		\Bigl( \frac{ma}{\mathcal{H}} \Bigr)^{\!2s}
		\Bigl( \frac{k}{\mathcal{H}} \Bigr)^{4n-4s}
	+ \ell^{(\lambda)}_{2n+1,s}
		\Bigl( \frac{ma}{\mathcal{H}} \Bigr)^{\!2s}
		\Bigl( \frac{k}{\mathcal{H}} \Bigr)^{4n-4s+2} \Biggr\} \, .
\label{radUapproxForm}
\end{equation}
The approximation to first subleading order in small mass now consists of 
keeping just $s=0$ and $s=1$ terms,
\begin{align}
U^{(\lambda)} ={}& C_0^{(\lambda)} 
	\Bigl( \frac{k}{\mathcal{H}} \Bigr)^\lambda
	\Biggl\{ \sum_{n=0}^{\infty}  
	 \ell^{(\lambda)}_{n,0}
		\Bigl( \frac{k}{\mathcal{H}} \Bigr)^{2n}
	+ \Bigl( \frac{ma}{\mathcal{H}} \Bigr)^{\!2} \times
	\sum_{n=0}^{\infty} \ell^{(\lambda)}_{n+2,1}
		\Bigl( \frac{k}{\mathcal{H}} \Bigr)^{2n} \Biggr\} \, .
\end{align}
What remains is to solve for the needed $\ell$-coefficients by using
\eqref{Cexpansion} and the recurrence relation \eqref{radRecurrence},
\begin{align}
\ell^{(\lambda)}_{n,0} ={}& \frac{(-1)^n}{(\lambda \!+\! 1)^{(2n)}} \, ,
\\
\ell^{(\lambda)}_{n,1} ={}&
	-\frac{(-1)^{n}}{(\lambda \!+\! 1)^{(2n)}} \times
	\frac{(n \!-\! 1)}{3} \Bigl[ 4n^2 + n(6\lambda \!-\! 5) 
		+ 3\lambda(\lambda \!-\! 1) \Bigr] \, ,
\end{align}
and to resum the power series in \eqref{radUapproxForm} using these coefficients.
The two linearly independent solutions ($\lambda_1=1, \lambda_2=0$) 
that we find are
\begin{align}
v_{R1} ={}& \frac{1}{k} \sin\Bigl( \frac{k}{\mathcal{H}} \Bigr)
	+ \Bigl( \frac{ma}{\mathcal{H}} \Bigr)^2 \times
	\Biggl\{ - \Bigl[ 1 \!-\! \frac{\mathcal{H}^2}{k^2} \Bigr]
	\frac{\mathcal{H}^2}{4k^3}
	\sin\Bigl( \frac{k}{\mathcal{H}} \Bigr) 
	+ \Bigl[ 1 \!-\! \frac{3\mathcal{H}^2}{2k^2} \Bigr] 	
	\frac{\mathcal{H}}{6k^2}\cos\Bigl( \frac{k}{\mathcal{H}} \Bigr)\Biggr\} \, ,
\label{vR1}
\\
v_{R2} ={}& \cos\Bigl( \frac{k}{\mathcal{H}} \Bigr)
	+ \Bigl( \frac{ma}{\mathcal{H}} \Bigr)^2 \times
	\Biggl\{ -\frac{\mathcal{H}^2}{4k^2} \cos\Bigl( \frac{k}{\mathcal{H}} \Bigr)
	- \Bigl[ 1 \!-\! \frac{3\mathcal{H}^2}{2k^2} \Bigr]
	\frac{\mathcal{H}}{6k} \sin\Bigl( \frac{k}{\mathcal{H}} \Bigr) \Biggr\} \, ,
\label{vR2}
\end{align}
where we have picked the normalizations $C_0^{(1)}\!=\!1/k$ and
$C_0^{(0)}\!=\!1$ for convenience, and so that the $k\!\rightarrow\!0$ limit
is well defined for both functions. These two functions are exactly the ones 
in curly brackets in \eqref{CTBDradApprox} that were found by expanding the 
exact solution \eqref{radCTBD} in small mass.

The CTBD mode function in radiation period is some linear combination 
of \eqref{vR1} and~\eqref{vR2},
\begin{equation}
u_R(k,\eta) = A_R(k,m) v_{R1}(k,\eta) + B_R(k,m) v_{R2}(k,\eta) \, .
\label{radCTBDlinCombination}
\end{equation}
Since CTBD mode functions are assumed to satisfy the Wronskian normalization
\eqref{Wronskian}, it is easy to compute that the coefficients above must satisfy
\begin{equation}
\Im \Bigl[ A_R^*(k,m) B_R(k,m) \Bigr] = \frac{1}{2} \, .
\label{radCoeffProperty}
\end{equation}
This is in fact an exact relation between these coefficients, valid to all orders in 
$m$, which we have already calculated from the exact solution in 
\eqref{CoeffRelation}.
We cannot say more about these coefficients just based on the equation of motion,
but luckily we do not have to for the purposes of computing the backreaction
energy-momentum tensor, \eqref{radCoeffProperty} will be the
only property needed.

Later we will need an IR expansion of the mode functions \eqref{vR1}
and \eqref{vR2} which we include here,
\begin{align}
v_{R1} \approx{}& \mathcal{H}^{-1} \Biggl[ 1
	- \frac{1}{20} \Bigl( \frac{ma}{\mathcal{H}} \Bigr)^{\!2}
	+ \mathcal{O}\Bigl( \frac{ma}{\mathcal{H}} \Bigr)^{\!4} \Biggr]
	+ \mathcal{O}\Bigl( \frac{k}{\mathcal{H}} \Bigr)^{\!2} \, ,
\label{vR1_IR}
\\
v_{R2} \approx{}& \Biggl[ 1
	- \frac{1}{12} \Bigl( \frac{ma}{\mathcal{H}} \Bigr)^{\!2}
	+ \mathcal{O}\Bigl( \frac{ma}{\mathcal{H}} \Bigr)^{\!4} \Biggr]
	+ \mathcal{O}\Bigl( \frac{k}{\mathcal{H}} \Bigr)^{\!2} \, .
\label{vR2_IR}
\end{align}
%


\subsection{Matter era}
\label{subsec: Matter period}

Here we apply the method introduced in the previous subsection from the
start since the exact solution for the mode function is not known.
During matter period ($\epsilon\!=\!3/2$) the background satisfies
$a\mathcal{H}^2=a_2\mathcal{H}_2^2$. The equation of motion for the
modes \eqref{modeEOM}, when a variable substitution
\begin{equation}
z=\frac{2k}{\mathcal{H}} \, ,
\end{equation}
is put into the form
\begin{equation}
\Biggl[ \frac{\partial^2}{\partial z^2} + 1
	+ \frac{\widetilde{m}^6}{16k^6} z^4 
	- \frac{2(1 \!-\! 6\xi)}{z^2} \Biggr] U = 0 \, ,
\label{mattEOM}
\end{equation}
where we have defined a mass parameter
\begin{equation}
\widetilde{m} = \Bigl[ m a_2 \mathcal{H}_2^2 \Bigr]^{1/3} \, ,
\end{equation}
which satisfies
\begin{equation}
\frac{\widetilde{m}^3}{\mathcal{H}^3} = \frac{ma}{\mathcal{H}} \ll 1 \, .
\end{equation}
We do not know the exact solutions of the equation of motion \eqref{mattEOM}.
That is why we will resort to the approximation scheme for the small mass 
expansion developed in the previous subsection.

As before, we use the Frobenius method to obtain a power series solution
to the equation,
\begin{equation}
U^{(\lambda)} = \sum_{n=0}^{\infty} C_n^{(\lambda)} z^{n+\lambda} \, .
\label{UmattSeries}
\end{equation}
Organizing the equation in powers of $z$ yields
\begin{align}
0 ={}& \Bigl[ \lambda(\lambda \!-\! 1) \!-\! 2(1 \!-\! 6\xi) \Bigr] 
	C_0^{(\lambda)} z^{\lambda-2}
	+ \Bigl[ (\lambda \!+\! 1) \lambda \!-\! 2(1 \!-\! 6\xi) \Bigr] 
	C_1^{(\lambda)} z^{\lambda-1}
\nonumber \\
&	+ \biggl\{ \Bigl[ (\lambda \!+\! 2)(\lambda \!+\! 1) \!-\! 2(1 \!-\! 6\xi) \Bigr]
		C_2^{(\lambda)} + C_0^{(\lambda)} \biggr\} z^{\lambda}
\nonumber \\
&	+ \biggl\{ \Bigl[ (\lambda \!+\! 3)(\lambda \!+\! 2) \!-\! 2(1 \!-\! 6\xi) \Bigr]
		C_3^{(\lambda)} + C_1^{(\lambda)} \biggr\} z^{\lambda+1}
\nonumber \\
&	+ \biggl\{ \Bigl[ (\lambda \!+\! 4)(\lambda \!+\! 3) \!-\! 2(1 \!-\! 6\xi) \Bigr]
		C_4^{(\lambda)} + C_2^{(\lambda)} \biggr\} z^{\lambda+2}
\nonumber \\
&	+ \biggl\{ \Bigl[ (\lambda \!+\! 5)(\lambda \!+\! 4) \!-\! 2(1 \!-\! 6\xi) \Bigr]
		C_5^{(\lambda)} + C_3^{(\lambda)} \biggr\} z^{\lambda+3}
\nonumber \\
&	+ \sum_{n=0}^{\infty} \biggl\{ 
	\Bigl[ (\lambda \!+\! 6 \!+\! n)(\lambda \!+\! 5 \!+\! n)
		\!-\! 2 (1 \!-\! 6\xi) \Bigr] C_{n+6}^{(\lambda)} 
	+ C_{n+4}^{(\lambda)}
	+ \frac{\widetilde{m}^6}{16k^6} C_n^{(\lambda)} \biggr\} \, .
\end{align}
Coefficients multiplying different powers of $z$ must vanish independently.
The order $z^{\lambda-2}$ gives the indicial polynomial, 
\begin{equation}
\lambda(\lambda \!-\! 1) - 2(1 \!-\! 6\xi) = 0 \, ,
\label{mattIndicial}
\end{equation}
whose solutions are
\begin{align}
\lambda_1 = \frac{1}{2} \Bigl[ 1 + \sqrt{1 \!+\! 8(1 \!-\! 6\xi)} \Bigr]
	\equiv \frac{1}{2} + \nu \, ,
\label{lambda1}
\\
\lambda_2 = \frac{1}{2} \Bigl[ 1 - \sqrt{1 \!+\! 8(1 \!-\! 6\xi)} \Bigr]
	\equiv \frac{1}{2} - \nu \, ,
\label{lambda2}
\end{align}
and $C_0^{(\lambda)}$ is the overall normalization constant.
Note that in \eqref{lambda1} and \eqref{lambda2} above we have
introduced the definition of $\nu$,
\begin{equation}
\nu = \sqrt{\frac{1}{4} + 2(1\!-\!6\xi)} \, ,
\label{nu}
\end{equation}
which would be the index of the Hankel functions in CTBD mode function
of the matter period in the massless limit.
The next order requires that $C_1^{(\lambda)}=0$, and in fact all the 
odd coefficients must vanish,
\begin{equation}
C^{(\lambda)}_{2n+1} = 0 \, ,\qquad n\in\mathbb{N} \, .
\end{equation}
Orders $z^{\lambda}$ and $z^{\lambda+2}$ give
\begin{equation}
C_2^{(\lambda)} = \frac{-C_0^{(\lambda)}}
	{2(2\lambda\!+\!1)} \, ,
\qquad
C_4^{(\lambda)} = \frac{C_0^{(\lambda)}}
	{4(2\lambda\!+\!3)} \, ,
\end{equation}
which serve as initial conditions for the recurrence relation
\begin{equation}
C_{2n+6}^{(\lambda)} = \frac{- C_{2n+4}^{(\lambda)}
			- \frac{\widetilde{m}^6}{16k^6} C_{2n}^{(\lambda)}}
	{2( 3\!+\! n) (2\lambda \!+\! 5 \!+\! 2n)} \, ,
	\qquad (n\!\in\!\mathbb{N}_0) \, .
\label{mattRecurrence}
\end{equation}
The indicial polynomial \eqref{mattIndicial} was used to simplify the denominators
of the above expressions for the coefficients.

In a similar fashion as for the case of radiation period in the previous section,
the coefficients in the expansion \eqref{UmattSeries} can be seen
from \eqref{mattRecurrence} to have the form
\begin{align}
C^{(\lambda)}_{6n} ={}&
	C_0 \sum_{s=0}^{n} \ell_{3n,s} 
		\Bigl( \frac{\widetilde{m}^6}{16k^6} \Bigr)^s \, ,
\\
C^{(\lambda)}_{6n+2} ={}&
	C_0 \sum_{s=0}^{n} \ell_{3n+1,s} 
		\Bigl( \frac{\widetilde{m}^6}{16k^6} \Bigr)^s \, ,
\\
C^{(\lambda)}_{6n+4} ={}&
	C_0 \sum_{s=0}^{n} \ell_{3n+2,s} 
		\Bigl( \frac{\widetilde{m}^6}{16k^6} \Bigr)^s \, ,
\end{align}
where we will not need to solve for all the $\ell$-coefficients. Plugging these into
the power series \eqref{UmattSeries}, and reorganizing gives
\begin{equation}
U^{(\lambda)} = C_0^{(\lambda)} 
	\Bigl( \frac{2k}{\mathcal{H}} \Bigr)^{\!\lambda}
	\sum_{n=0}^{\infty} \sum_{s=0}^{n}
	\Bigl( \frac{2ma}{\mathcal{H}} \Bigr)^{\!2s}
	\Biggl[ \ell^{(\lambda)}_{3n,s}  
	+ \ell^{(\lambda)}_{3n+1,s} \Bigl( \frac{2k}{\mathcal{H}} \Bigr)^{\!2}
	+ \ell^{(\lambda)}_{3n+2,s} \Bigl( \frac{2k}{\mathcal{H}} \Bigr)^{\!4}
	\Biggr] \Bigl( \frac{2k}{\mathcal{H}} \Bigr)^{\!6(n-s)} \, ,
\end{equation}
which is straightforward to approximate in the $ma\!\ll\!\mathcal{H}$ limit,
by keeping only $s\!=\!0$ and $s\!=\!1$ terms,
\begin{equation}
U^{(\lambda)} \approx C_0^{(\lambda)} 
	\Bigl( \frac{2k}{\mathcal{H}} \Bigr)^{\!\lambda} \Biggl\{ 
	\sum_{n=0}^{\infty} \ell_{n,0}^{(\lambda)} 
	\Bigl( \frac{2k}{\mathcal{H}} \Bigr)^{\!2n}
	+ 4\Bigl( \frac{ma}{\mathcal{H}} \Bigr)^2
	\sum_{n=0}^{\infty} \ell^{(\lambda)}_{n+3,1}
	\Bigl( \frac{2k}{\mathcal{H}} \Bigr)^{2n} \Biggr\} \, .
\label{mattUapproxSymbolic}
\end{equation}
Now the $\ell$-coefficients we need can be found from \eqref{mattRecurrence},
and they are
\begin{align}
\ell_{n,0}^{(\lambda)} ={}& \frac{(-1)^n}
	{4^n \, n! \left( \frac{1}{2} \!+\! \lambda \right)^{(n)}} \, ,
\\
\ell_{n,1}^{(\lambda)} ={}& \frac{(-1)^n}
	{4^n \, n! \left( \frac{1}{2} \!+\! \lambda \right)^{(n)}} \times
	\frac{4n(n \!-\! 1)(n \!-\! 2)}{15} \Bigl[ 
	 24 \!-\! 39n \!+\! 12n^2 \!-\! 50\lambda 
		\!+\! 30n\lambda \!+\! 20\lambda^2 \Bigr] \, .
\end{align}
Resumming the series \eqref{mattUapproxSymbolic} yields
the two linearly independent solutions to first subleading order in small
$ma/\mathcal{H}$,
\begin{align}
v_{M1} ={}& \Gamma(1\!+\!\nu) k^{-\nu} \mathcal{H}^{-1/2} \,
	J_{\nu} \Bigl( \frac{2k}{\mathcal{H}} \Bigr)
\nonumber \\
& + \Bigl( \frac{ma}{\mathcal{H}} \Bigr)^{\!2} \times
	\frac{\Gamma(1 \!+\! \nu)}{30} k^{-1-\nu} \mathcal{H}^{1/2}
	\Biggl\{ \biggl[ -6 \!+\! (1\!-\!\nu)(2\!-\!\nu) \frac{\mathcal{H}^2}{k^2} 
		\biggr] J_{1+\nu} \Bigl( \frac{2k}{\mathcal{H}} \Bigr)
\nonumber \\
&	\hspace{2.5cm}
	 + (2\!-\!\nu) \biggl[ 3 \!-\! (1\!-\!\nu)(2\!+\!\nu) \frac{\mathcal{H}^2}{k^2}
		 \biggr] \frac{\mathcal{H}}{k} J_{2+\nu} 
		\Bigl( \frac{2k}{\mathcal{H}} \Bigr) \Biggr\} 
	+ \mathcal{O}\Bigl( \frac{ma}{\mathcal{H}} \Bigr)^{\!4} \, ,
\label{vM1}
\\
v_{M2} ={}& \Gamma(1\!-\!\nu) k^{\nu} \mathcal{H}^{-1/2} \,
	J_{-\nu} \Bigl( \frac{2k}{\mathcal{H}} \Bigr)
\nonumber \\
&	+ \Bigl( \frac{ma}{\mathcal{H}} \Bigr)^{\!2} \times
	\frac{\Gamma(1\!-\!\nu)}{30} k^{-1+\nu} \mathcal{H}^{1/2}
	\Biggl\{ \biggl[ -6 \!+\! (1\!+\!\nu)(2\!+\!\nu) \frac{\mathcal{H}^2}{k^2} 
		\biggr] J_{1-\nu}\Bigl( \frac{2k}{\mathcal{H}} \Bigr) 
\nonumber \\
&	\hspace{2.5cm}
	+ (2\!+\!\nu) \biggl[ 3 \!-\! (1\!+\!\nu)(2\!-\!\nu) \frac{\mathcal{H}^2}{k^2}
		 \biggr] \frac{\mathcal{H}}{k} J_{2-\nu} 
		\Bigl( \frac{2k}{\mathcal{H}} \Bigr)  \Biggr\}
	+ \mathcal{O}\Bigl( \frac{ma}{\mathcal{H}} \Bigr)^{\!4} \, ,
\label{vM2}
\end{align}
where a convenient overall normalization was chosen, 
$C_0^{(1/2+\nu)} \!=\! (2k)^{-1/2-\nu}$ and 
$C_0^{(1/2-\nu)}~\!=\!~(2k)^{-1/2+\nu}$,
and $\nu$ was defined in \eqref{lambda1}.

The CTBD mode function in matter period is some linear combination of the two
independent solutions above,
\begin{equation}
u_M(k,\eta) = A_M(k,m) v_{M1}(k,\eta) + B_M(k,m) v_{M2}(k,\eta) \, .
\end{equation}
We cannot determine them just from the equation of motion, but they satisfy
\begin{equation}
\Im \Bigl[ A^*_M(k,m) B_M(k,m) \Bigr] = \frac{1}{2\nu} \, ,
\label{mattCoeffProperty}
\end{equation}
which follows from the Wronskian normalization \eqref{Wronskian}, and the 
Wronskian of the functions \eqref{vM1} and \eqref{vM2}. 
We expect this to be an exact relation in the same way as the analogous one for
radiation period \eqref{radCoeffProperty}, but we have not checked it explicitly.
We will not need
any other properties of the coefficients other than \eqref{mattCoeffProperty}
in order to compute the energy-momentum tensor of the backreaction in
matter period.

Later we will need also an IR expansion of the mode functions \eqref{vM1}
and \eqref{vM2}, which we give here,
\begin{align}
v_{M1} \approx{}& \mathcal{H}^{-1/2-\nu} \Biggl\{ 
	\biggl[ 1 - \frac{1}{3(3\!+\!\nu)} \Bigl( \frac{ma}{\mathcal{H}} \Bigr)^{\!2}
	+ \mathcal{O}\Bigl( \frac{ma}{\mathcal{H}} \Bigr)^{\!4} \biggr]
	+ \mathcal{O}\Bigl( \frac{k}{\mathcal{H}} \Bigr)^{\!2} \Biggr\} \, ,
\\
v_{M2} \approx{}& \mathcal{H}^{-1/2+\nu} \Biggl\{ 
	\biggl[ 1 - \frac{1}{3(3\!-\!\nu)} \Bigl( \frac{ma}{\mathcal{H}} \Bigr)^{\!2}
	+ \mathcal{O}\Bigl( \frac{ma}{\mathcal{H}} \Bigr)^{\!4} \biggr]
	+ \mathcal{O}\Bigl( \frac{k}{\mathcal{H}} \Bigr)^{\!2} \Biggr\} \, .
\end{align}
%


\section{Energy density and pressure}
\label{sec: Energy density and pressure}

This section is devoted to computing the leading contributions to integrals
\eqref{integralsI}, using the rationale from Section \ref{sec: Energy density and 
pressure integrals}, and the mode functions derived in Section \ref{sec: Mode 
functions}. The computation is made for all three cosmological eras
(Fig. \ref{expansion history}), up until the onset of DE domination. 
The final answers are leading contributions in the small ratios of physical
parameters (which satisfy a hierarchy from Fig. \ref{Hubble rate}). Also,
the computations are restricted to the regions long enough after the transition
periods so that this hierarchy can be exploited. At the end of each subsection
the minimally coupled limit is discussed, and compared to \cite{Aoki:2014dqa},
as well as the limits of small nonminimal coupling which is the main focus of this
work.


\subsection{Inflationary era}
\label{subsec: Inflationary era}

For exact de Sitter inflationary era we can actually evaluate the integrals 
\eqref{integralsI} for the energy-momentum tensor exactly, and there is no
need to resorting to approximations. First we compute the IR part of
\eqref{integralsI}, using the CTBD mode function \eqref{CTBDinf},
\begin{align}
\mathcal{I}_0^{IR}
	={}& \int\limits_{k_0}^{\mu} \! dk\, k^2 \bigl| u_I(k,\eta) \bigr|^2
\nonumber \\
={}& \frac{\mu^2}{4} 
	+ \frac{\mathcal{H}^2}{4} \Bigl( \nu_I^2 \!-\! \tfrac{1}{4} \Bigr)
	\ln\Bigl( \frac{\mu}{\mathcal{H}} \Bigr)
	+ \frac{\mathcal{H}^2}{8} \Bigl( \nu_I^2 \!-\! \tfrac{1}{4} \Bigr)
	\biggl[ 2\ln2 \!-\! 1 \!-\! \psi\Bigl( - \tfrac{1}{2}\!-\!\nu_I \Bigr)
	\!-\! \psi\Bigl( -\tfrac{1}{2}\!+\!\nu_I \Bigr) \biggr]
\nonumber \\
&	+ \frac{2^{2\nu_I-3} \, \Gamma^2(\nu_I)}
	{\pi \left( \nu_I \!-\! \frac{3}{2} \right)} \mathcal{H}^{2\nu_I-1}
	k_0^{3-2\nu_I} 
	\Bigl[ 1 + \mathcal{O}\Bigl( \frac{k_0}{\mathcal{H}} \Bigr)^{\!2} \Bigr] \, ,
\\
\mathcal{I}_1^{IR}
	={}& \int\limits_{k_0}^{\mu} \! dk\, k^4 \bigl| u_I(k,\eta) \bigr|^2
\nonumber \\
={}& \frac{\mu^4}{8} 
	+ \frac{\mathcal{H}^2\mu^2}{8} \Bigl( \nu_I^2 \!-\! \tfrac{1}{4} \Bigr)
	+ \frac{3\mathcal{H}^2}{16} \Bigl( \nu_I^2 \!-\! \tfrac{1}{4} \Bigr)
	\Bigl( \nu_I^2 \!-\! \tfrac{9}{4} \Bigr)
	\ln\Bigl( \frac{\mu}{\mathcal{H}} \Bigr)
\nonumber \\
&	+ \frac{\mathcal{H}^4}{64} \Bigl( \nu_I^2 \!-\! \tfrac{1}{4} \Bigr)
	\Bigl( \nu_I^2 \!-\! \tfrac{9}{4} \Bigr)
	\biggl[ 12\ln2 \!-\! 7 \!-\! 6\psi\Bigl( - \tfrac{3}{2}\!-\!\nu_I \Bigr)
	\!-\! 6\psi\Bigl( -\tfrac{3}{2}\!+\!\nu_I \Bigr) \biggr]
\nonumber \\
&	+ \frac{2^{2\nu_I-3} \, \Gamma^2(\nu_I)}
	{\pi \left( \nu_I \!-\! \frac{5}{2} \right)} \mathcal{H}^{2\nu_I-1}
	k_0^{5-2\nu_I} 
	\Bigl[ 1 + \mathcal{O}\Bigl( \frac{k_0}{\mathcal{H}} \Bigr)^{\!2} \Bigr] \, .
\end{align}
Plugging these integrals into expressions \eqref{RHOinI} and \eqref{PinI}
gives us the IR contributions to energy density and pressure during inflationary
period. Combining these with the UV contributions \eqref{rhoUVconstEpsilon}
and \eqref{pUVconstEpsilon} specialized to $\epsilon\!=\!0$ (including the 
conformal anomaly \eqref{CAeps} as well), 
the dependence on the
artificially introduced UV cutoff cancels as promised, 
and the physical renormalized quantity remains,
\begin{align}
\rho_Q ={}& \frac{H_I^4}{32\pi^2} \Biggl\{ 
	- (1 \!-\! 6\xi)^2 + \frac{1}{30} 
	- 2(1 \!-\! 6\xi) \Bigl[ \frac{1\!-\!C_0}{2} \Bigr]
	\Bigl( \frac{m}{H} \Bigr)^{\!2}
	- \Bigl[ \frac{1\!+\!2C_0}{4} \Bigr]
	\Bigl( \frac{m}{H_I} \Bigr)^{\!4}  \Biggr\} 
\nonumber \\
&	+ \frac{4^{\nu_I-2}\, \Gamma^2(\nu_I)}
		{\pi^3 \left( \nu_I\!-\!\frac{3}{2} \right)}
	\biggl[ \nu_I \Bigl( \nu_I \!-\! \tfrac{3}{2} \!+\! 6\xi \Bigr)
		+ \Bigl( \frac{m}{H_I} \Bigr)^{\!2} \biggr] H_I^4
	\Bigl( \frac{a_0H_I}{k_0} \Bigr)^{2\nu_I-3} \, ,
\label{rhoINF}
\\
p_Q ={}& \frac{H_I^4}{32\pi^2} \Biggl\{ 
	(1 \!-\! 6\xi)^2 - \frac{1}{30} 
	+ 2(1 \!-\! 6\xi) \Bigl[ \frac{1\!-\!C_0}{2} \Bigr]
	\Bigl( \frac{m}{H_I} \Bigr)^{\!2}
	+ \Bigl[ \frac{1\!+\!2C_0}{4} \Bigr]
	\Bigl( \frac{m}{H_I} \Bigr)^{\!4}  \Biggr\} 
\nonumber \\
&	+ \frac{4^{\nu_I-2}\, \Gamma^2(\nu_I)}
		{\pi^3 \left( \nu_I\!-\!\frac{3}{2} \right)} 
		 \nu_I \Bigl[ \nu_I \!-\! \tfrac{3}{2} 
		\!-\! 4\xi (\nu_I \!-\! 2) \Bigr] H_I^4
	\Bigl( \frac{a_0H_I}{k_0} \Bigr)^{2\nu_I-3} \, ,
\label{pINF}
\end{align}
where
\begin{equation}
C_0 = 2\ln2 - 1 - \psi\Bigl( -\tfrac{1}{2} \!+\! \nu_I \Bigr)
	- \psi\Bigl( -\tfrac{1}{2} \!-\! \nu_I \Bigr) \, ,
\end{equation}
and $\nu_I$ is given in \eqref{nuI}.
In the expressions above we have reverted to using the physical Hubble rate
$H_I(=\!a\mathcal{H})$, assumed to be constant during inflation 
($\epsilon\!=\!0$), in order to make the expression more transparent.

The IR cutoff $k_0$ in \eqref{rhoINF} and \eqref{pINF}
(and in the results to follow) can be given physical meaning by relating it to the 
Hubble scale at the beginning of inflation $\mathcal{H}_0$. In 
\cite{Janssen:2009nz} (and \cite{Glavan:2014uga}),
the IR regularization method employed was matching the inflationary period
onto a pre-inflationary radiation-dominated period, where IR issues are absent
(due to vanishing Ricci scalar), so the Hubble rate at the beginning of inflation
was explicitly introduced. By specializing that result to $\epsilon\!=\!0$, and
comparing it to the one above, we see that $k_0\!\sim\!\mathcal{H}_0$
of course. In this paper we will be dealing with the nonminimal coupling
restricted to $0 \! \le \! -\xi \!\ll\! 1$, in which case on de Sitter the two scales
coincide to leading order in $\xi$, $k_0 \!=\! \mathcal{H}_0$. 
Therefore, from now on we will be making this 
identification.


\subsubsection{Minimally coupled limit}

Setting the nonminimal coupling $\xi$ to zero, and working in the small
mass limit gives
\begin{align}
\rho_Q ={}& - \frac{119H_I^4}{960\pi^2}
	+ \frac{3 H_I^4}{16\pi^2} 
	\biggl[ 1 - e^{-\frac{2}{3} \frac{m^2}{H_I^2} N} \biggr] \, ,
\\
p_Q ={}&  \frac{119H_I^4}{960\pi^2}
	- \frac{3 H_I^4}{16\pi^2} 
	\biggl[ 1 - e^{-\frac{2}{3}\frac{m^2}{H_I^2}N} \biggr] \, ,
\end{align}
for the backreaction energy density and pressure,
which is a standard result.
In the expressions above $N$ stands for the number of e-foldings from
the beginning of inflation, $N\!=\!\ln(a/a_0)$.
 The first term corresponds to the energy density
and pressure of an exactly massless scalar in a CTBD state during de Sitter inflation
\cite{Habib:1999cs,Glavan:2013mra}. The second term is a contribution from 
$m^2\langle\phi^2\rangle$, whose behavior is well known 
for (slow roll) inflationary backgrounds 
\cite{Starobinsky:1986fx,Finelli:2008zg,Finelli:2010sh}. 
For an extremely long inflation the energy density
and pressure saturate to
\begin{equation}
\rho_Q = \frac{61H_I^4}{960\pi^2} \, , \qquad
	p_Q = - \frac{61H_I^4}{960\pi^2} \, , \qquad
	N_I\gg \Bigl( \frac{m}{H_I} \Bigr)^{\!-2} \, ,
\end{equation}
and contribute just a tiny correction to the effective cosmological constant
(determined by the expansion rate).
This limit was used in \cite{Ringeval:2010hf} in the context of late-time
quantum backreaction.
For a ``short'' inflation the backreaction at the end of inflation
can be approximated to be
\begin{equation}
\rho_Q = - \frac{119H_I^4}{960\pi^2}
	+ \frac{ H_I^2m^2}{8\pi^2} N_I \, , \qquad
	p_Q = \frac{119H_I^4}{960\pi^2}
	- \frac{ H_I^2m^2}{8\pi^2} N_I \, , \qquad
	N_I\ll \Bigl( \frac{m}{H_I} \Bigr)^{\!-2} \, ,
\end{equation}
where here and henceforth $N_I$ is the total number of e-foldings of inflation.
But $N_I$ does not have to be small, in fact it can still be very large if $m$ is 
very small compared to $H_I$.
This limit proved better when constructing a DE model
based on backreaction \cite{Aoki:2014dqa}.


\subsubsection{Limit $(m/H_I)^2\ll|\xi|\ll1$}

\begin{figure}[h!]
\includegraphics[width=10cm]{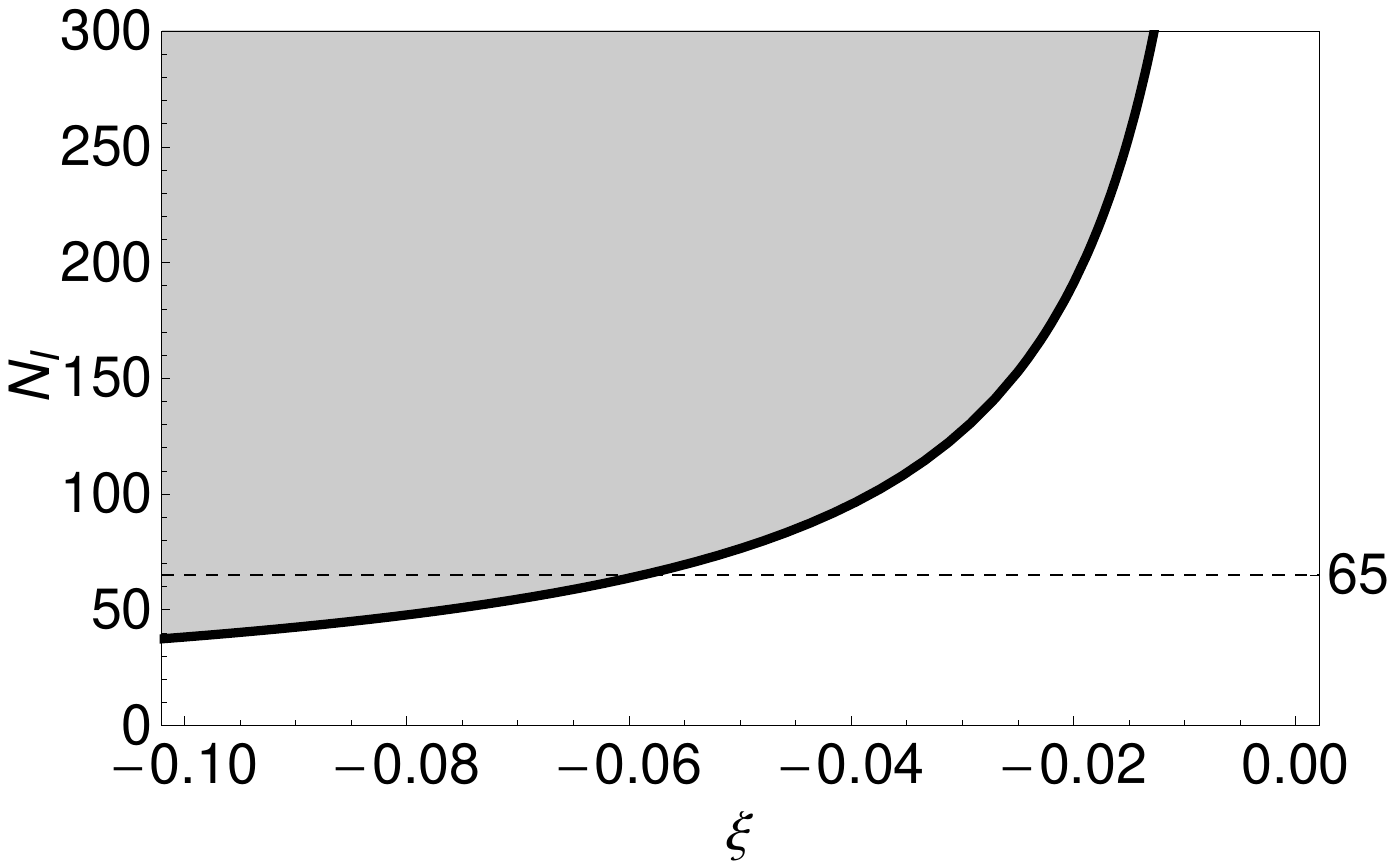}
\vskip-0.5cm
\caption{Parameter space of nonminimal coupling $\xi$, and the 
total number of e-foldings of inflation $N_I$. The bold curve corresponds to
the condition $\rho_Q/\rho_B\!=\!1$ at the end of inflation 
($H_I\!=\!10^{13}\,{\rm GeV}$), 
and the shaded region to situations when
quantum backreaction starts to dominate even before the end of inflation.
This part of parameter space is not examined in this work, but rather
concentrate on the white region where quantum backreaction stays perturbative
during inflation. The dashed line represents the requirement on the minimal
duration of inflation. It follows that we restrict ourselves to considering
$-0.05\lesssim\xi\le0$ in this work.}
\label{InflConstraint}
\end{figure}

This is effectively a massless limit of the full result \eqref{rhoINF}
and \eqref{pINF}, and coincides with the results in \cite{Janssen:2009nz},
\begin{equation}
\rho_Q \approx - \frac{3H_I^4}{32\pi^2} \, e^{8|\xi|N_I}
	\Bigl[ 1 + \mathcal{O}(\xi) \Bigr] \, ,
	\qquad p_Q \approx - \rho_Q \, .
\label{backINF}
\end{equation}
In the end we will be interested in this range of parameters during inflation,
when we try to construct a model in which the quantum backreaction is small
throughout the expansion history, and becomes large only at the onset of 
the DE-dominated period (Fig. \ref{expansion history}). We see that the 
backreaction \eqref{backINF} is negative and grows in amplitude exponentially
with $N$ during 
inflation, and how much
it grows depends on the value of nonminimal coupling and the duration of inflation.
Since we want the backreaction to remain perturbative during inflation, this 
imposes a constraint on the $\xi-N_I$ parameter space depicted in Fig.
\ref{InflConstraint}, which was derived by requiring $\rho_Q/\rho_B\!<\!1$.
This implies for the number of e-foldings,
\begin{equation}
N_I \le \frac{1}{8|\xi|} 
	\ln\biggl[ 4\pi \Bigl( \frac{E_P}{\hbar H_I} \Bigr)^2 \biggr] \, ,
\label{boundInf}
\end{equation}
where the dimensionful units were restored, and 
$E_P\!=\!(\hbar c^5/G_{\!N})^{1/2}$ is the Planck energy,
and the inflationary Hubble scale is taken to be $\hbar H_I \!\sim\! 10^{13} 
\,{\rm GeV}$.

Although strong backreaction in inflation would be 
very interesting to study in its own right (especially since its energy density
has a negative sign which would work towards slowing down inflation), here
we restrict ourself to studying just the DE scenarios, for which we assume small
backreaction at the end of inflation.


\subsection{Radiation era}

Some time after the transition to the radiation period the hierarchy of scales
$k_0 \!\ll\!\mathcal{H}\!\ll\!\mathcal{H}_1$ is reached (together with the assumed
$(ma)^2/\mathcal{H}^2\!\ll\! 1$). The relevant contribution to 
integrals \eqref{integralsI} is
\begin{equation}
\mathcal{I}_n \approx \int\limits_{k_0}^{\mu} \! dk \, k^{2+2n}
	Z_{\text{Bog.}}(k,\eta) \, ,
\label{radI}
\end{equation}
as was established in subsection \ref{subsubsec: Decelerating period}, with
\begin{equation}
k_0 \!\ll\! \mathcal{H} \!\ll\! \mu \!\ll\! \mathcal{H}_1 \, ,
\label{radHierarchy}
\end{equation}
and the integrand, as defined in 
\eqref{IntBog}, is
\begin{equation}
Z_{\text{Bog.}}(k,\eta) = 2|\beta_R(k)|^2 |u_R(k,\eta)|^2
	+ \alpha_R(k) \beta^*(k) u_R^2(k,\eta)
	+ \alpha_R^*(k) \beta_R(k) [u_R^2(k,\eta)]^* \, .
\label{radIntegrand}
\end{equation}
The Bogolyubov coefficients in this integrand are determined by the fast transition
from the inflationary period to the radiation one. For the scales integrated over 
in \eqref{radI} they are well approximated by the sudden transition ones
\eqref{suddenAlpha} and \eqref{suddenBeta},
\begin{align}
\alpha_R(k) ={}& -i \Bigl[ u_I(k,\eta_1) u_R'^*(k,\eta_1)
		- u_I'(k,\eta_1) u_R^*(k,\eta_1) \Bigr] \, ,
\\
\beta_R(k) ={}& i \Bigl[ u_I(k,\eta_1) u_R'(k,\eta_1)
		- u_I'(k,\eta_1) u_R(k,\eta_1) \Bigr] \, .
\end{align}
The inflationary CTBD mode function in Bogolyubov coefficients above 
is given by \eqref{CTBDinf}, and the radiation CTBD mode function 
by \eqref{radCTBDlinCombination}. For the range of integration in 
\eqref{radI}, the mode functions inside of Bogolyubov coefficients are in fact
very well described by the small momentum limit (on top of small mass limit).
We use this to simplify the integrand before integration.

Using the IR expansion \eqref{uI_IR} for the inflationary CTBD mode function, 
it follows that
\begin{equation}
u_I'(k,\eta_1) \approx \bigl( \nu_I \!-\! \tfrac{1}{2} \bigr)
	\mathcal{H}_1 u_I(k,\eta_1) \, ,
\end{equation}
and the Bogolyubov coefficients simplify to
\begin{align}
\alpha_R(k) \approx{}& - \bigl| u_I(k,\eta_1) \bigr|
	\Bigl[ u_R'^*(k,\eta_1)
	- \bigl( \nu_I \!-\! \tfrac{1}{2} \bigr) \mathcal{H}_1 u_R^*(k,\eta_1) \Bigr]
	\equiv - \bigl| u_I(k,\eta_1) \bigr| \widetilde{\beta}_R^*(k) \, ,
\\
\beta_R(k) \approx{}&  \bigl| u_I(k,\eta_1) \bigr|
	\Bigl[ u_R'(k,\eta_1)
	- \bigl( \nu_I \!-\! \tfrac{1}{2} \bigr) \mathcal{H}_1 u_R(k,\eta_1) \Bigr]
	\equiv  \bigl| u_I(k,\eta_1) \bigr| \widetilde{\beta}_R(k) \, ,
\end{align}
and to leading order satisfy
\begin{equation}
\alpha_R(k) \approx - \beta_R^*(k) \, ,
\end{equation}
which implies the following simplification for the integrand \eqref{radIntegrand},
\begin{equation}
Z_{\text{Bog.}}(k,\eta) \approx 4 \bigl| u_I(k,\eta_1) \bigr|^2
	\biggl\{ \Im\Bigl[ \widetilde{\beta}_R^* \, u_R(k,\eta) \Bigr] \biggr\}^2 \, .
\end{equation}
This integrand is further simplified by considering the small mass limit
$ma/\mathcal{H} \!\ll\! 1$, using the approximate mode functions \eqref{vR1}
and \eqref{vR2},
\begin{align}
\Im \Bigl[ \widetilde{\beta}_R^*(k) u_R(k,\eta) \Bigr] \approx{}&
	\Im \Bigl[ A_R^*(k,m) B_R(k,m) \Bigr]  \times
	\biggl\{ v_{R2}'(k,\eta_1) v_{R1}(k,\eta)
	- v_{R1}'(k,\eta_1) v_{R2}(k,\eta) 
\nonumber \\
&	- \bigl( \nu_I \!-\! \tfrac{1}{2} \bigr) \mathcal{H}_1
	\Bigl[ v_{R2}(k,\eta_1) v_{R1}(k,\eta) - v_{R1}(k,\eta_1) v_{R2}(k,\eta) \Bigr]
	\biggr\}
\nonumber \\
={}& \frac{1}{2} \biggl\{ v_{R2}'(k,\eta_1) v_{R1}(k,\eta)
	- v_{R1}'(k,\eta_1) v_{R2}(k,\eta) 
\nonumber \\
&	- \bigl( \nu_I \!-\! \tfrac{1}{2} \bigr) \mathcal{H}_1
	\Bigl[ v_{R2}(k,\eta_1) v_{R1}(k,\eta) - v_{R1}(k,\eta_1) v_{R2}(k,\eta) \Bigr]
	\biggr\} \, ,
\end{align}
where the property \eqref{radCoeffProperty} was used, and it is the only place
where we need to refer to coefficients $A_R$ and $B_R$, no matter how
complicated they may be. Furthermore, we may expand this expression to
leading order in $\mathcal{H}_1$ because of hierarchy \eqref{radHierarchy},
\begin{align}
\Im \Bigl[ \widetilde{\beta}_R^*(k) u_R(k,\eta) \Bigr] \approx{}&
	 - \frac{1}{2} \bigl( \nu_I \!-\! \tfrac{1}{2} \bigr) \Biggl\{ 
	\frac{\mathcal{H}}{k} \sin\Bigl( \frac{k}{\mathcal{H}} \Bigr)
	+ \Bigl( \frac{ma}{\mathcal{H}} \Bigr)^{\!2} 
	\Bigl( \frac{\mathcal{H}}{k} \Bigr)^{\!2}
	\biggl[ - \Bigl( 1 \!-\! \frac{\mathcal{H}^2}{k^2} \Bigr)
		\frac{\mathcal{H}}{4k} \sin\Bigl( \frac{k}{\mathcal{H}} \Bigr)
\nonumber \\
&	+ \frac{1}{6} \Bigl( 1 \!-\! \frac{3\mathcal{H}^2}{2k^2} \Bigr)
	\cos\Bigl( \frac{k}{\mathcal{H}} \Bigr) \biggr]
	+ \mathcal{O}\Bigl( \frac{ma}{\mathcal{H}} \Bigr)^{\!4} \Biggr\}
	\Bigl( \frac{\mathcal{H}_1}{\mathcal{H}} \Bigr)
	\Bigl[ 1 + \mathcal{O}\Bigl( \frac{\mathcal{H}}{\mathcal{H}_1} \Bigr) 
	\Bigr] \, ,
\end{align}
so that now, after plugging in \eqref{CTBDinf}, the full integrand is 
well approximated by
\begin{align}
Z_{\text{Bog.}} \approx{}&
	\frac{4^{\nu_I-1}}{\pi} \Gamma^2(\nu_I) 
	\bigl( \nu_I \!-\! \tfrac{1}{2} \bigr)^2
	\Bigl( \frac{\mathcal{H}_1}{\mathcal{H}} \Bigr)^{\!2}
	\mathcal{H}_1^{2\nu_I-1} k^{-2\nu_I}
	\Biggl\{ \frac{\mathcal{H}^2}{k^2} 
	\sin^2 \Bigl( \frac{k}{\mathcal{H}} \Bigr) 
\nonumber \\
&	+ \Bigl( \frac{ma}{\mathcal{H}} \Bigr)^{\!2} \times
	\frac{2\mathcal{H}^3}{k^3} \sin\Bigl( \frac{k}{\mathcal{H}} \Bigr)
	\biggl[ -\Bigl( 1 \!-\! \frac{\mathcal{H}^2}{k^2} \Bigr) \frac{\mathcal{H}}{4k}
	\sin\Bigl( \frac{k}{\mathcal{H}} \Bigr)
	+ \frac{1}{6} \Bigl( 1 \!-\! \frac{3\mathcal{H}^2}{2k^2} \Bigr)
	\cos\Bigl( \frac{k}{\mathcal{H}} \Bigr) \biggr]
\Biggr\} \, .
\label{radIntegrandApprox}
\end{align}

Finally, we can perform the integrals \eqref{radI} using this approximated 
integrand. The result we expand according to the hierarchy \eqref{radHierarchy},
\begin{align}
\mathcal{I}_0 \approx{}& 
	\frac{\Gamma^2(\nu_I)}{2^{3-2\nu_I}\pi} 
	\bigl( \nu_I \!-\! \tfrac{1}{2} \bigr)^2
	\Biggl\{ 
	\frac{1}{\left( \nu_I \!-\! \frac{3}{2} \right)}
	\biggl[ 1 - \frac{1}{10} \Bigl( \frac{ma}{\mathcal{H}} \Bigr)^{\!2}
	+ \mathcal{O}\Bigl( \frac{ma}{\mathcal{H}} \Bigr)^{\!4} \biggr]
	\Bigl( \frac{\mathcal{H}_0}{\mathcal{H}} \Bigr)^{\!3-2\nu_I}
	\Bigl[ 1 
	+ \mathcal{O} \Bigl( \frac{\mathcal{H}_0}{\mathcal{H}} \Bigr)^{\!2} \Bigr]
\nonumber \\
&	- \frac{\Gamma(1\!-\!2\nu_I)}{2^{1-2\nu_I}} \sin(\pi\nu_I)
	\biggl[ 1 - \frac{8 \Gamma(-3\!-\!2\nu_I)}{3\Gamma(1\!-\!2\nu_I)} 
	\bigl( \nu_I \!-\! \tfrac{1}{2} \bigr)
	(2\nu_I^2 \!+\! 4\nu_I \!+\! 3) 
	\Bigl( \frac{ma}{\mathcal{H}} \Bigr)^{\!2}
	+ \mathcal{O}\Bigl( \frac{ma}{\mathcal{H}} \Bigr)^{\!4} \biggr] 
\nonumber \\
&	
	- \frac{1}{2\left( \nu_I \!-\! \frac{1}{2} \right)}
	\Bigl( \frac{\mathcal{H}}{\mu} \Bigr)^{2\nu_I-1}
	\Bigl[ 1 + \mathcal{O} \Bigl( \frac{\mathcal{H}}{\mu} \Bigr)^{\!2} \Bigr] 
	\Biggr\} \mathcal{H}^2 
	\Bigl( \frac{\mathcal{H}_1}{\mathcal{H}} \Bigr)^{\!2\nu_I+1}
	\Bigl[ 1 + \mathcal{O} \Bigl( \frac{\mathcal{H}}{\mathcal{H}_1} \Bigr) \Bigr]
\, ,
\end{align}
\begin{align}
\mathcal{I}_1 \approx{}& 
	\frac{\Gamma^2(\nu_I)}{2^{3-2\nu_I}\pi} 
	\bigl( \nu_I \!-\! \tfrac{1}{2} \bigr)^2
	\Biggl\{ 
	\frac{1}{\left( \nu_I \!-\! \frac{5}{2} \right)}
	\biggl[ 1 - \frac{1}{10} \Bigl( \frac{ma}{\mathcal{H}} \Bigr)^{\!2}
	+ \mathcal{O}\Bigl( \frac{ma}{\mathcal{H}} \Bigr)^{\!4} \biggr]
	\Bigl( \frac{\mathcal{H}_0}{\mathcal{H}} \Bigr)^{\!5-2\nu_I}
	\Bigl[ 1 
	+ \mathcal{O} \Bigl( \frac{\mathcal{H}_0}{\mathcal{H}} \Bigr)^{\!2} \Bigr]
\nonumber \\
&	+ \frac{\Gamma(3\!-\!2\nu_I)}{2^{3-2\nu_I}} \sin(\pi\nu_I)
	\biggl[ 1 - \frac{8 \Gamma(-1\!-\!2\nu_I)}{3\Gamma(3\!-\!2\nu_I)} 
	\bigl( \nu_I \!-\! \tfrac{3}{2} \bigr)
	(2\nu_I^2 \!+\! 1) 
	\Bigl( \frac{ma}{\mathcal{H}} \Bigr)^{\!2} 
	+ \mathcal{O}\Bigl( \frac{ma}{\mathcal{H}} \Bigr)^{\!4}\biggr] 
\nonumber \\
&	- \frac{1}{2\left( \nu_I \!-\! \frac{3}{2} \right)}
	\Bigl( \frac{\mathcal{H}}{\mu} \Bigr)^{2\nu_I-3}
	\Bigl[ 1 + \mathcal{O} \Bigl( \frac{\mathcal{H}}{\mu} \Bigr)^{\!2} \Bigr] 
	\Biggr\} \mathcal{H}^4
	\Bigl( \frac{\mathcal{H}_1}{\mathcal{H}} \Bigr)^{\!2\nu_I+1}
	\Bigl[ 1 + \mathcal{O} \Bigl( \frac{\mathcal{H}}{\mathcal{H}_1} \Bigr) \Bigr]
\, .
\end{align}
Then plugging them into \eqref{RHOinI}  and \eqref{PinI} gives the 
backreaction energy density and pressure (in the small mass limit),
\begin{align}
\rho_Q \approx{}&
	\frac{\Gamma(\nu_I)H_I^4}{2^{5-2\nu_I}\pi^3}
	\bigl( \nu_I \!-\! \tfrac{1}{2} \bigr) e^{(2\nu_I \!-\! 3)N_I}
	\Biggl\{ 
	\frac{\left( \nu_I \!-\! \frac{1}{2} \right) \Gamma(\nu_I)}
		{\left( \nu_I \!-\! \frac{3}{2} \right)} \,
	 \biggl[ 6\xi \Bigl( \frac{a_1}{a} \Bigr)^{\!4}
	+ (1 \!-\! 3\xi) \Bigl( \frac{m}{H_I} \Bigr)^{\!2} \biggr]
\nonumber \\
&	\hspace{0.5cm}
	- \frac{\left( \nu_I \!-\! \frac{1}{2} \right) \Gamma(\nu_I)}
		{\left( \nu_I \!-\! \frac{3}{2} \right)} 
		\Bigl( \frac{\mathcal{H}_0}{\mu} \Bigr)^{\! 2\nu_I - 3}
	\Bigl( \frac{a_1}{a} \Bigr)^{\!4}
	- \sqrt{\pi} (\nu_I \!-\! 1) \Gamma\bigl( \tfrac{3}{2} \!-\! \nu_I \bigr)
	 \biggl[ (1 \!-\! 6\xi)\Bigl( \frac{a_1}{a} \Bigr)^{\! 4} 
\nonumber \\
&	\hspace{0.5cm}
	- \frac{ (1 \!-\! 6\xi) \left( \nu_I \!-\! \frac{3}{2} \right) 
		+ 3\nu_I \left( \nu_I \!+\! \frac{3}{2} \right)
		+ 2(1\!-\!6\xi)\nu_I^2 \left( \nu_I \!+\! \frac{3}{2} \right)}
		{6\nu_I(\nu_I \!-\!1)\left( \nu_I \!+\! \frac{1}{2} \right)
			\left( \nu_I \!+\! \frac{3}{2} \right)} 
	\Bigl( \frac{m}{H_I} \Bigr)^{\!2} \biggr] 
	\Bigl( \frac{\mathcal{H}_0}{\mathcal{H}} \Bigr)^{\! 3 \!-\! 2\nu_I}
\Biggr\} \, ,
\label{rhoQm}
\\
p_Q \approx{}&
	\frac{\Gamma(\nu_I)H_I^4}{2^{5-2\nu_I}\pi^3}
	\bigl( \nu_I \!-\! \tfrac{1}{2} \bigr) e^{(2\nu_I \!-\! 3)N_I}
	\Biggl\{ 
	\frac{\left( \nu_I \!-\! \frac{1}{2} \right) \Gamma(\nu_I)}
		{\left( \nu_I \!-\! \frac{3}{2} \right)} \,
	 \biggl[ 2\xi \Bigl( \frac{a_1}{a} \Bigr)^{\!4}
	- (1 \!-\! 3\xi) \Bigl( \frac{m}{H_I} \Bigr)^{\!2} \biggr]
\nonumber \\
&	\hspace{0.5cm}
	- \frac{\left( \nu_I \!-\! \frac{1}{2} \right) \Gamma(\nu_I)}
		{3\left( \nu_I \!-\! \frac{3}{2} \right)} 
		\Bigl( \frac{\mathcal{H}_0}{\mu} \Bigr)^{\! 2\nu_I - 3}
	\Bigl( \frac{a_1}{a} \Bigr)^{\!4}
	+ \frac{2\sqrt{\pi}}{3} 
	(\nu_I \!-\! 1) \Gamma\bigl( \tfrac{3}{2} \!-\! \nu_I \bigr)
	 \biggl[ (1 \!-\! 6\xi)(\nu_I\!-\!2)\Bigl( \frac{a_1}{a} \Bigr)^{\! 4} 
\nonumber \\
&	\hspace{0.5cm}
	- \frac{ (1 \!-\! 6\xi) \left( \nu_I \!-\! \frac{3}{2} \right) 
		+ 3\nu_I \left( \nu_I \!+\! \frac{3}{2} \right)
		+ 2(1\!-\!6\xi)\nu_I^2 \left( \nu_I \!+\! \frac{3}{2} \right)}
		{6(\nu_I \!-\!1)\left( \nu_I \!+\! \frac{1}{2} \right)
			\left( \nu_I \!+\! \frac{3}{2} \right)} 
	\Bigl( \frac{m}{H_I} \Bigr)^{\!2} \biggr] 
	\Bigl( \frac{\mathcal{H}_0}{\mathcal{H}} \Bigr)^{\! 3 \!-\! 2\nu_I}
\Biggr\} \, .
\label{pQm}
\end{align}
We have given some parts of the expressions above in terms of physical quantities 
for the sake of clarity. The comment of the dependence on the arbitrary
cutoff $\mu$ is warranted. It is bound to cancel with the same contribution 
from the remaining part of integration interval, and it does not
contribute to the full result. The reason we kept it 
explicitly is because we want to take the minimally coupled limit, which we do 
and discuss in the next subsection.


\subsubsection{Minimally coupled limit}

The minimally coupled limit consists in taking $\xi\!=\!0$, and
then expanding $\nu_I$, defined in \eqref{nuI}, for small mass in \eqref{rhoQm}
and \eqref{pQm},
\begin{align}
\rho_Q \approx{}& 
	\frac{3 H_I^4}{16\pi^2} 
	\biggl[ 1 - e^{- \frac{2}{3}\frac{m^2}{H_I^2} N_I } \biggr]
	+ \frac{H_I^4}{8\pi^2} \biggl[ 
		\ln\Bigl( \frac{a}{a_1} \Bigr) - \frac{1}{2} 
		+ \ln\Bigl( \frac{2\mu}{a_1 H_I} \Bigr)
		+ \gamma_E \biggr] \Bigl( \frac{a_1}{a} \Bigr)^4 \, ,
\label{RHOradMIN}
\\
p_Q \approx{}& 
	- \frac{3 H_I^4}{16\pi^2} 
	\biggl[ 1 - e^{- \frac{2}{3}\frac{m^2}{H_I^2} N_I } \biggr]
	+ \frac{H_I^4}{24\pi^2} \biggl[ 
		\ln\Bigl( \frac{a}{a_1} \Bigr) - \frac{3}{2} 
		+ \ln\Bigl( \frac{2\mu}{a_1 H_I} \Bigr)
		+ \gamma_E \biggr] \Bigl( \frac{a_1}{a} \Bigr)^4 \, .
\label{PradMIN}
\end{align}
Note that the dependence on $\mu$ has persisted in the final answer.
This does not signal that the answer is wrong, but rather that the UV
does not contribute a suppressed contribution. In the exactly massless limit
of this result the first terms drop out from the energy density and pressure above,
and what remains is exactly the result from \cite{Glavan:2013mra},
where the massless minimally coupled case was studied from the start.
What effectively cuts off the UV radiation-like contribution is the finite
time of transition $\tau$ between the inflationary and radiation period.

We have assumed here that the radiation period does not last excessively long,
more precisely,
\begin{equation}
N_R \ll \Bigl( \frac{m}{H_I} \Bigr)^{\!-2} \, ,
\end{equation}
where $N_R$ is the total number of e-foldings of radiation period,
which will be satisfied by the requirements in the end. This first
terms in \eqref{RHOradMIN} and \eqref{PradMIN} coincide with
the ones computed in \cite{Aoki:2014dqa}, since they derive from 
$m^2\langle \hat{\phi}^2 \rangle$ term. Note that this term
is not the dominant one for a very small mass.


\subsubsection{Limit $(m/H)^2\sim|\xi|\ll1$}

The leading order contribution in this limit is
\begin{align}
\rho_Q \approx{}& - \frac{3 H_I^4}{32\pi^2} \, e^{8|\xi|N_I} 
	\Biggl[ \Bigl( \frac{a_1}{a} \Bigr)^{\!4} 
	- \frac{1}{6|\xi|} \Bigl( \frac{m}{H_I} \Bigr)^{\!2} \Biggr] \, ,
\label{radRHOlimit2}
\\
p_Q \approx{}&  -\frac{3 H_I^4}{32\pi^2} \, e^{8|\xi|N_I} 
	\Biggl[ \frac{1}{3} \Bigl( \frac{a_1}{a} \Bigr)^{\!4} 
	+ \frac{1}{6|\xi|} \Bigl( \frac{m}{H_I} \Bigr)^{\!2} \Biggr] \, ,
\label{radPlimit2}
\end{align}
where we have assumed that the radiation period will not last longer (in e-foldings)
than the inflationary period ($N_I\!>\!N_R\sim60$), which will be true in the end
for the ranges of nonminimal couplings of interest in this work. There are two 
qualitatively different contributions to the energy density and pressure above.
The first contribution is radiation-like and it redshifts just as the background does.
It is easy to see that if the constraint from Figure \ref{InflConstraint} is satisfied,
this contribution will never dominate in radiation period. 

The second contribution is of the CC type, and its energy density has a positive
sign. For suitable choices of parameters one can get this contribution to be the
dominant one in the backreaction during radiation period. And since it does
not redshift away the ratio $\rho_Q/\rho_B$ grows, and it might grow to order 
one for suitable masses and small enough nonminimal couplings. 
But we are not interested in this scenario happening in radiation period.
What we are interested in is realizing it in matter period, which we turn to next.
During radiation period we require the CC-type contribution to be negligible,
in which case \eqref{radRHOlimit2} and \eqref{radPlimit2} reduce to the
massless limit of \cite{Glavan:2014uga},
\begin{equation}
\rho_Q \approx - \frac{3 H_I^4}{32\pi^2} \, e^{8|\xi|N_I} 
	\Bigl( \frac{a_1}{a} \Bigr)^{\!4}  \, ,
\qquad
p_Q \approx  \frac{1}{3} \rho_Q \, .
\end{equation}
%


\subsection{Matter era}
\label{Matter era}

After the transition to the matter era, the following hierarchy of scales is reached
(Fig.~\ref{Hubble rate}),
\begin{equation}
\mathcal{H}_0 \ll \mathcal{H} \ll \mu \ll \mathcal{H}_2 \ll \mathcal{H}_1 \, ,
\label{mattHierarchy}
\end{equation}
where $\mu$ is a fiducial scale, introduced for the sake of isolating the relevant
contribution to integrals \eqref{integralsI},
\begin{equation}
\mathcal{I}_n \approx \int\limits_{\mathcal{H}_0}^{\mu}
	\! dk\, k^{2+2n} Z_{\text{Bog.}}(k,\eta) \, ,
\end{equation}
as discussed in Section \ref{sec: Energy density and pressure integrals}.
The integrand here is
\begin{equation}
Z_{\text{Bog.}} = 2\bigl| \beta_M(k) \bigr|^2 \bigl| u_M(k,\eta) \bigr|^2
	+ \alpha_M(k) \beta_M^*(k) u_M^2(k,\eta)
	+ \alpha_M^*(k) \beta_M(k) [u_M^2(k,\eta)]^* \, ,
\label{mattIntegrand}
\end{equation}
where the Bogolyubov coefficients are well approximated by the sudden transition
ones \eqref{suddenAlpha} and \eqref{suddenBeta} determined by two fast 
transitions -- from inflation to radiation, and from radiation to matter,
\begin{align}
\alpha_M(k) ={}& -i \Bigl[ \alpha_R(k) u_R(k,\eta_2)
		+ \beta_R(k) u_R^*(k,\eta_2) \Bigr] u_M'^*(k,\eta_2)
\nonumber \\
&	\hspace{2cm}	
	+ i \Bigl[ \alpha_R(k) u_R'(k,\eta_2) 
		+ \beta_R(k) u_R'^*(k,\eta_2) \Bigr] u_M^*(k,\eta_2) \, ,
\\
\beta_M(k) ={}& i \Bigl[ \alpha_R(k) u_R(k,\eta_2)
		+ \beta_R(k) u_R^*(k,\eta_2) \Bigr] u_M'(k,\eta_2)
\nonumber \\
&	\hspace{2cm}	
	+ i \Bigl[ \alpha_R(k) u_R'(k,\eta_2) 
		+ \beta_R(k) u_R'^*(k,\eta_2) \Bigr] u_M(k,\eta_2) \, .
\end{align}
Bogolyubov coefficients in radiation period $\alpha_R$ and $\beta_R$,
appearing in the expression above, were already approximated in the
previous subsection, where it was found that $\alpha_R \!\approx\! - \beta_R^*$.
Applying this here gives
\begin{align}
\alpha_M(k) \approx{}&
	-2 \Im\Bigl[ \beta_R^*(k) u_R(k,\eta_2) \Bigr] u_M'^*(k,\eta_2)
	+ 2 \Im \Bigl[ \beta_R^*(k) u_R'(k,\eta_2) \Bigr] u_M^*(k,\eta_2) \, ,
\\
\beta_M(k) \approx{}&
	2 \Im\Bigl[ \beta_R^*(k) u_R(k,\eta_2) \Bigr] u_M'(k,\eta_2)
	- 2 \Im \Bigl[ \beta_R^*(k) u_R'(k,\eta_2) \Bigr] u_M(k,\eta_2) \, ,
\end{align}
from where we see that again
\begin{equation}
\alpha_M(k) \approx - \beta_M^*(k) \, .
\end{equation}
Now the integrand \eqref{mattIntegrand} simplifies to
\begin{equation}
Z_{\text{Bog.}}(k,\eta) \approx 4 \biggl\{ \Im \Bigl[ 
	\beta_M^*(k) u_M(k,\eta) \Bigr] \biggr\}^2 \, .
\end{equation}
which we can write out as
\begin{align}
\Im\Bigl[ \beta_M^*(k) u_M(k,\eta) \Bigr] ={}&
	2 \Im\Bigl[ \beta_R^*(k) u_R(k,\eta_2) \Bigr] 
	\Im\Bigl[ u_M'^*(k,\eta_2) u_M(k,\eta) \Bigr]
\nonumber \\
&	- 2 \Im\Bigl[ \beta_R^*(k) u_R'(k,\eta_2) \Bigr]
	\Im\Bigl[ u_M^*(k,\eta_2) u_M(k,\eta) \Bigr] \, .
\end{align}
Part of this integrand was already approximated in \eqref{radIntegrandApprox}.
Because of the hierarchy of scales \eqref{mattHierarchy} in matter period,
we may expand this further in the IR limit,
\begin{align}
\Im\Bigl[ \beta_R^*(k) u_R(k,\eta_2) \Bigr]
	\approx{}& - \frac{2^{\nu_I-2}}{\sqrt{\pi}}
	\Gamma(\nu_I)
	\bigl( \nu_I \!-\! \tfrac{1}{2} \bigr)
	\mathcal{H}_1^{\nu_I+1/2}\mathcal{H}_2^{-1} k^{-\nu_I}
	\biggl[ 1 \!-\! \frac{1}{20} 
	\Bigl( \frac{ma_2}{\mathcal{H}_2} \Bigr)^{\!2} \biggr] \, ,
\\
\Im\Bigl[ \beta_R^*(k) u_R'(k,\eta_2) \Bigr]
	\approx{}& - \frac{2^{\nu_I-2}}{\sqrt{\pi}}
	\Gamma(\nu_I)
	\bigl( \nu_I \!-\! \tfrac{1}{2} \bigr)
	\mathcal{H}_1^{\nu_I+1/2} k^{-\nu_I}
	\biggl[ 1 \!-\! \frac{1}{8} 
	\Bigl( \frac{ma_2}{\mathcal{H}_2} \Bigr)^{\!2} \biggr] \, .
\end{align}
In the remaining part of the integrand we first use the property 
\eqref{mattCoeffProperty} to express it solely in terms of mode functions
\eqref{vM1} and \eqref{vM2},
\begin{align}
 \Im\Bigl[ u_M'^*(k,\eta_2) u_M(k,\eta) \Bigr]
={}& \Im\Bigl[ A_M^*(k,m) B_M(k,m) \Bigr] 
	\!\times\! \Bigl[ v_{M1}(k,\eta_1) v_{M2}(k,\eta)
		- v_{M2}(k,\eta_2) v_{M1}(k,\eta) \Bigr]
\nonumber \\
={}& \frac{1}{2\nu} \Bigl[ v_{M1}(k,\eta_1) v_{M2}(k,\eta)
		- v_{M2}(k,\eta_2) v_{M1}(k,\eta) \Bigr] \, ,
\end{align}
and then we use the hierarchy \eqref{mattHierarchy} to simplify it further
by expanding it to leading order in $\mathcal{H}_2$,
\begin{align}
& \Im\Bigl[ u_M^*(k,\eta_2) u_M(k,\eta) \Bigr] 
	\approx
	\frac{\Gamma(\nu)}{2} \mathcal{H}_2^{-1/2+\nu} 
	\mathcal{H}^{-1/2} k^{-\nu} 
\nonumber \\
&	\hspace{1cm}
	\times \Biggl\{ 
	J_\nu \Bigl( \frac{2k}{\mathcal{H}} \Bigr)
	+ \Bigl( \frac{ma}{\mathcal{H}} \Bigr)^2	\times
	\frac{\mathcal{H}}{30k} \biggl[ 
	\Bigl( -6 \!+\! (1\!-\!\nu)(2\!-\!\nu) \frac{\mathcal{H}^2}{k^2} \Bigr) 
		J_{1+\nu} \Bigl( \frac{2k}{\mathcal{H}} \Bigr) 
\nonumber \\
&	\hspace{5.3cm}
	+ (2\!-\!\nu) 
	\Bigl( 3 \!-\! (1\!-\!\nu)(2\!+\!\nu) \frac{\mathcal{H}^2}{k^2} \Bigr)
		\frac{\mathcal{H}}{k} J_{2+\nu} \Bigl( \frac{2k}{\mathcal{H}} \Bigr) 
	\biggr] \Biggr\} \, ,
\\
& \Im\Bigl[ u_M'^*(k,\eta_2) u_M(k,\eta) \Bigr] 
	\approx
	- \frac{1}{2} \bigl( \nu \!-\! \tfrac{1}{2} \bigr)
	\mathcal{H}_2 \times
	\Im\Bigl[ u_M^*(k,\eta_2) u_M(k,\eta) \Bigr] \, .
\end{align}
The full integrand is now given approximately as
\begin{align}
Z_{\text{Bog.}}(k,\eta) \approx{}& \frac{2^{2\nu_I-2}}{\pi} 
	\Gamma^2(\nu_I)
	\bigl( \nu_I \!-\! \tfrac{1}{2} \bigr)^2 
	\bigl( \nu \!+\! \tfrac{3}{2} \bigr)^2
	\mathcal{H}_1^{2\nu_I-1} k^{-2\nu_I}
	\biggl\{ \Im \Bigl[ u_M^*(k,\eta_2) u_M(k,\eta) \Bigr] \biggr\}^2
\nonumber \\
={}& \frac{2^{2\nu_I-4}}{\pi} \Gamma^2(\nu_I) \Gamma^2(\nu)
	\bigl( \nu_I \!-\! \tfrac{1}{2} \bigr)^2
	\bigl( \nu \!+\! \tfrac{3}{2} \bigr)^2
	\mathcal{H}_1^{2\nu_I+1} \mathcal{H}_2^{2\nu-1}
	\mathcal{H}^{-1} k^{-2\nu_I-2\nu}
\nonumber \\
& \times \Biggl\{ J_{\nu}^2\Bigl( \frac{2k}{\mathcal{H}} \Bigr)
	+ \Bigl( \frac{ma}{\mathcal{H}} \Bigr)^2 \times
	\frac{\mathcal{H}}{15k} J_\nu\Bigl( \frac{2k}{\mathcal{H}} \Bigr)
	\biggl[ \Bigl( - 6 \!+\! (1\!-\!\nu)(2\!-\!\nu) \frac{\mathcal{H}^2}{k^2} \Bigr)
	J_{1+\nu} \Bigl( \frac{2k}{\mathcal{H}} \Bigr) 
\nonumber \\
&	\hspace{4cm}
	+ (2\!-\!\nu) 
	\Bigl( 3 \!-\! (1\!-\!\nu)(2\!+\!\nu)\frac{\mathcal{H}^2}{k^2} \Bigr) 
	\frac{\mathcal{H}}{k} J_{2+\nu}\Bigl( \frac{2k}{\mathcal{H}} \Bigr) \biggr] 
	 \Biggr\} \, .
\label{mattZapprox}
\end{align}

Given the approximation \eqref{mattZapprox}, we can now perform integrals
\eqref{integralsI}, and expand the result according to the hierarchy 
\eqref{mattHierarchy},
\begin{align}
\mathcal{I}_0 \approx{}&
	\frac{\Gamma^2(\nu_I)\Gamma^2(\nu)}{2^{5-2\nu_I}\pi}
	\bigl( \nu_I \!-\! \tfrac{1}{2} \bigr)^2
	\bigl( \nu \!+\! \tfrac{3}{2} \bigr)^2
	\times \Biggl\{ 
	\frac{1}{\left( \nu_I \!-\!\tfrac{3}{2} \right) \Gamma^2(1\!+\!\nu)}
	\Bigl( \frac{\mathcal{H}_1}{\mathcal{H}_0} \Bigr)^{2\nu_I-3}
	\biggl[ 1 - \frac{2}{3(3\!+\!\nu)} 
		\Bigl( \frac{ma}{\mathcal{H}} \Bigr)^{\!2} \biggr]
\nonumber \\
&	\hspace{1cm}
	- \frac{2^{-3+2\nu+2\nu_I} \Gamma\left( \frac{3}{2}\!-\!\nu_I \right)
			\Gamma(\nu_I\!+\!\nu\!-\!1)}
		{\sqrt{\pi} \, \Gamma(\nu_I \!+\! \nu \!-\! \frac{1}{2})
		\Gamma(\nu_I\!+\!2\nu\!-\!\frac{1}{2})}
	\Bigl( \frac{\mathcal{H}_1}{\mathcal{H}_2} \Bigr)^{\!2\nu_I-3}
	\Bigl( \frac{\mathcal{H}_2}{\mathcal{H}} \Bigr)^{\!2\nu_I-3}
	\Bigl[ 1 + \mathcal{O}\Bigl( \frac{ma}{\mathcal{H}} \Bigr)^{\!2} 
		\Bigr] \Biggr\}
\nonumber \\
& \times	\Bigl( \frac{\mathcal{H}_1}{\mathcal{H}_2} \Bigr)^{\!4}
	\Bigl( \frac{\mathcal{H}_2}{\mathcal{H}} \Bigr)^{2\nu_I-3} 
	\mathcal{H}^2 \, ,
\end{align}
\begin{align}
\mathcal{I}_1 \approx{}&
	\frac{\Gamma^2(\nu_I)\Gamma^2(\nu)}{2^{5-2\nu_I}\pi}
	\bigl( \nu_I \!-\! \tfrac{1}{2} \bigr)^2
	\bigl( \nu \!+\! \tfrac{3}{2} \bigr)^2
	\times \Biggl\{ 
	\frac{1}{\left( \nu_I \!-\!\tfrac{5}{2} \right) \Gamma^2(1\!+\!\nu)}
	\Bigl( \frac{\mathcal{H}_1}{\mathcal{H}_0} \Bigr)^{2\nu_I-3} 
	\mathcal{H}_0^2
	\biggl[ 1 - \frac{2}{3(3\!+\!\nu)} 
		\Bigl( \frac{ma}{\mathcal{H}} \Bigr)^{\!2} \biggr]
\nonumber \\
&	\hspace{1cm}
	+ \frac{2^{-5+2\nu+2\nu_I} \Gamma\left( \frac{5}{2}\!-\!\nu_I \right)
			\Gamma(\nu_I\!+\!\nu\!-\!2)}
		{\sqrt{\pi} \, \Gamma(\nu_I \!+\! \nu \!-\! \frac{3}{2})
		\Gamma(\nu_I\!+\!2\nu\!-\!\frac{3}{2})}
	\Bigl( \frac{\mathcal{H}_1}{\mathcal{H}_2} \Bigr)^{\!2\nu_I-3}
	\Bigl( \frac{\mathcal{H}_2}{\mathcal{H}} \Bigr)^{\!2\nu_I-3}
	\mathcal{H}^2 
	\Bigl[ 1 + \mathcal{O}\Bigl( \frac{ma}{\mathcal{H}} \Bigr)^{\!2} 
		\Bigr] \Biggr\}
\nonumber \\
&	\times \Bigl( \frac{\mathcal{H}_1}{\mathcal{H}_2} \Bigr)^{\!4}
	\Bigl( \frac{\mathcal{H}_2}{\mathcal{H}} \Bigr)^{2\nu_I-3} 
	\mathcal{H}^2 \, .
\end{align}
Plugging these two integrals into \eqref{RHOinI} and \eqref{PinI}
yields the backreaction energy density and pressure in matter era,
\begin{align}
\rho_Q \approx{}&
	\frac{\Gamma^2(\nu_I) H_I^4}{2^{5-2\nu_I} } 
	\bigl( \nu_I \!-\! \tfrac{1}{2} \bigr)^2
	\bigl( \nu \!+\! \tfrac{3}{2} \bigr)^2
	e^{(2\nu_I-3)N_I}
\nonumber \\
&	\Biggl\{ \biggl[ - \frac{\left( \nu\!-\!\frac{3}{2} \right)
			\left( \nu\!+\!\frac{1}{2} \right)}
		{8\nu(\nu_I\!-\!\frac{3}{2})}
		\, e^{-4N_R} \Bigl( \frac{a_2}{a} \Bigr)^{\!3}
	+ \frac{\left( \nu^2 \!-\! \nu \!+\! \frac{9}{4} \right)}
		{12\nu^2\left( \nu_I\!-\!\frac{3}{2} \right)}
	\Bigl( \frac{m}{H_I} \Bigr)^{\!2} \biggr]
	\Bigl( \frac{a_2}{a} \Bigr)^{\frac{3}{2}-\nu}
\nonumber \\
&	- \frac{(1\!-\!6\xi)(\nu_I\!+\!\nu\!-\!3)\, 
			\Gamma\!\left( \frac{3}{2}\!-\!\nu_I \right)
			\Gamma^2(\nu) \,
			\Gamma(2\nu_I\!+\!2\nu\!-\!4)}
		{\Gamma^2\!\left( \nu_I \!+\! \nu\!-\!\frac{3}{2} \right)
			\Gamma\!\left( \nu_I \!+\! 2\nu \!-\! \frac{1}{2} \right)}	
	\, e^{-4N_R} \Bigl(\frac{a_2}{a} \Bigr)^{\!\frac{9}{2}-\nu}
	\Bigl( \frac{\mathcal{H}_0}{\mathcal{H}} \Bigr)^{2\nu_I-3} \Biggr\} \, ,
\label{mattRHOfull}
\end{align}
\begin{align}
p_Q \approx{}&
	\frac{\Gamma^2(\nu_I) H_I^4}{2^{5-2\nu_I} } 
	\bigl( \nu_I \!-\! \tfrac{1}{2} \bigr)^2
	\bigl( \nu \!+\! \tfrac{3}{2} \bigr)^2
	e^{(2\nu_I-3)N_I}
\nonumber \\
&	\Biggl\{ \biggl[  \frac{\left( \nu\!-\!\frac{3}{2} \right)^2
			\left( \nu\!+\!\frac{1}{2} \right)}
		{24\nu(\nu_I\!-\!\frac{3}{2})}
		\, e^{-4N_R} \Bigl( \frac{a_2}{a} \Bigr)^{\!3}
	- \frac{\left( \nu\!+\!\frac{3}{2} \right)
	\left( \nu^2 \!-\! \nu \!+\! \frac{9}{4} \right)}
		{36\nu^2\left( \nu_I\!-\!\frac{3}{2} \right)}
	\Bigl( \frac{m}{H_I} \Bigr)^{\!2} \biggr]
	\Bigl( \frac{a_2}{a} \Bigr)^{\frac{3}{2}-\nu}
\nonumber \\
&	+ \frac{(1\!-\!6\xi)(\nu_I\!+\!\nu\!-\!3)^2\, 
			\Gamma\!\left( \frac{3}{2}\!-\!\nu_I \right)
			\Gamma^2(\nu) \,
			\Gamma(2\nu_I\!+\!2\nu\!-\!4)}
		{3\,\Gamma^2\!\left( \nu_I \!+\! \nu\!-\!\frac{3}{2} \right)
			\Gamma\!\left( \nu_I \!+\! 2\nu \!-\! \frac{1}{2} \right)}	
	\, e^{-4N_R} \Bigl(\frac{a_2}{a} \Bigr)^{\!\frac{9}{2}-\nu}
	\Bigl( \frac{\mathcal{H}_0}{\mathcal{H}} \Bigr)^{2\nu_I-3} \Biggr\} \, .
\label{mattPfull}
\end{align}
The terms containing factors $\mathcal{H}_0/\mathcal{H}$ must be negligible
in order for us to have control over the approximation. 
The reason behind this lies in the IR regularization employed in this 
computation. We have introduced a sharp IR cutoff $k_0 \!=\! \mathcal{H}_0$
in comoving momentum space arguing that it is an approximation of the 
contribution coming from a full state (from all scales) that is smoothly suppressed
in the deep IR. The deep IR scales are assumed to contribute subdominantly,
and hence they were dropped right from the start by the introduction of this
cutoff. But once $\mathcal{H}\!<\!\mathcal{H}_0$ is reached by the cosmological
evolution it means that there are no more modes in the IR except the ones
below the cutoff scale $k_0$, so they are the only ones contributing
relevantly to the backreaction (this was treated in the massless limit
in \cite{Glavan:2014uga}). This is where our approximation breaks down.
The physical requirement that we make on the model is that the inflation 
lasts long enough for the expressions above to be reliable. This will indeed
be true for the cases of interest in this work.


\subsubsection{Minimally coupled limit}

Setting $\xi\!=\!0$ in \eqref{mattRHOfull} and \eqref{mattPfull}
produces the minimally coupled limit,
\begin{align}
\rho_Q \approx{}& \frac{3H_I^4}{32\pi^2} e^{(2\nu_I-3)N_I-4N_R}
	\Bigl( \frac{a_2}{a} \Bigr)^{\!3}
	+ \frac{3H_I^4}{16\pi^2} \biggl[ 1
	- e^{-\frac{2}{3} \frac{m^2}{H_I^2}N_I} \biggr] \, ,
\label{mattRHOmin}
\\
p_Q \approx{}& - \frac{3H_I^4}{16\pi^2} \biggl[ 1
	- e^{-\frac{2}{3} \frac{m^2}{H_I^2}N_I} \biggr] \, .
\label{mattPmin}
\end{align}
There are two types of contributions to the backreaction energy density and
pressure here. The first one scales like nonrelativistic matter, and redshifts
at the same rate as the background fluid driving the expansion.
It is a contribution that was found in the massless minimally coupled case
\cite{Glavan:2013mra}. The second contribution is of the CC type,
making it interesting in the context of DE scenarios. There are two limits one
can discuss regarding the second term, which we do in the following.

In the limit of very long inflation, $N_I\!\gg\! (m/H_I)^{-2}$, 
the CC-type contribution saturates to a maximum value during inflation,
and remains constant throughout expansion (provided $m\!\ll\! H$ always).
It is easy to see that it contributes dominantly to \eqref{mattRHOmin}
and \eqref{mattPmin} (since it does not redshift away),
\begin{equation}
\rho_Q \approx \frac{3H_I^4}{16\pi^2} \, , \qquad
	p_Q \approx - \rho_Q \, .
\end{equation}
This is a scenario that was suggested in \cite{Ringeval:2010hf},
although the rigorous computation has not been performed, 
which we supply in this paper.
It was found that in order for this limit
to work as a DE model (requiring $\rho_Q/\rho_B\!\sim\!1$),
{\it i.e.} for the backreaction to have the right value at late-time matter era,
one must considerably lower the inflationary Hubble scale
$(\hbar H_I)\!\lesssim\!6\!\times\! 10^{-3} \, {\rm eV}$
(corresponding to energy scale of inflation
$E_I \!\lesssim\! 5\! \times\! 10^3 \,{\rm GeV} $). 
The conditions $N_I\!\gg\!(m/H_I)^{-2}$ for very long inflation,
and $m/H_{\text{today}}\!<\!1$ then imply that
 $N_I\!\gtrsim\!10^{60}$.

In the limit of ``short'' inflation, $N_I \!\lesssim\! (m/H_I)^2$, the CC-type 
contribution in \eqref{mattRHOmin} and \eqref{mattPmin} does not
have enough time to reach its maximum value in inflation, but the value it has
at the end of inflation freezes throughout subsequent expansion 
(provided $m\!\ll\! H$ always). For short enough inflation the backreaction 
energy density and pressure at late times in matter era are
\begin{align}
\rho_Q \approx \frac{H_I^2m^2}{8\pi^2}N_I \, ,
	\qquad
p_Q \approx{} - \rho_Q \, .
\end{align}
The matter-like contribution never becomes important \cite{Glavan:2013mra}
compared to the background, and that is why we have neglected it above.
This is the result and the scenario suggested in \cite{Aoki:2014dqa}. 
It was found that in order to work as a DE scenario, one does not have to
lower the inflationary scale, $(\hbar H_I)\!\sim\!10^{13}\text{GeV}$, but still
a very long inflation is required, $N_I\!\gtrsim10^{13}$
(together with the mass being lighter than the Hubble rate today,
$m\!\lesssim\!10^{-33}\text{eV}$). The computations of the quantum backreaction 
in \cite{Aoki:2014dqa} were performed for inflationary and radiation periods,
but a rigorous computation for the matter period was missing
(even though the predicted result was correct), which is supplied
by the limit taken in this subsection.



\subsubsection{Limit $(m/H)^2\sim|\xi|\ll1$}

In this limit the leading contributions to the quantum backreaction energy density
and pressure are
\begin{align}
\rho_Q \approx{}& - \frac{3H_I^4}{32\pi^2} \,
	e^{8|\xi|N_I} \biggl[
	e^{-4N_R} \Bigl( \frac{a_2}{a} \Bigr)^{\!3}
	-  \frac{1}{6|\xi|}   
	\Bigl( \frac{m}{H_I} \Bigr)^{\!2} \biggr] \, ,
\label{matter era:rhoQ}
\\
p_Q \approx{}& - \frac{3H_I^4}{32\pi^2} \,
	e^{8|\xi|N_I} \biggl[
	  \frac{1}{6|\xi|}  
	\Bigl( \frac{m}{H_I} \Bigr)^{\!2} \biggr]  \, .
\label{matter era:pQ}
\end{align}
One can recognize two types of contributions -- first one that survives in the 
massless limit and scales like nonrelativistic matter, and another that behaves like 
the CC and depends on the mass. We want to discuss in which cases the CC-type
contribution dominates the backreaction, and under which conditions can it be
large enough to influence the background dynamics.

Firstly, since the first matter-like contribution behaves like a tracer solution
in matter and radiation era (see \eqref{radRHOlimit2} and \eqref{radPlimit2}),
its ratio compared to the background is determined by the ratio
at the end of inflation, which then freezes for subsequent evolution.
So, if the conditions of Figure \ref{InflConstraint} are met, if the backreaction
is small at the end of inflation ($\rho_Q/\rho_B\!\ll\!1$), 
this term will never be important. 

The second CC-type contribution does not redshift away and can become 
comparable to the background, under the condition
that $m\lesssim H_{\text{today}}$ (otherwise the field becomes heavy
and starts contributing like nonrelativistic matter). 
A more precise constraint on the scalar field mass is to compare it to the Hubble
rate at the onset of DE domination which we take to be the time of equality
of the CC and matter defined by $\rho_{\text{M}}\!=\!\rho_{\text{CC}}$.
It follows quickly from the first Friedmann equation and the density parameters
of CC and matter today, $\Omega_{\text{CC}}\!=\!0.68$ and 
$\Omega_{\text{M}}\!=\!0.32$ \cite{Adam:2015rua,Ade:2015xua}, that
\begin{equation}
H_{\text{DE}} = H_{\text{today}} \times 
	\sqrt{2\Omega_{\text{CC}}}  
	= 1.7 \!\times\! 10^{-33} {\rm eV} \, ,
\label{HDE}
\end{equation}
corresponding to $0.25$ e-foldings in the past of today (redshift $z\!=\!0.29$).
The measure of the strength of the backreaction at late times is the ratio,
\begin{equation}
\frac{\rho_Q}{\rho_B}
	= \frac{e^{8|\xi|N_I} G_NH_I^2}{24\pi |\xi|}
	\Bigl( \frac{m}{H_{\text{DE}}} \Bigr)^2 \, ,
\label{ratio}
\end{equation}
When this becomes of order one the backreaction starts dominating.
Up until that moment the quantum backreaction
behaves like a cosmological constant with positive energy density,
so its tendency would be to speed up the expansion rate. The exact details of
this process are a matter of performing the full self-consistent evolution,
but at least the initial tendency is clear.

\begin{figure}[h]
\includegraphics[width=10cm]{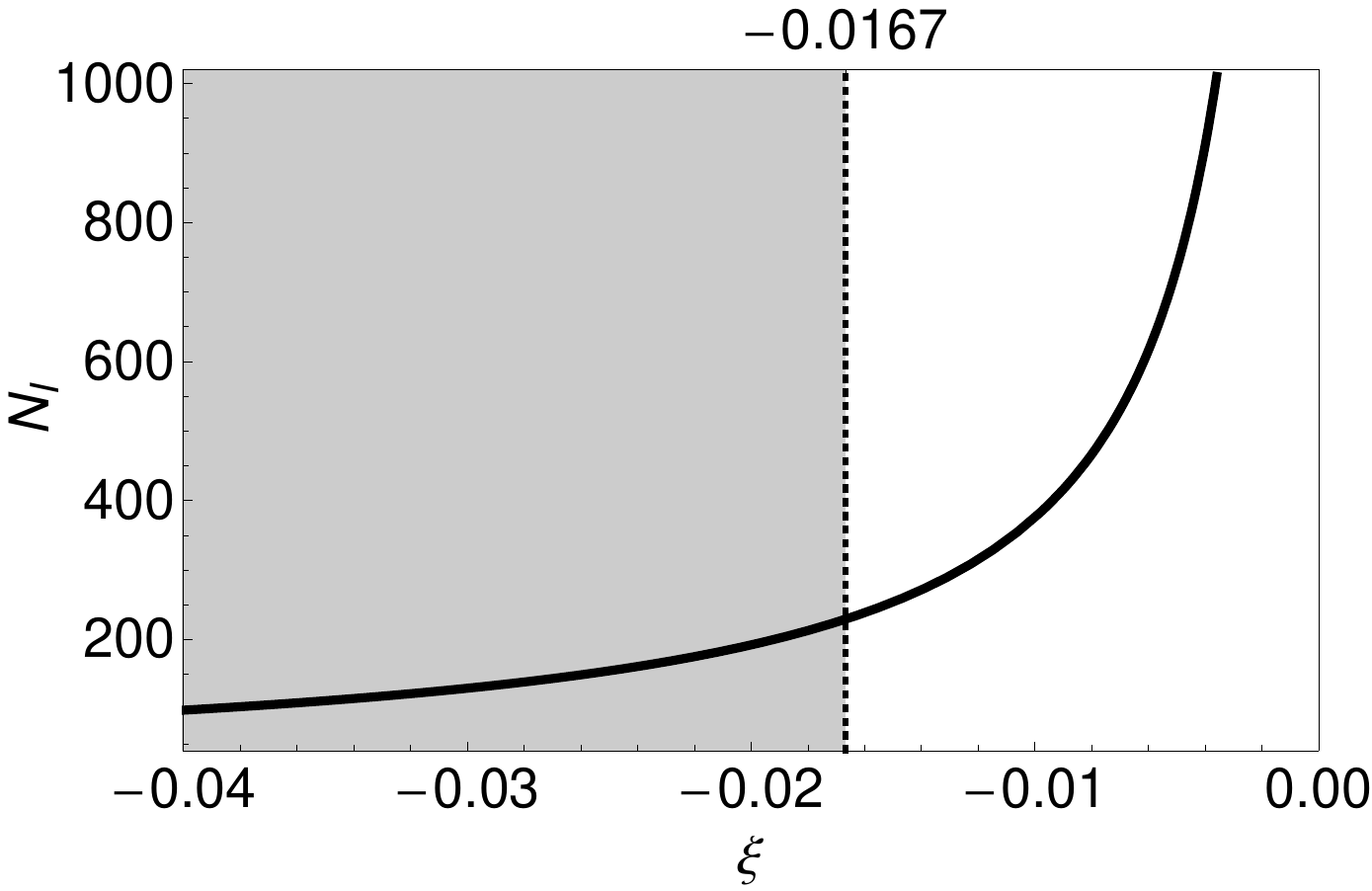}
\vskip-0.5cm
\caption{The relation between the nonminimal coupling parameter $\xi$
and the total number of e-foldings of inflation $N_I$ determined by
condition $\rho_Q/\rho_B\!=\!1$, and 
$(m/H_{\text{DE}})^2\!=\!0.1$, represented by the
bold curve. The shaded region is excluded by the requirements that quantum 
backreaction stays perturbative until late times (Fig.~\ref{InflConstraint}).}
\label{NImin}
\end{figure}

This model has three parameters -- $m$, $\xi$, and $N_I$ -- 
that are not completely independent. The mass is constrained to be smaller 
than the Hubble rate at the start of DE domination,
\begin{equation}
\frac{m}{H_{\text{DE}}}\!<\!1 \, .
\label{ExpansionRatio}
\end{equation} 
The nonminimal coupling $\xi$
and the duration of inflation $N_I$ are constrained in Fig. \ref{InflConstraint}
by the requirement that backreaction remains perturbative during inflation.
The crucial requirement for the model to work is that the ratio \eqref{ratio}
is of order one. This condition determines the number of e-foldings as a
function of the nonminimal coupling $\xi$ and the ratio \eqref{ExpansionRatio},
\begin{equation}
N_I = \frac{1}{8|\xi|} \ln \biggl[ 24\pi|\xi| 
	\Bigl( \frac{H_{\text{DE}}}{m} \Bigr)^2
	\Bigl(\frac{E_P}{\hbar H_I} \Bigr)^{\!2} \biggr] \, ,
\label{NIpredicted}
\end{equation}
where $E_P\!=\!(\hbar c^5/G_{\!N})^{1/2}$ is the Planck energy.
The plot of $N_I(\xi)$ for the case $(m/H_{\text{DE}})^2\!=\!0.1$ is shown in 
Fig.~\ref{NImin}.
Part of the curve in Fig.~\ref{NImin}
does not lie in the allowed region of Fig. \ref{InflConstraint}, and hence the model
does not work for the whole considered range of nonminimal couplings
$-0.1 \!<\! \xi \!<\! 0$, but still $0>\!\xi\!\sim\!-10^{-3}$ is allowed.
Of course, the limits on nonminimal coupling depend on \eqref{ExpansionRatio}.
In fact, we can derive this bound on nonminimal coupling dependent
on the ratio $m/H_{\text{DE}}$
by requiring that the predicted number of e-foldings \eqref{NIpredicted}
satisfies the constraint from inflation \eqref{boundInf}, which gives
\begin{equation}
\xi > - \frac{1}{6} \Bigl( \frac{m}{H_{\text{DE}}} \Bigr)^2 \, ,
\label{condition on xi}
\end{equation}
which does not depend on the inflationary scale $H_I$. 
Since for the case depicted in Fig.~\ref{NImin} 
$(m/H_{\text{DE}})^2\!=\!0.1$, 
the results should be trusted to the 10\%
level due to possible subleading mass corrections.
\vskip+0.3cm
%


\section{Discussion and outlook}
\label{sec: Discussion and outlook}

  In this paper we investigate the evolution of quantum fluctuations of a 
very light, non-minimally coupled, spectator scalar 
field~(\ref{action:massive nonminimal scalar}) 
throughout the history of the Universe, from the beginning of 
inflation and throughout 
inflation, radiation and matter era.
When the field couples to a fixed classical homogeneous and isotropic cosmological  
background (characterised by the Hubble rate as 
a function of time, $H\!=\!H(t)$), and when gravity is assumed non-dynamical ({\it i.e} the quantum gravitational effects 
are turned off), 
the relevant scalar field equation~(\ref{Klein-Gordon equation:1})  
is linear and can thus in principle be solved exactly, at least with the help of numerical methods.
Since here we are interested in the (one-loop quantum) backreaction 
of the scalar quantum fluctuations on
the background space-time, numerically solving our problem 
turns out to be tedious
(but can be done as in~\cite{Suen:1987gu}). 
Instead, here we resort to approximate analytical treatment.
Indeed, as illustrated in Fig.~\ref{expansion history},
it is convenient to split the history of the Universe into relatively long epochs, during which
the principal slow roll parameter, $\epsilon\!=\!-\dot H/H^2$, is to a good approximation constant, 
and relatively short transition periods,
during which $\epsilon$ rapidly changes. 
We show that, provided the transition periods are sufficiently short (for the 
transition in question that means that the characteristic time scale must be shorter than the Hubble time at the transition),
they can be treated in a sudden transition approximation. This is so because the main contribution 
to the energy-momentum tensor, that determines the backreaction on the background space-time, comes from the infrared modes
for which the sudden transition approximation applies. 

It is important to investigate how the results depend on the choice of initial state
(which when chosen naively suffers from IR divergences), and the IR regularization
method. We have argued in subsection~\ref{subsec: Choice of state} that the
three viable regularization method must give at least qualitatively the same 
answer, which was supported by comparisons of explicit computations done
using different schemes. Therefore,
the results of our analysis are {\it quite generic}, 
{\it i.e.} to a large extent independent on the choice of the initial state.

It is also important to emphasize that, in the process of calculating the one-loop energy-momentum tensor, 
we have used dimensional regularization to remove all divergences and that our final result for the 
renormalized energy-momentum tensor is finite and cutoff independent.
The cutoff independence is not trivial to 
achieve, since in the process of dimensional regularization one has to introduce an 
ultraviolet cutoff splitting the UV and IR parts. Regularization and renormalization procedures for the UV part
are outlined in Appendix~\ref{app: Regularization and renormalization}.
The cutoff dependence
introduced this way is only fiducial and cancels completely when the two parts 
are added up, resulting in a cutoff independent, fully regulated, finite answer.

It is now a good moment to state the most important results of this work. The one-loop late-time contribution 
to the energy density and pressure in matter era, in the limit when $|\xi|\ll 1$
and $m/H_{\text{DE}}\!<\!1$ are of the form
(see~(\ref{matter era:rhoQ}--\ref{matter era:pQ})),
\begin{equation}
\rho_Q \approx \frac{\hbar}{c^3} \times \frac{H_I^2}{8\pi^2 }
	 \Bigl( \frac{mc^2}{\hbar} \Bigr)^{\!2}  \!\times\!
	\frac{e^{8|\xi|N_I}}{8|\xi|} \, , \qquad p_Q \approx - \rho_Q \, ,
\label{summary:rhoQ:matter era:1}
\end{equation}
where $m$ is the scalar field mass, and $\xi$ its nonminimal coupling to
the Ricci scalar (see action \eqref{action:massive nonminimal scalar}),
$\hbar H_I\!\sim\!10^{13} {\rm GeV}$ is the inflationary Hubble rate.
Note that seemingly the one-loop energy density and pressure in
\eqref{summary:rhoQ:matter era:1} are proportional to $\hbar^{-1}$.
This is a consequence of the convention for writing the quadratic part
of the potential in terms of particle mass $m$. But $mc^2/\hbar$,
which appears in action \eqref{action:massive nonminimal scalar} 
instead of $m$ when dimensionful units are restored, should 
be considered as constant, and not as singular in the classical limit 
$\hbar\!\rightarrow \! 0$. This means \eqref{summary:rhoQ:matter era:1} is
proportional to $\hbar$ in the usual one-loop sense.

Since $\rho_Q$ is approximately constant and $p_Q \!\approx\! - \rho_Q$, this contribution can be 
{\it a good candidate for dark energy}. The principal goal of the upcoming work~\cite{GlavanProkopecStarobinsky:2015}
-- in which we study in the Gaussian approximation the self-consistent one-loop backreaction 
in the model~(\ref{action:massive nonminimal scalar}) -- 
is to establish whether this na\^ive proposition
is justified. Namely, the perturbative treatment employed in this work fails to be reliable when
the backreaction becomes significant, which is precisely when it becomes interesting from the dark energy point of 
view. The contribution~(\ref{summary:rhoQ:matter era:1})
can be compared with the energy density at the onset of DE domination, 
$\rho_B\!=\!3M_{\rm P}^2H^2_{\text{DE}}$
(where $\hbar H_{\rm DE}\!=\! 1.7\!\times\!10^{-33}{\rm eV}$
was defined in~\eqref{HDE}), to give,

\begin{equation}
\frac{\rho_Q}{\rho_B} \approx  
	\frac{1}{3\pi} \Bigl( \frac{\hbar H_I}{E_{\rm P}} \Bigr)^{\!2}
	\times \Bigl( \frac{m c^2}{\hbar H_{\text{DE}}} \Bigr)^{\!2}
	\times
	\frac{e^{8|\xi|N_I}}{8|\xi|} \, ,
\label{summary:rhoQ:matter era:2}
\end{equation}
where the last factor represents the enhancement factor due to the inflationary particle production.
In section~\ref{Matter era} we show (see Eq.~(\ref{condition on xi})) 
that the expression~(\ref{summary:rhoQ:matter era:2}) is valid provided the following inequalities are satisfied,
\begin{equation}
\frac{m c^2}{\hbar H_{\text{DE}}}<1 \, ,
	\qquad
	0>\xi>-\frac{1}{6} \Bigl( \frac{m c^2}{\hbar H_{\text{DE}}} \Bigr)^2
\,.
\label{ultralight mass and small xi}
\end{equation}
The enhancement factor $e^{8|\xi|N_I}/(8|\xi|)$ in Eq.~(\ref{summary:rhoQ:matter era:2}) needs to be sufficiently large to 
compensate the loop suppression factor, 
$(\hbar H_I)^2/[3\pi E_{\rm P}^2] \!=\! 
G_NH_I^2\hbar/[3\pi c^5] \!\sim\! 10^{-13}$,
and the factor $[m c^2/ (\hbar H_{\text{DE}})]^2$. The number of e-foldings of inflation required for that follows 
immediately from~(\ref{summary:rhoQ:matter era:2}),
\begin{equation}
N_I = \frac{1}{8|\xi|} \ln \biggl[ 24\pi|\xi| 
	\Bigl( \frac{\hbar H_{\text{DE}}}{mc^2} \Bigr)^{\!2}
	\Bigl(\frac{E_{\rm P}}{\hbar H_I} \Bigr)^{\!2} \biggr] \, ,
\label{number of efolds required}
\end{equation}
where $E_{\rm P}\!\simeq\!1.2\times 10^{19}~{\rm GeV}$ is the Planck energy. 
This number of e-foldings, also shown in Fig.~\ref{NImin}, must be consistent with the requirement 
that particle production does not lead to dominant contribution to the energy density
during inflation, {\it i.e.} must be consistent with the bound shown in Fig.~\ref{InflConstraint}.
Albeit these two conditions significantly restrict the allowed parameter space for $|\xi|$, one can show 
that, provided $\xi\!<\!0$ and $|\xi| \!\ll\! 1$, there still exists an ample set of allowed choices for $\xi$.

 In the following we discuss how our result~(\ref{number of efolds required}) for the required number of e-foldings 
in our model with negative nonminimal coupling $\xi$ compares with other models, in particular with the minimally coupled 
case (when $\xi\!=\!0$) studied in Ref.~\cite{Ringeval:2010hf,Aoki:2014dqa}.
In the minimally coupled case, the amplification factor
in \eqref{summary:rhoQ:matter era:1} is simply $N_I$, such that 
\begin{equation}
N_I=3\pi
	\Bigl( \frac{\hbar H_{\text{DE}}}{mc^2} \Bigr)^2
	\Bigl(\frac{E_{\rm P}}{\hbar H_I} \Bigr)^{\!2} 
\,.
\label{number of efolds required:2}
\end{equation}
While the number of e-foldings implied by 
Eq.~(\ref{number of efolds required}) is typically hundreds or thousands, the  
number of e-foldings  implied by the minimally coupled case~(\ref{number of efolds required:2}) 
is of the order or greater than $10^{13}$ (for $\hbar H_I\!\simeq\! 10^{13}~{\rm GeV}$),
which is many orders of magnitude larger. There is a simple explanation for this large difference:
while the rate of particle production in the minimally coupled case generates secular effects in the one-loop 
energy-momentum tensor that grow linearly in time,
due to the tachyonic scalar mass generated by a negative $\xi$ during inflation particle production rate in the
nonminimally coupled case generates an exponentially growing 
contribution to the one-loop energy-momentum tensor.
This, at first sight small difference in the model, has thus very important consequences 
for model building and arguably favors scalar models with negative non-minimal coupling as the model of choice 
for dark energy from inflationary quantum fluctuations.

 This paper is not the first to investigate the possibility that 
vacuum fluctuations of matter fields may be 
responsible for dark energy and here we make a cursory 
overview of these earlier works. 
Apart from our earlier papers~\cite{Glavan:2013mra,Glavan:2014uga},
in which we investigated the late-time one-loop quantum backreaction 
from one-loop inflationary fluctuations of 
massless and minimally and nonminimally coupled  scalars,
 and of gravitons,
probably the closest to our work are the papers of 
Ringeval, Suyama, Takahashi, Yamaguchi and Yokoyama~\cite{Ringeval:2010hf},
of Aoki and Iso~\cite{Aoki:2014dqa}, 
results of which we discussed in the previous paragraph, and
of Parker and 
Raval~\cite{Parker:1999td,Parker:1999ac,Parker:1999fc,
	Parker:2000pr,Parker:2001ws},
and of Parker and Vanzella~\cite{Parker:2002xa,Parker:2003as}.
Refs.~\cite{Ringeval:2010hf,Aoki:2014dqa} investigated the 
late-time one-loop quantum backreaction 
from one-loop inflationary fluctuations of very light minimally coupled scalars. 
The conclusion of Ref.~\cite{Ringeval:2010hf}
is that, provided inflation lasts long enough and it is at the right 
scale, the model can account for the observed dark energy,
while the conclusion of Ref.~\cite{Aoki:2014dqa}
(which improves on~\cite{Ringeval:2010hf} by performing a computation
similar to the one presented in this work) is that inflation 
can occur at the grand-unified scale and provided 
it lasts for about $10^{13}$ e-foldings (see~(\ref{number of efolds required:2})) 
scalar field fluctuations can account for the dark energy. 
However, in~\cite{Aoki:2014dqa}, a careful removal of all cut-off dependences, 
and a careful construction of approximate
mode functions in matter era are not accounted for, which we properly include in this paper.
 
  In a series of papers that appeared soon after the original supernovae results, 
Parker and Raval~\cite{Parker:1999td,Parker:1999ac,
	Parker:1999fc,Parker:2000pr,Parker:2001ws}
used the effective gravitational action obtained by integrating out the matter fields. 
The method used was the Schwinger-DeWitt proper time method and 
$\zeta$ function regularization~\cite{Parker:1984dj, Jack:1985mw,Bekenstein:1981xe},
which is an ultraviolet expansion of the effective action (that holds at 
short geodesic distances), and
the final effective action is presented as an expansion in powers of $1/M^2$, 
where $M$ is the effective mass of the field
(in the case of a non-minimally coupled scalar, 
$M^2\!=\!m^2\!+\!(\xi\!-\!\tfrac{1}{6}) R$, 
where $m$ is the tree-level scalar mass and 
$R$ denotes the Ricci scalar curvature).  In their effective action 
Parker and Raval maintain the terms of the order $M^4$, $M^2$, $M^0$, 
$M^4\ln(M^2/\mu^2)$, $M^2\ln(M^2/\mu^2)$, and 
$M^0\ln(M^2/\mu^2)$, thus neglecting the inverse powers of $M^2$.
Strictly speaking, this expansion applies (and hence the truncation 
is reasonable) when 
$M^2\!\gg\! \|{\rm Riem}\|, \|{\rm Ricc}\|, \| \square \|$, where $\|\cdot \|$ 
denote a suitably chosen norm.
A careful look at those papers reveals that the analysis was conducted 
strictly speaking where the expansion
does not apply, {\it i.e.} in the region of parameter space where 
$M^2$ is of the order or smaller than the components of e 
Ricci tensor, making the conclusions questionable
(in fact abundant particle production in the Parker-Raval model occurs 
when $m^2 \!\sim\! (\tfrac{1}{6}\!-\!\xi)R$,
which is precisely where $M^2\sim 0$, at which point the expansion 
used to construct the model is unreliable).
In contrast, the analysis in this work is performed in the opposite 
regime, namely in the regime when 
$m^2 \!\ll\! \|{\rm Riem}\|, \|{\rm Ricc}\|$. This is not just a technical 
point, but an essential 
assumption required to get abundant particle production during 
inflation that we need in order to get 
large quantum backreaction discussed in this paper.  
Making a more detailed comparison with the first paper~\cite{Parker:1999td} 
reveals that a large late-time backreaction was obtained when 
$(\xi\!-\!\tfrac{1}{6})\!<\!0$, {\it i.e.} when $\xi$ is close 
to, but smaller than conformal coupling, $\xi_c\!=\!\tfrac{1}{6}$. This is to be 
contrasted with our results,
which indicate that a large quantum backreaction is obtained only when 
both conditions, (1) $\xi\!<\!0$ and 
(2) there is a sufficiently long inflationary period preceding radiation 
era, are satisfied. Furthermore, the quantum backreaction 
in~\cite{Parker:1999td} becomes large (during matter era)
at a particular redshift $z_j$ given by 
$(1\!+\!z_j)^3 \!=\! m^2M_{\rm P}^2/[\rho_{m0}(1/6-\xi)]
=m^2\Omega_{m0}/[3H_0^2(\tfrac{1}{6}\!-\!\xi)]$ (here $\rho_{m0}$ 
denotes the matter density today and 
$\Omega_{m0}\!=\!\rho_{m0}/[3M_{\rm P}^2H_0] \!\simeq\! 0.3$ and 
$H_0$ is the Hubble parameter today), 
at which moment a large particle production occurs
due to an instability. We see no sign of this kind of instability, 
albeit in fairness to the reader we 
note that our analysis is perturbative and therefore we might 
not be able to see such an instability. 
In the remaining 
papers~\cite{Parker:1999ac,Parker:1999fc,Parker:2000pr,Parker:2001ws} 
the same effective action is used, and hence the same comparative analysis applies.


The results presented in this work provide a very good motivation for 
further studies of the quantum backreaction effects in connection to
the dark energy problem.
In particular, the model presented here
invites for more detailed studies that would:
\begin{itemize}
 \item
 treat the backreaction self-consistently,
\item
 confront the model with the currently existing dark energy data,
\item
 make forecasts that would test the model against the upcoming data, 
\item
examine the clustering properties of dark energy in this model, and
\item
address the issue of a very light scalar field mass required for the model to work.
\end{itemize}

By self-consistent treatment we mean solving self-consistently the 
quantum-corrected Friedmann equations 
(with the one-loop backreaction included) together with the scalar field equations. 
The perturbative treatment
executed in this work fails as soon as the one-loop backreaction terms 
become comparable with the background contributions $\rho_B$ and $p_B$. 
The idea of the upcoming work~\cite{GlavanProkopecStarobinsky:2015}
is to extend the Starobinsky's stochastic 
formalism~\cite{Starobinsky:1986fx,{Starobinsky:1994bd}} 
for inflation 
to subsequent epochs and solve the resulting scalar field equations 
together with the Friedmann equations (that include the backreaction from the 
field fluctuations). 
This will allow us to get detailed predictions on how the (global) Hubble parameter 
depends on time when the backreaction starts dominating, 
and study its dependence 
on the principal parameters of our model: $m$, $\xi$ and $N_I$. 
Of course the choice of these parameters 
is already quite limited by the constraints discussed at length in this 
work. The results of the self-consistent study 
can then be used 
to confront the model with the existing (and upcoming) data that already today put 
rather strict constraints on the recent evolution of dark energy.

While in the initial work in Ref.~\cite{GlavanProkopecStarobinsky:2015}
we intend to study the dynamics of the Universe in the Gaussian approximation, 
in the follow up work 
we intend to generalize that work and calculate (perturbatively) the 
higher order (non-Gaussian) 
correlators, which can be used to further test the model. Namely, one of the hot topics in current studies of dark energy models
involves the question whether dark energy clusters and if it does, 
how much. We expect 
that our model makes very specific predictions  
on how large non-Gaussian features in dark energy are today and how they evolve in recent times.
 
 Next, we would like to make theoretical improvements of the 
model discussed in this work.
In order for our model to work, the scalar field mass has to be very light.
More precisely, it has to satisfy~(\ref{ultralight mass and small xi}), 
which means that 
$m$ has to be smaller than the Hubble scale at the onset of DE domination, 
$(\hbar H_{\rm DE} ) \!\simeq\! 1.7\!\times\! 10^{-33}{\rm eV}$.
Unless one has a mechanism for generation of such a tiny mass scale,
this very light scalar mass remains a mystery of the model.
In future work we intend to investigate (dynamical) mechanisms that 
can shed light on the question why the scalar mass is so tiny today.

To conclude, in light of the results presented in this work it is clear that 
the possibility that inflationary quantum fluctuations could be 
responsible for the observed dark energy should be taken seriously. 
However, it is also clear that much more work is needed to put that 
idea on more solid foundations.


\section*{Acknowledgments}
\label{sec: Acknowledgments}

T.T. would like to thank Institute for Theoretical Physics at Utrecht University 
for the hospitality during the visit, where a part of this work has been done. 
The work of T.T.  is partially supported by JSPS KAKENHI Grant Number 15K05084  and 
MEXT KAKENHI Grant Number 15H05888. 
This work is part of the D-ITP consortium, a program of the Netherlands
Organisation for Scientific Research (NWO) that is funded by the Dutch
Ministry of Education, Culture and Science (OCW).
During the final stages of the project D.G. was supported by the grant
2014/14/E/ST9/00152 (National Science Centre, Poland).


\appendix


\section{Regularization and renormalization}
\label{app: Regularization and renormalization}

This appendix summarizes the dimensional regularization of the energy-momentum
tensor integrals \eqref{RHOintegral} and \eqref{Pintegral}, and the renormalization
procedure by introducing counterterms. It is comprised of well known results
\cite{Birrell:1982ix,Parker:2009uva}.

The first task is to isolate the divergences in \eqref{RHOintegral} and
\eqref{Pintegral}. In order to do that we examine just the UV parts of the
integrals,
\begin{align}
\rho_Q^{\text{UV},0} ={}&
	\frac{a^{-D}}{(4\pi)^{\frac{D-1}{2}} 
			\, \Gamma \! \left( \frac{D-1}{2} \right)}
	\int\limits_{\mu}^{\infty} \!\! dk  \, k^{D-2}
	\Biggl\{ 2k^2 |U|^2 
	- \frac{1}{2} \bigl[ D \!-\! 2 \!-\! 4\xi(D \!-\! 1) \bigr]
		\mathcal{H}' |U|^2
\nonumber \\
&	+ 2m^2a^2 |U|^2
	- \frac{1}{2} \bigl[ D \!-\! 2 \!-\! 4\xi(D \!-\! 1) \bigr] \mathcal{H}
		\frac{\partial}{\partial\eta} |U|^2
	+ \frac{1}{2} \frac{\partial^2}{\partial\eta^2} |U|^2 \Biggr\} \, ,
\label{RHOuv}
\\
p_Q^{\text{UV},0} ={}&
	\frac{\delta_{ij} a^{-D}}{(4\pi)^{\frac{D-1}{2}} 
			\, \Gamma \! \left( \frac{D-1}{2} \right)}
	\int\limits_{\mu}^{\infty} \!\! dk  \, k^{D-2}
	\Biggl\{ \frac{2k^2}{(D \!-\! 1)} |U|^2 
	- \frac{1}{2} \bigl[ D \!-\! 2 \!-\! 4\xi(D \!-\! 1) \bigr]
		\mathcal{H}' |U|^2
\nonumber \\
&	- \frac{1}{2} \bigl[ D \!-\! 2 \!-\! 4\xi(D \!-\! 1) \bigr] \mathcal{H}
		\frac{\partial}{\partial\eta} |U|^2
	+ \frac{(1 \!-\! 4\xi)}{2} \frac{\partial^2}{\partial\eta^2} |U|^2 \Biggr\} \, ,
\label{Puv}
\end{align}
where the UV scale $\mu$ is assumed to be larger than any other physical 
scale (such as the curvature and the mass of the scalar). 
The momentum in the integrands is then
much larger than any other physical scale appearing, whic is exploited in the following subsection.


\subsection{Mode function in the UV}

Therefore, we do
not need the exact momentum dependence in order to isolate the divergences,
its asymptotic expansion in powers of $1/k$ (where $k \!=\! \|\vec{k}\|$) 
is enough in order to accomplish 
the task. For that reason we first solve the equation of motion \eqref{modeEOM},
\begin{equation}
U''(k,\eta) + \Bigl[ k^2 + \mathcal{M}^2(\eta) \Bigr] U(k,\eta) = 0 \, ,
	\qquad k\rightarrow\infty \, ,
\label{EOMapp}
\end{equation}
in the UV limit. The WKB method is well suited for that. Here we implement
a somewhat simpler method, which is directly expanding the solution in
inverse powers of momenta,
\begin{equation}
U(k,\eta) = \frac{e^{-ik\eta}}{\sqrt{2k}} \Biggl[ 
	1 + \frac{iU_1(\eta)}{k} + \frac{U_2(\eta)}{k^2}
	+ \frac{iU_3(\eta)}{k^3} + \frac{U_4(\eta)}{k^4}
	+ \frac{iU_5(\eta)}{k^5} + \mathcal{O}(k^{-6}) \Biggr] \, ,
\label{UansatzUV}
\end{equation}
where we have assumed the mode function to be a pure positive-frequency one 
in the UV. The coefficient functions $U_i$ are all real, which does not seem to be 
a general ansatz, but it can be shown that their imaginary parts can always
be absorbed into a time-independent phase, which we can always add since it
does not show up in physical quantities. 

It is enough to determine $U_1$-$U_4$ for the purpose of isolating UV divergences
in \eqref{RHOuv} and \eqref{Puv}. Equation of motion they satisfy are obtained
by plugging ansatz \eqref{UansatzUV} into \eqref{EOMapp}, and organizing
it order by order in momenta,
\begin{align}
U_1'(\eta) ={}& - \frac{1}{2} \mathcal{M}^2(\eta) \, ,
\label{eom1}
\\
U_2'(\eta) ={}& \frac{1}{2} \Bigl[ U_1''(\eta) 
	+ \mathcal{M}^2(\eta) U_1(\eta) \Bigr] \, ,
\label{eom2}
\\
U_3'(\eta) ={}&  - \frac{1}{2} \Bigl[ U_2''(\eta) 
	+ \mathcal{M}^2(\eta) U_2(\eta) \Bigr] \, ,
\label{eom3}
\\
U_4'(\eta) ={}& \frac{1}{2} \Bigl[ U_3''(\eta) 
	+ \mathcal{M}^2(\eta) U_3(\eta) \Bigr] \, .
\label{eom4}
\end{align}
We also have to impose the Wronskian normalization \eqref{Wronskian},
which we do order by order as well,
\begin{align}
0 ={}& 2 U_2(\eta) + U_1^2(\eta) - U_1'(\eta) \, ,
\label{wr1}
\\
0 ={}& 2 U_4(\eta) + 2 U_3(\eta) U_1(\eta) + U_2^2(\eta)
	- U_3'(\eta) + U_1(\eta) U_2'(\eta) - U_1'(\eta) U_2(\eta) \, .
\label{wr2}
\end{align}
The solutions to equations of motion \eqref{eom1}-\eqref{eom4}, and 
constraints \eqref{wr1} and \eqref{wr2} are,
\begin{align}
U_1(\eta) ={}& - \frac{1}{2} \biggl[ 
	\int_{\eta_0}^{\eta}\! d\tau \mathcal{M}^2(\tau) \biggr] \, ,
\\
U_2(\eta) ={}& - \frac{\mathcal{M}^2(\eta)}{4}
	- \frac{1}{8}\biggl[ \int_{\eta_0}^{\eta}\! d\tau \mathcal{M}^2(\tau) 
	\biggr]^2 \, ,
\\
U_3(\eta) ={}& 
	\frac{1}{8} \Bigl( [\mathcal{M}^2(\eta)]' - [\mathcal{M}^2(\eta_0)]' \Bigr)
	+ \frac{\mathcal{M}^2(\eta)}{8} 
	\biggl[ \int_{\eta_0}^{\eta}\! d\tau \mathcal{M}^2(\tau) \biggr]
\nonumber \\
&	+ \frac{1}{48} \biggl[ \int_{\eta_0}^{\eta}\! d\tau \mathcal{M}^2(\tau) 
	\biggr]^3
	+ \frac{1}{8} \biggl[ \int_{\eta_0}^{\eta}\! d\tau \mathcal{M}^4(\tau) 
	\biggr]
\\
U_4(\eta) ={}& \frac{[\mathcal{M}^2(\eta)]''}{16} 
	+ \frac{5\mathcal{M}^4(\eta)}{32} 
	+ \frac{1}{16} \Bigl( [\mathcal{M}^2(\eta)]' - [\mathcal{M}^2(\eta_0)]' \Bigr)
		\biggl[ \int_{\eta_0}^{\eta}\! d\tau \mathcal{M}^2(\tau) \biggr]
\nonumber \\
&	+ \frac{\mathcal{M}^2(\eta)}{32} 
	\biggl[ \int_{\eta_0}^{\eta}\! d\tau \mathcal{M}^2(\tau) \biggr]^2
	+ \frac{1}{384} 
		\biggl[ \int_{\eta_0}^{\eta}\! d\tau \mathcal{M}^2(\tau) \biggr]^4
\nonumber \\
&	+ \frac{1}{16} \biggl[\int_{\eta_0}^{\eta}\! d\tau \mathcal{M}^2(\tau) \biggr]
		\biggl[ \int_{\eta_0}^{\eta}\! d\tau \mathcal{M}^4(\tau) \biggr] \, .
\end{align}
The dependence on an arbitrary time $\eta_0$ (which corresponds
to integration constants from \eqref{eom1}-\eqref{eom4}) is physically
irrelevant in the sense that it can be absorbed into the time-independent
overall phase which has no physical meaning. In fact, what we need to
calculate \eqref{RHOuv} and \eqref{Puv} is just the modulus squared,
\begin{equation}
|U(k,\eta)|^2 = \frac{1}{2k} \Biggl\{ 1 
	+ \frac{1}{k^2} \Bigl[ 2U_2(\eta) + U_1^2(\eta) \Bigr]
	+ \frac{1}{k^4} \Bigl[ 2U_4(\eta) + 2U_1(\eta) U_3(\eta)
		+ U_2^2(\eta) \Bigr] + \mathcal{O}(k^{-6}) \Biggr\} \, ,
\end{equation}
which is independent of all the integration constants,
\begin{equation}
|U(k,\eta)|^2 = \frac{1}{2k} \Biggl\{ 1 - \frac{\mathcal{M}^2(\eta)}{2k^2}
	+ \frac{1}{8k^4} \Bigl[ 3\mathcal{M}^4(\eta)
		+ [\mathcal{M}^2(\eta)]'' \Bigr] + \mathcal{O}(k^{-6}) \Biggr\} \, .
\label{U2expansion}
\end{equation}
now it is straightforward to evaluate integrals \eqref{RHOuv} and \eqref{Puv},
\begin{align}
\rho_Q^{UV,0} ={}& \frac{1}{16\pi^2a^4} \biggl( - \mu^4
	- \mu^2 \Bigl[ (1 \!-\! 6\xi)\mathcal{H}^2 + (ma)^2 \Bigr] \biggr)
\nonumber \\
&	+ \frac{\mu^{D-4}}{(D \!-\! 4)} 
	\times \frac{a^{-D}}{(4\pi)^{\frac{D-1}{2}} 
		\Gamma\left( \frac{D-1}{2} \right)} \Biggl\{ 
	\frac{(ma)^4}{8} 
	+ \frac{(D \!-\! 6)}{16} \Bigl[ D \!-\! 2 \!-\! 4\xi(D \!-\! 1) \Bigr] 
		(ma)^2 \mathcal{H}^2
\nonumber \\
&	+ \frac{1}{128} \Bigl[ D \!-\! 2 \!-\! 4\xi(D \!-\! 1) \Bigr]^2
	\Bigl[ 8 \mathcal{H}''\mathcal{H} - 4(\mathcal{H}')^2
		- 3(D \!-\! 2)^2\mathcal{H}^4 \Bigr] \Biggr\}
	+ \mathcal{O}(\mu^{-2}) \, ,
\label{rhoUV0}
\end{align}
\begin{align}
p_Q^{UV,0} ={}& \frac{1}{48\pi^2a^4} \biggl( - \mu^4
	+ \mu^2 \Bigl[ (1 \!-\! 6\xi)(2\mathcal{H}' - \mathcal{H}^2) 
	+ (ma)^2 \Bigr] \biggr)
\nonumber \\
&	+ \frac{\mu^{D-4}}{(D \!-\! 4)} 
	\times \frac{a^{-D}}{(4\pi)^{\frac{D-1}{2}(D\!-\!1)} 
		\Gamma\left( \frac{D-1}{2} \right)} \Biggl\{ 
	-\frac{3(ma)^4}{8} 
	- \frac{(D \!-\! 6)}{16} \Bigl[ D \!-\! 2 \!-\! 4\xi(D \!-\! 1) \Bigr] 
		(ma)^2 (2\mathcal{H}' \!+\! \mathcal{H}^2)
\nonumber \\
&	+ \frac{1}{128} \Bigl[ D \!-\! 2 \!-\! 4\xi(D \!-\! 1) \Bigr]^2
	\Bigl[ -8\mathcal{H}''' + 8 \mathcal{H}''\mathcal{H} - 4(\mathcal{H}')^2
\nonumber \\
&	\hspace{3cm}
		+ 12(D \!-\! 2)^2\mathcal{H}'\mathcal{H}^2
		-3(D\!-\!2)^2 \mathcal{H}^4 \Bigr] \Biggr\}
	+ \mathcal{O}(\mu^{-2}) \, .
\label{pUV0}
\end{align}
%


\subsection{Counterterms}

The action for the counterterms, which include a cosmological constant one,
a Newton's constant one, and a higher-derivative one, is given by
\begin{equation}
S_{\text{ct}} = \int\! d^D \! x \sqrt{-g} 
	\Biggl\{ \frac{G_{\!N} \Delta G_{\!N}^{-1}}{16\pi G_{\!N}} R
	- \frac{2 \Delta\Lambda}{16\pi G_N} 
	+ (\alpha \!+\! \Delta\alpha) R^2 \Biggr\} \, ,
\end{equation}
from where its their contribution to the energy-momentum tensor follows,
\begin{equation}
T_{\mu\nu}^{\text{ct}} = \frac{-2}{\sqrt{-q}} 
	\frac{\delta S_{\text{ct}}}{\delta g^{\mu\nu}}
	= \frac{\Delta\Lambda}{8\pi G_{\! N}} g_{\mu\nu}
	- \frac{\Delta G_{\!N}^{-1}}{8\pi} G_{\mu\nu}
	+ (\alpha \!+\! \Delta\alpha) {}^{\scriptscriptstyle{(1)}\!}H_{\mu\nu} \, ,
\end{equation}
where
\begin{equation}
{}^{\scriptscriptstyle{(1)}\!}H_{\mu\nu}
	= 4 \nabla_\mu \nabla_\nu R - 4g_{\mu\nu} \square R
	+ g_{\mu\nu} R^2 - 4 R_{\mu\nu} R \, .
\end{equation}
Specialized to FLRW space-time the counterterms contribution to the 
energy-momentum tensor is diagonal,
\begin{align}
\rho_{\text{ct}} ={}& \frac{\Delta\Lambda}{8\pi G_{\!N}}
	+ \frac{\Delta G_{\!N}^{-1}}{16\pi} 
		(D \!-\! 1)(D \!-\! 2) \frac{\mathcal{H}^2}{a^2} 
\nonumber \\
&	+ (\alpha \!+\! \Delta\alpha) \frac{(D \!-\! 1)^2}{a^4}
	\Bigl[ -8\mathcal{H}''\mathcal{H} + 4(\mathcal{H}')^2
	- 8(D \!-\! 4) \mathcal{H}'\mathcal{H}^2 
	- (D \!-\! 2)(D \!-\! 10) \mathcal{H}^4 \Bigr] \, ,
\label{rhoCT}
\\
p_{\text{ct}} ={}& - \frac{\Delta\Lambda}{8\pi G_{\!N}}
	+ \frac{\Delta G_{\!N}^{-1}}{16\pi} 
		\frac{(D \!-\! 2)}{a^2} 
		\Bigl[ 2 \mathcal{H}' + (D \!-\! 3)\mathcal{H}^2 \Bigr]
\nonumber \\
&	+ (\alpha \!+\! \Delta\alpha) \frac{(D \!-\! 1)}{a^4} 
	\Bigl[ 8 \mathcal{H}''' + 8(2D \!-\! 9)\mathcal{H}''\mathcal{H}
	+ 4(3D \!-\! 11)(\mathcal{H}')^2
\nonumber \\
&	\hspace{3cm} 
	+ 12(D^2 \!-\! 10D \!+\! 20) \mathcal{H}'\mathcal{H}^2
	+ (D \!-\! 2)(D \!-\! 5)(D \!-\! 10) \mathcal{H}^4 \Bigr] \, .
\label{pCT}
\end{align}
We choose the coefficients of the counterterms to be
\begin{align}
\Delta\Lambda ={}& - \frac{G_{\!N} m^4}
		{4(4\pi)^{\frac{D-3}{2}} \, \Gamma\!\left( \frac{D-1}{2} \right)}
	\frac{\mu^{D-4}}{(D\!-\!4)} \, ,
\\
\Delta G_{\! N}^{-1} ={}&
	\frac{(D \!-\! 6) [D \!-\! 2 \!-\! 4\xi(D \!-\! 1)]m^2}
	{8 (4\pi)^{\frac{D-3}{2}} (D\!-\!2) \, \Gamma\!\left( \frac{D-1}{2} \right)}
	\frac{\mu^{D-4}}{(D \!-\! 4)} \, ,
\\
\Delta\alpha ={}& 
	\frac{[D \!-\! 1 \!-\! 4\xi(D \!-\! 1)]^2}
	{128(D \!-\! 1)^2 (4\pi)^{\frac{D-1}{2}} 
		\, \Gamma\!\left( \frac{D-1}{2} \right)}
	\frac{\mu^{D-4}}{(D \!-\! 4)} \, .
\end{align}
in order to absorb the divergences from \eqref{rhoUV0} and \eqref{pUV0}.
The finite parts of these coefficients were picked for convenience
so as to cancel as much of finite parts of the bare expectation value.


\subsection{Renormalized UV contribution}

The renormalized UV contribution to energy density and pressure is obtained by 
adding the counterterms' contribution \eqref{rhoCT} and \eqref{pCT}
to the bare contribution \eqref{rhoUV0} and \eqref{pUV0},
\begin{align}
\rho_Q^{UV} ={}& \frac{1}{16\pi^2a^4} \biggl\{ - \mu^4 
	- \mu^2 \Bigl[ (1 \!-\! 6\xi) \mathcal{H}^2 + (ma)^2 \Bigr] \biggr\}
	- \frac{m^4}{32\pi^2} \ln(a) 
	+ \frac{(1 \!-\! 6\xi) m^2}{16\pi^2a^2} \mathcal{H}^2 \ln(a)
\nonumber \\
&	+ \frac{(1 \!-\! 6\xi)^2}{32\pi^2 a^4} 
	\biggl\{ \Bigl[ \ln(a)+ \widetilde{\alpha} \Bigr]
	\Bigl[ -2\mathcal{H}''\mathcal{H} \!+\! 
		(\mathcal{H}')^2 \!+\! 3 \mathcal{H}^4 \Bigr]
	-2 (\mathcal{H}'\mathcal{H}^2 \!+\! \mathcal{H}^4) \biggr\} \, ,
\label{rhoQUV}
\end{align}
\begin{align}
p_Q^{UV} ={}& \frac{1}{48\pi^2a^4} \biggl\{ - \mu^4
	+ \mu^2 \Bigl[ (1 \!-\! 6\xi)(2\mathcal{H}' \!-\! \mathcal{H}^2)
		+ (ma)^2 \Bigr] \biggr\}
\nonumber \\
&	+ \frac{m^4}{32\pi^2} \Bigl[ \ln(a) + \frac{1}{3} \, \Bigr]
	 - \frac{(1 \!-\! 6\xi) m^2}{48\pi^2a^2} \Bigl[ 
	(2\mathcal{H}' \!+\! \mathcal{H}^2) \ln(a) + \mathcal{H}^2 \Bigr]
\nonumber \\
& + \frac{(1 \!-\! 6\xi)^2}{96\pi^2a^4} \biggl\{ 
	\Bigl[ \ln(a) + \widetilde{\alpha} \Bigr]
	\Bigl[ 2\mathcal{H}''' \!-\! 2\mathcal{H}''\mathcal{H}
	\!+\! (\mathcal{H}')^2 \!-\! 12\mathcal{H}'\mathcal{H}^2
	\!+\! 3\mathcal{H}^4 \Bigr]
\nonumber \\
&	\hspace{2cm}
	+ 4\mathcal{H}''\mathcal{H} \!+\! 3(\mathcal{H}')^2
	\!+\! 6 \mathcal{H}'\mathcal{H}^2 \!-\! 5\mathcal{H}^4 \biggr\} \, ,
\label{pQUV}
\end{align}
where $\widetilde{\alpha}=1152\alpha/(1 \!-\! 6\xi)^2$. Specialized to 
constant $\epsilon$ FLRW backgrounds these contributions are
\begin{align}
\rho_Q^{UV} ={}& \frac{1}{16\pi^2a^4} \biggl\{ - \mu^4 
	- \mu^2 \Bigl[ (1 \!-\! 6\xi) \mathcal{H}^2 + (ma)^2 \Bigr] \biggr\}
	- \frac{m^4}{32\pi^2} \ln(a) 
	+ \frac{(1 \!-\! 6\xi) m^2}{16\pi^2a^2} \mathcal{H}^2 \ln(a)
\nonumber \\
&	+ \frac{(1 \!-\! 6\xi)^2}{32\pi^2} (2 \!-\! \epsilon)
	\frac{\mathcal{H}^4}{a^4}
	\biggl\{ 3 \epsilon\Bigl[ \ln(a)+ \widetilde{\alpha} \Bigr] - 2 \biggr\} \, ,
\label{rhoUVconstEpsilon}
\end{align}
\begin{align}
p_Q^{UV} ={}& \frac{1}{48\pi^2a^4} \biggl\{ - \mu^4
	+ \mu^2 \Bigl[ (1 \!-\! 6\xi)(1 \!-\! 2\epsilon) \mathcal{H}^2
		+ (ma)^2 \Bigr] \biggr\}
	+ \frac{m^4}{32\pi^2} \Bigl[ \ln(a) + \frac{1}{3} \, \Bigr]
\nonumber \\
&	 - \frac{(1 \!-\! 6\xi) m^2}{48\pi^2} \frac{\mathcal{H}^2}{a^2} \Bigl[ 
	(3 \!-\! 2\epsilon) \ln(a) + 1 \Bigr]
\nonumber \\
& + \frac{(1 \!-\! 6\xi)^2}{96\pi^2} (2 \!-\! \epsilon) \frac{\mathcal{H}^4}{a^4} 
	\biggl\{ 
	-3\epsilon(3 \!-\! 4\epsilon) \Bigl[ \ln(a) + \widetilde{\alpha} \Bigr]
	+ (6 \!-\! 11\epsilon) \biggr\} \, .
\label{pUVconstEpsilon}
\end{align}

Had we performed the renormalization procedure on arbitrary curved backgrounds,
we would have found another contribution that breaks the classical conformal
invariance -- the conformal anomaly \cite{Capper:1974ic,Dowker:1976zf},
\begin{align}
\rho_{\text{CA}} ={}& \frac{1}{2880\pi^2a^4} 
	\Bigl[ 2\mathcal{H}''\mathcal{H} \!-\! (\mathcal{H}')^2 \Bigr] \, ,
\\
p_{\text{CA}} ={}& - \frac{1}{8640\pi^2a^4}
	\Bigl[ 2\mathcal{H}''' \!-\! 2\mathcal{H}''\mathcal{H} 
		\!+\! (\mathcal{H}')^2 \Bigr] \, ,
\label{CA}
\end{align}
which should be added to the contributions above.
Specialized to constant $\epsilon$ backgrounds it is
\begin{equation}
\rho_{\text{CA}} = \frac{(1 \!-\! \epsilon)^2 \mathcal{H}^4}{960 \pi^2 a^4} \, ,
	\qquad p_{\text{CA}} = 
	- \frac{(1 \!-\! \epsilon)^2(3 \!-\! 4\epsilon)\mathcal{H}^4}
		{2880\pi^2a^4} \, .
\label{CAeps}
\end{equation}
%



\begin{thebibliography}{99}


\bibitem{Perlmutter:1998np}
  S.~Perlmutter {\it et al.} [Supernova Cosmology Project Collaboration],
  ``Measurements of Omega and Lambda from 42 high redshift supernovae,''
  Astrophys.\ J.\  {\bf 517} (1999) 565
  [astro-ph/9812133].


\bibitem{Riess:1998cb}
  A.~G.~Riess {\it et al.} [Supernova Search Team Collaboration],
  ``Observational evidence from supernovae for an accelerating universe and a cosmological constant,''
  Astron.\ J.\  {\bf 116} (1998) 1009
  [astro-ph/9805201].


\bibitem{Adam:2015rua}
  R.~Adam {\it et al.} [Planck Collaboration],
  ``Planck 2015 results. I. Overview of products and scientific results,''
  arXiv:1502.01582 [astro-ph.CO].


\bibitem{Ade:2015xua}
  P.~A.~R.~Ade {\it et al.} [Planck Collaboration],
  ``Planck 2015 results. XIII. Cosmological parameters,''
  arXiv:1502.01589 [astro-ph.CO].


\bibitem{Amendola:2011bo}
  L.~Amendola and S.~Tsujikawa, 
  ``Dark Energy: Theory and Observations,'' 
  Cambridge University Press (2011).


\bibitem{Clifton:2011jh} 
  T.~Clifton, P.~G.~Ferreira, A.~Padilla and C.~Skordis,
 ``Modified Gravity and Cosmology,''
 Phys.\ Rept.\  {\bf 513}, 1 (2012)
  [arXiv:1106.2476 [astro-ph.CO]].


\bibitem{Peebles:1998qn}
  P.~J.~E.~Peebles and A.~Vilenkin,
  ``Quintessential inflation,''
  Phys.\ Rev.\ D {\bf 59} (1999) 063505
  [astro-ph/9810509].


\bibitem{GarciaBellido:2011de}
  J.~Garcia-Bellido, J.~Rubio, M.~Shaposhnikov and D.~Zenhausern,
  ``Higgs-Dilaton Cosmology: From the Early to the Late Universe,''
  Phys.\ Rev.\ D {\bf 84} (2011) 123504
  [arXiv:1107.2163 [hep-ph]].


\bibitem{Kolb:2005me}
  E.~W.~Kolb, S.~Matarrese, A.~Notari and A.~Riotto,
  ``Primordial inflation explains why the universe is accelerating today,''
  hep-th/0503117.


\bibitem{Kolb:2005da}
  E.~W.~Kolb, S.~Matarrese and A.~Riotto,
  ``On cosmic acceleration without dark energy,''
  New J.\ Phys.\  {\bf 8} (2006) 322
  [astro-ph/0506534].


\bibitem{Hirata:2005ei}
  C.~M.~Hirata and U.~Seljak,
  ``Can superhorizon cosmological perturbations explain the acceleration of the Universe?,''
  Phys.\ Rev.\ D {\bf 72} (2005) 083501
  [astro-ph/0503582].


\bibitem{Glavan:2013mra}
  D.~Glavan, T.~Prokopec and V.~Prymidis,
  ``Backreaction of a massless minimally coupled scalar field from inflationary quantum fluctuations,''
  Phys.\ Rev.\ D {\bf 89} (2014) 2,  024024
  [arXiv:1308.5954 [gr-qc]].


\bibitem{Aoki:2014ita}
  H.~Aoki, S.~Iso and Y.~Sekino,
  ``Evolution of vacuum fluctuations generated during and before inflation,''
  Phys.\ Rev.\ D {\bf 89} (2014) 10,  103536
  [arXiv:1402.6900 [hep-th]].


\bibitem{Glavan:2014uga}
  D.~Glavan, T.~Prokopec and D.~C.~van der Woude,
  ``Late-time quantum backreaction from inflationary fluctuations of a nonminimally coupled massless scalar,''
  Phys.\ Rev.\ D {\bf 91} (2015) 2,  024014
  [arXiv:1408.4705 [gr-qc]].


\bibitem{Finelli:2008zg}
  F.~Finelli, G.~Marozzi, A.~A.~Starobinsky, G.~P.~Vacca and G.~Venturi,
  ``Generation of fluctuations during inflation: Comparison of stochastic and field-theoretic approaches,''
  Phys.\ Rev.\ D {\bf 79} (2009) 044007
  [arXiv:0808.1786 [hep-th]].


\bibitem{Finelli:2010sh}
  F.~Finelli, G.~Marozzi, A.~A.~Starobinsky, G.~P.~Vacca and G.~Venturi,
  ``Stochastic growth of quantum fluctuations during slow-roll inflation,''
  Phys.\ Rev.\ D {\bf 82} (2010) 064020
  [arXiv:1003.1327 [hep-th]].


\bibitem{Ringeval:2010hf}
  C.~Ringeval, T.~Suyama, T.~Takahashi, M.~Yamaguchi and S.~Yokoyama,
  ``Dark energy from primordial inflationary quantum fluctuations,''
  Phys.\ Rev.\ Lett.\  {\bf 105} (2010) 121301
  [arXiv:1006.0368 [astro-ph.CO]].


\bibitem{Aoki:2014dqa}
  H.~Aoki and S.~Iso,
  ``Evolution of Vacuum Fluctuations of an Ultra-Light Massive Scalar Field generated during and before Inflation,''
  arXiv:1411.5129 [gr-qc].


\bibitem{Parker:1999td}
  L.~Parker and A.~Raval,
  ``Nonperturbative effects of vacuum energy on the recent expansion of the universe,''
  Phys.\ Rev.\ D {\bf 60} (1999) 063512
   [Phys.\ Rev.\ D {\bf 67} (2003) 029901]
  [gr-qc/9905031].


\bibitem{Parker:1999ac}
  L.~Parker and A.~Raval,
  ``Vacuum effects of ultralow mass particle account for recent acceleration of universe,''
  Phys.\ Rev.\ D {\bf 60} (1999) 123502
   [Phys.\ Rev.\ D {\bf 67} (2003) 029902]
  [gr-qc/9908013].


\bibitem{Parker:1999fc}
  L.~Parker and A.~Raval,
  ``Vacuum driven metamorphosis,''
  gr-qc/9908069.


\bibitem{Parker:2000pr}
  L.~Parker and A.~Raval,
  ``New quantum aspects of a vacuum dominated universe,''
  Phys.\ Rev.\ D {\bf 62} (2000) 083503
   [Phys.\ Rev.\ D {\bf 67} (2003) 029903]
  [gr-qc/0003103].


\bibitem{Parker:2001ws}
  L.~Parker and A.~Raval,
  ``A New Look at the Accelerating Universe,''
  Phys.\ Rev.\ Lett.\  {\bf 86} (2001) 749.


\bibitem{Birrell:1982ix}
  N.~D.~Birrell and P.~C.~W.~Davies,
  ``Quantum Fields in Curved Space,''  Cambridge Monographs on Mathematical Physics (1984).


\bibitem{Fulling:1989nb}
  S.~A.~Fulling,
  ``Aspects of Quantum Field Theory in Curved Space-time,''
  London Math.\ Soc.\ Student Texts {\bf 17} (1989) 1.


\bibitem{Mukhanov:2007zz}
  V.~Mukhanov and S.~Winitzki,
  ``Introduction to quantum effects in gravity,'' 
  Cambridge University Press, Cambridge (2007)


\bibitem{Parker:2009uva}
  L.~E.~Parker and D.~Toms,
  ``Quantum Field Theory in Curved Spacetime : Quantized Field and Gravity,''
  Cambridge Monographs on Mathematical Physics, Cambridge (2009)


\bibitem{Chernikov:1968zm}
  N.~A.~Chernikov and E.~A.~Tagirov,
  ``Quantum theory of scalar fields in de Sitter space-time,''
  Annales Poincare Phys.\ Theor.\ A {\bf 9} (1968) 109.


\bibitem{Parker:1969au}
  L.~Parker,
  ``Quantized fields and particle creation in expanding universes. 1.,''
  Phys.\ Rev.\  {\bf 183} (1969) 1057.

\bibitem{Parker:1971pt}
  L.~Parker,
  ``Quantized fields and particle creation in expanding universes. 2.,''
  Phys.\ Rev.\ D {\bf 3} (1971) 346
   [Phys.\ Rev.\ D {\bf 3} (1971) 2546].

\bibitem{Zeldovich:1971mw}
  Y.~B.~Zeldovich and A.~A.~Starobinsky,
  ``Particle production and vacuum polarization in an anisotropic gravitational field,''
  Sov.\ Phys.\ JETP {\bf 34} (1972) 1159
   [Zh.\ Eksp.\ Teor.\ Fiz.\  {\bf 61} (1971) 2161].


\bibitem{Janssen:2009nz}
  T.~M.~Janssen and T.~Prokopec,
  ``Regulating the infrared by mode matching: A Massless scalar in expanding spaces with constant deceleration,''
  Phys.\ Rev.\ D {\bf 83} (2011) 084035
  [arXiv:0906.0666 [gr-qc]].


\bibitem{Tsamis:1992xa}
  N.~C.~Tsamis and R.~P.~Woodard,
  ``The Structure of perturbative quantum gravity on a De Sitter background,''
  Commun.\ Math.\ Phys.\  {\bf 162} (1994) 217.


\bibitem{Tsamis:2002qk}
  N.~C.~Tsamis and R.~P.~Woodard,
  ``Plane waves in a general Robertson-Walker background,''
  Class.\ Quant.\ Grav.\  {\bf 20} (2003) 5205
  [astro-ph/0206010].


\bibitem{Bunch:1978yq}
  T.~S.~Bunch and P.~C.~W.~Davies,
  ``Quantum Field Theory in de Sitter Space: Renormalization by Point Splitting,''
  Proc.\ Roy.\ Soc.\ Lond.\ A {\bf 360} (1978) 117.


\bibitem{Kulsrud:1957zz}
  R.~M.~Kulsrud,
  ``Adiabatic Invariant of the Harmonic Oscillator,''
  Phys.\ Rev.\  {\bf 106} (1957) 205.


\bibitem{Ford:1977in}
  L.~H.~Ford and L.~Parker,
  ``Infrared Divergences in a Class of Robertson-Walker Universes,''
  Phys.\ Rev.\ D {\bf 16} (1977) 245.


\bibitem{gradshteyn2007}
	Gradshteyn, I. S. and Ryzhik, I. M.
	``Table of integrals, series, and products, Seventh Edition,''
	Elsevier/Academic Press, Amsterdam, 2007


\bibitem{NIST:DLMF}
  ``NIST Digital Library of Mathematical Functions'',
	http://dlmf.nist.gov/, Release 1.0.10 of 2015-08-07


\bibitem{Arfken}
	G.~B.~Arfken and H.~J.~Weber,
	``Mathematical methods for physicists; 6th ed.,''
	Elsevier, Oxford (2005)


\bibitem{Habib:1999cs}
  S.~Habib, C.~Molina-Paris and E.~Mottola,
  ``Energy momentum tensor of particles created in an expanding universe,''
  Phys.\ Rev.\ D {\bf 61} (2000) 024010
  [gr-qc/9906120].


\bibitem{Starobinsky:1986fx}
  A.~A.~Starobinsky,
  ``Stochastic De Sitter (inflationary) Stage In The Early Universe,''
  Lect.\ Notes Phys.\  {\bf 246} (1986) 107.


\bibitem{Suen:1987gu}
  W.~M.~Suen and P.~R.~Anderson,
  ``Reheating in the Higher Derivative Inflationary Models,''
  Phys.\ Rev.\ D {\bf 35} (1987) 2940.


\bibitem{GlavanProkopecStarobinsky:2015}
	D.~Glavan, T.~Prokopec and A.~A.~Starobinsky
	``Stochastic approach to late-time quantum backreaction,''
	{\sl in preparation}


\bibitem{Parker:2002xa}
  L.~Parker, W.~Komp and D.~Vanzella,
  ``Cosmological acceleration through transition to constant scalar c curvature,''
  Astrophys.\ J.\  {\bf 588} (2003) 663
  [astro-ph/0206488].

\bibitem{Parker:2003as}
  L.~Parker and D.~A.~T.~Vanzella,
  ``Acceleration of the universe, vacuum metamorphosis, and the large time asymptotic form of the heat kernel,''
  Phys.\ Rev.\ D {\bf 69} (2004) 104009
  [gr-qc/0312108].


\bibitem{Parker:1984dj}
  L.~Parker and D.~J.~Toms,
  ``New Form for the Coincidence Limit of the Feynman Propagator, or Heat Kernel, in Curved Space-time,''
  Phys.\ Rev.\ D {\bf 31} (1985) 953.

\bibitem{Jack:1985mw}
  I.~Jack and L.~Parker,
  ``Proof of Summed Form of Proper Time Expansion for Propagator in Curved Space-time,''
  Phys.\ Rev.\ D {\bf 31} (1985) 2439.


\bibitem{Bekenstein:1981xe}
  J.~D.~Bekenstein and L.~Parker,
  ``Path Integral Evaluation of Feynman Propagator in Curved Space-time,''
  Phys.\ Rev.\ D {\bf 23} (1981) 2850.


\bibitem{Starobinsky:1994bd}
  A.~A.~Starobinsky and J.~Yokoyama,
  ``Equilibrium state of a selfinteracting scalar field in the De Sitter background,''
  Phys.\ Rev.\ D {\bf 50} (1994) 6357
  [astro-ph/9407016].


\bibitem{Capper:1974ic}
  D.~M.~Capper and M.~J.~Duff,
  ``Trace anomalies in dimensional regularization,''
  Nuovo Cim.\ A {\bf 23} (1974) 173.


\bibitem{Dowker:1976zf}
  J.~S.~Dowker and R.~Critchley,
  ``The Stress Tensor Conformal Anomaly for Scalar and Spinor Fields,''
  Phys.\ Rev.\ D {\bf 16} (1977) 3390.



\end{thebibliography}





\end{document}